\documentclass[11pt,a4paper,reqno]{amsart}
\usepackage[scale=0.809,marginratio={1:1, 1:1},ignoreall]{geometry}

\usepackage{amsmath,amssymb}

\usepackage{setspace}

\usepackage{proof-at-the-end}
\usepackage{hyperref}

\usepackage{enumitem}
\usepackage[round]{natbib}

\theoremstyle{plain}
\newtheorem{thm}{Theorem}[section]

\theoremstyle{definition}

\newtheorem{example}[thm]{Example}
\newtheorem{Def}[thm]{Definition}

\theoremstyle{remark}
\newtheorem*{Rem}{Remark}

\numberwithin{equation}{section}

\DeclareMathOperator{\spt}{spt}

\DeclareMathOperator*{\argmax}{arg\;\!max}
\newcommand{\bdsb}{\boldsymbol}
\newcommand{\cplm}[1]{{#1}^{\mathsf{c}}}
\newcommand{\supnorm}[1]{\|{#1}\|_{\infty}}

\newcommand{\rcvx}[1]{{#1}^{\mathbb{Q}}}
\newcommand{\witilde}{\widetilde}
\newcommand{\wihat}{\widehat}
\newcommand{\nc}{\mathrm{nc}}
\newcommand{\co}{\mathrm{co}}

\begin{document}


\title[Evolutionarily Stable Preferences Against Multiple Mutations]{Evolutionarily Stable Preferences Against Multiple Mutations in Multi-player Games}
\author{Yu-Sung Tu}
\address{Institute of Economics\\
         Academia Sinica\\
         Taipei 115201, Taiwan}
\email[Yu-Sung Tu]{yusungtu@gate.sinica.edu.tw}
\author{Wei-Torng Juang} 
\address{Department of Economics\\
         Soochow University\\
         Taipei 100006, Taiwan} 
\email[Wei-Torng Juang]{wjuang@scu.edu.tw}

\keywords{Evolution of preferences, indirect evolutionary approach, multi-mutation stability, multi-player games, strict union equilibrium}

\begin{abstract}
We use the indirect evolutionary approach to study evolutionarily stable preferences against multiple mutations
in single- and multi-population matching settings, respectively.
Players choose strategies to maximize their subjective preferences,
which may be inconsistent with the material payoff function giving them the actual fitness values.
In each of the two settings, $n$-player games are played,
and we provide necessary and sufficient conditions for multi-mutation stability.
These results definitively reveal the connection between the order of stability and the level of efficiency.
\end{abstract}
\maketitle


\section{Introduction}


Traditionally, economic analysis treats individual preferences as exogenous. 
To address this issue, 
the \emph{indirect evolutionary approach}, pioneered by \citet{wGut-mYaa:erbssg} and \citet{wGut:eaecbri}, 
established a rigorous theoretical foundation by distinguishing subjective preferences from objective material payoffs. 
While individuals strategically choose actions to maximize their subjective preferences, 
natural selection ultimately operates on the resulting material payoffs.
This framework has successfully explained the persistence of non-materialistic preferences that appear inconsistent with material self-interest, 
such as altruism, vengeance, punishment, fairness, and reciprocity.\footnote{%
See, for example, \citet{wGut-mYaa:erbssg}, \citet{wGut:eaecbri}, \citet{hBes-wGut:iaes}, \citet{sHuc-jOec:ieaefa}, and \citet{eOst:caesn}.}

Despite its rich insights, most existing literature on preference evolution focuses on symmetric two-player interactions within a single population,\footnote{%
Exceptions can be found: \citet[p.~61]{svW:eneup} gives a definition of stability for two separate populations and shows some illustrative examples;
\citet{ysTu-wtJuang:epmp} study the evolution of general preferences in multiple populations under various degrees of observability.} 
requiring robustness against at most a single mutation at a time (e.g., \cite{eDek-jEly-oYil:ep}).\footnote{%
The exceptions include \citet{fHer-cKuz:esdo} and \citet{yHel-eMoh:cdp}, who consider stability against polymorphic groups of mutants.}
However, 
economic applications rarely conform to these simplifying assumptions. 
First, 
real-world economic interactions frequently involve multilateral partnerships across asymmetric populations where individuals occupy distinct institutional roles, 
such as workers, managers, and investors, or buyers, sellers, and platform intermediaries. 
Second, 
economic systems are regularly exposed to multifaceted systemic shocks or the simultaneous entry of diverse behavioral norms. 
Moreover, under \emph{neutral stability},\footnote{%
The concept of a \emph{neutrally stable strategy}, 
which means that mutants may coexist with the incumbents but they will not displace the incumbents, 
was introduced by \citet{jMay:etg}.} 
incumbents often coexist with invading mutants. 
This naturally raises a fundamental economic question: 
\emph{Can an incumbent population withstand subsequent waves of heterogeneous preference shocks,\footnote{%
Regarding standard evolutionary game theoretic models,
\citet{mvV:rii} studies the question of how a neutral mutant affects a neutrally stable strategy.}  
and what level of economic efficiency is required to guarantee long-term stability against such higher-order invasions? }

To address these questions, 
this paper establishes a comprehensive framework to analyze evolutionarily stable preferences 
against multiple mutations in both single-population and multi-population matching settings for general $n$-player games. 
Under preference observability, 
non-materialistic preferences can serve as effective commitment devices to gain fitness advantages, 
ensuring that stable outcomes inherit certain efficiency properties 
(e.g., \citet{eDek-jEly-oYil:ep} and \citet{ysTu-wtJuang:epmp}). 
Extending this logic to higher-order stability, 
we prove that the order of multi-mutation stability is fundamentally bound to the level of material Pareto efficiency. 
As the number of invading mutant types grows, the potential for complex coalitional interactions expands, 
requiring higher efficiency to prevent incumbent displacement.

Crucially, 
our analysis reveals striking structural differences between two-player and multi-player games. 
In two-player games, 
sufficiently efficient strict Nash equilibria remain robust against any multiple mutations. 
In multi-player games, however, 
even highly efficient strict Nash equilibria fail to guarantee multi-mutation stability 
as coalitional deviations by a subset of players may leave the deviators unpunished.  
Strikingly, stronger concepts like \emph{strictly strong Nash equilibrium} also prove insufficient in multi-population settings.\footnote{%
The concept of a strong Nash equilibrium was introduced by \citet{rAum:apgcng}.} 
We construct a three-population counterexample showing that a strictly strong Nash equilibrium maximizing total material payoff can still be destabilized 
because payoff reallocations within specific mutant coalitions can fully compensate individual losses.

To ensure stability against such complex combinations of arbitrary mutant coalitions, 
we introduce the concept of a \emph{strict union equilibrium}.\footnote{%
The concept of strict union equilibria was introduced by \citet{ysTu-wtJuang:epmp}.} 
This equilibrium concept extends strictly strong Nash equilibrium to a transferable utility setting 
by requiring that any coalition strictly decreases its collective material payoff upon deviation. 
We prove that a strict union equilibrium provides a universal shield against any arbitrary number of mutations in both matching settings. 
To obtain more precise results,
we introduce two analogues,
the \emph{strict $k$-type union equilibrium} and
the \emph{strict $(n - 1)$-union equilibrium},
which will be used to derive sufficient conditions for lower-order stability in the single- and multi-population cases, respectively. 
Notably, although our framework contains no pre-game bargaining or coalition formation mechanism, 
our stability characterization bridges non-cooperative preference evolution with cooperative game theory, 
establishing that multi-mutation stable outcomes mirror the core of a normal-form game with transferable utility 
(e.g., \cite{rAum:ccgwsp} and \citet{dRay-rVoh:eba}).

The key conceptual distinction between the single- and multi-population matching settings lies in self-matching. 
A preference type in the single-population case would meet itself. 
Consequently, 
stability requires only symmetric efficiency, even within a polymorphic population. 
By contrast, individuals in the multi-population case interact exclusively with opponents from distinct populations, 
necessitating a more general notion of efficiency to characterize stability. 
Furthermore, 
holding the number of invading mutant types constant across settings, 
multi-population matching yields a far wider array of strategic interactions, 
which renders lower-efficiency outcomes significantly more vulnerable to destabilization.

The paper proceeds as follows. 
Section~\ref{sec:basic} introduces essential notions and standard notation used throughout this work.
Section~\ref{sec:single} analyzes evolutionarily stable preferences against multiple mutations in a single-population matching setting, 
establishing necessary and sufficient conditions for various orders of stability. 
Section~\ref{sec:multi} extends this analysis to the multi-population case. 
Section~\ref{sec:conclusion} concludes. 
For completeness, 
Appendix~\ref{sec:sym-inv} addresses the existence of a type-homogeneous equilibrium, 
and Appendix~\ref{sec:alt-def} explores an alternative definition of multi-mutation stability under single-population matching. 
Additionally, Appendix~\ref{sec:counterexamples} 
provides supplementary discussion on the multi-mutation stability of strictly strong Nash equilibria. 
Finally, Appendix~\ref{sec:proofs} collects all proofs.


\section{Essential Notions and Notation}\label{sec:basic}


\subsection*{The Underlying Game and Preference Types} 

In this paper, we consider $n$-player games.
Let $N = \{1, \dots, n\}$ be a finite set of players,
and for each $i\in N$, let $A_i$ be a finite set of actions available to player~$i$.
For any finite set $S$,
we use $\Delta S$, or $\Delta(S)$, to denote the set of probability distributions over $S$.
An element $\bdsb{\sigma}\in \prod_{i\in N}\Delta A_i$ is known as a \emph{mixed-strategy profile}.
More generally,
a \emph{correlated strategy} for matched individuals is defined as an element $\varphi$ in $\Delta(\prod_{i\in N}A_i)$.
Every mixed-strategy profile $\bdsb{\sigma} = (\sigma_1, \dots, \sigma_n)$ can be interpreted as a correlated strategy $\varphi_{\bdsb{\sigma}}$ in the following way:
$\varphi_{\bdsb{\sigma}}(a_1, \dots, a_n) = \prod_{i\in N} \sigma_i(a_i)$ for all $(a_1, \dots, a_n)\in \prod_{i\in N}A_i$.
We will call this correlated strategy the \emph{induced correlated strategy of} $\bdsb{\sigma}$.\footnote{%
In a finite normal-form game,
the set of all mixed-strategy profiles can be put into one-to-one correspondence with the set of all induced correlated strategies.}

For $i\in N$, let $\pi_i\colon \prod_{j\in N}A_j\to \mathbb{R}$ be player~$i$'s material payoff function,
which can be extended by taking expectations to a continuous function
defined on the set $\prod_{j\in N}\Delta A_j$, or on the set $\Delta(\prod_{j\in N}A_j)$.
More precisely,
given a correlated strategy $\varphi\in \Delta(\prod_{j\in N}A_j)$,
player~$i$'s material payoff to $\varphi$ is $\pi_i(\varphi) = \sum_{\bdsb{a}\in \prod_{j\in N}A_j} \varphi(\bdsb{a}) \pi_i(\bdsb{a})$.
Besides,
we have $\pi_i(\bdsb{\sigma}) = \pi_i(\varphi_{\bdsb{\sigma}})$
for each $\bdsb{\sigma}\in \prod_{j\in N}\Delta A_j$ and its induced correlated strategy $\varphi_{\bdsb{\sigma}}$.
We interpret the value $\pi_i(\bdsb{a})$ as the reproductive fitness that player~$i$ obtains if an action profile $\bdsb{a}\in \prod_{j\in N}A_j$ is played.
The evolutionary success or failure of a preference type is determined by its average fitness.
Combining the $n$ players' material payoff functions,
we get the vector-valued material payoff function $\pi\colon \prod_{i\in N}A_i\to \mathbb{R}^n$
that assigns to each action profile $\bdsb{a}$ the $n$-tuple $(\pi_1(\bdsb{a}),\dots,\pi_n(\bdsb{a}))$ of fitness values,
which we call the \emph{pure-payoff profile}.
This material payoff function $\pi$ can also be extended to the set $\prod_{i\in N}\Delta A_i$,
or to the set $\Delta(\prod_{i\in N}A_i)$, through $\pi_1$, \dots,~$\pi_n$.
Under these,
we denote the underlying game by $(N, (A_i), (\pi_i))$, which may be either symmetric or asymmetric.\footnote{%
We use $(N, (A_i), (\pi_i))$ to denote the normal-form game $(N, (A_i)_{i\in N}, (\pi_i)_{i\in N})$.}
In addition, for notational brevity,
we will simply write $\mathcal{F}_{\nc}$ and $\mathcal{F}_{\co}$ to denote, respectively,
the \emph{noncooperative payoff region} $\pi( \prod_{i\in N}\Delta A_i )$ and the \emph{cooperative payoff region} $\pi\big( \Delta(\prod_{i\in N}A_i) \big)$.
The region $\mathcal{F}_{\co}$ achievable under correlated strategies is the convex hull of the payoff region $\mathcal{F}_{\nc}$ achievable under mixed-strategy profiles.\footnote{%
Clearly, the cooperative payoff region can also be formed by constructing the convex hull of the set $\pi(\prod_{i\in N}A_i)$ of pure-payoff profiles.}

Let $\Theta$ be the set of von Neumann--Morgenstern utility functions defined on $\prod_{i\in N}A_i$.
A utility function $\theta\in \Theta$, referred to as a \emph{preference type} or simply as a \emph{type},
is identified with a group of individuals who have such a preference relation and make the same decisions.
However,
different individuals having the same preferences may choose different strategies to maximize their preferences.
In such a case,
any two of these individuals' types can be represented by two different utility functions congruent modulo a positive affine transformation.
A preference type is said to be \emph{indifferent} if it is a constant utility function.
This means that an individual having indifferent preferences is indifferent among all alternatives.
In the stability analysis,
we allow for polymorphic incumbent populations;
not all individuals in an incumbent population have the same preferences over potential outcomes.
We assume that a population contains a finite number of preference types,
but the number of individuals in a population is imagined to be infinite.\footnote{%
Under this assumption, we can let the population share of a mutant type be arbitrarily small.}
A population will then be represented by a finite-support distribution defined over $\Theta$.

\subsection*{Multi-mutation Stability Concept} 

Throughout the paper, preferences are perfectly observable,
and all players are drawn independently to play $n$-player games depending on the setting of random matching.
We study in Section~\ref{sec:single} the single-population case:
individuals are drawn uniformly from a single population without identifying their player positions.
In Section~\ref{sec:multi}, we study the multi-population case:
individuals are drawn from $n$ populations, one from each population uniformly with its player role.
In any game round,
$n$ players choose strategies to maximize their subjective preferences induced by their types,
but their actual fitness values are assigned by the material payoff function $\pi$ according to players' chosen strategies.
We imagine that such a game is repeated independently an infinite number of times,
under the same distribution of types in a single population or in multiple populations.
Then the average fitness of a type can be determined,\footnote{%
\label{footnote:1018}
The calculation of average fitness will indicate that the type distribution is unchanged in the process of learning to play an equilibrium.
To justify it, we assume that behavioral adjustments are infinitely faster than the adjustment of type distribution,
which is supported by \citet[p.~21]{rSel:eleb}.} 
and the principle of survival of the fittest suggests that
a preference type will survive only if the type receives the highest average fitness in its population.

This paper examines the evolutionary stability of preferences against polymorphic groups of mutants that enter simultaneously, 
introducing new preference types and behavioral variants. 
One might argue that distinct mutant types are unlikely to appear simultaneously. 
Consequently, one might consider a weaker concept of stability, 
in which multiple mutations enter sequentially rather than simultaneously, 
provided that no mutant is strictly outperformed at any step in the sequence. 
However, 
such a sequential framework is functionally equivalent to our simultaneous approach once indifferent preference types are accounted for. 
Specifically, 
if mutant types possessing distinct yet indifferent preferences enter the environment, 
they can analytically be treated as entering sequentially. 
Because these mutants can mimic incumbent behavior at intermediate steps, 
they naturally coexist with the incumbents throughout the transition.

To explore the effects of multiple mutations on preference evolution,
the criteria for multi-mutation stability in the single- and multi-population cases are defined in alignment with 
\citet{eDek-jEly-oYil:ep} and \citet{ysTu-wtJuang:epmp}, respectively, 
in which only monomorphic groups of mutants are considered. 
The state of one or multiple populations is characterized by a \emph{configuration}, 
which comprises a finite-support distribution of types and a behavioral rule specifying a Nash equilibrium for each match.
Beyond the requirement that all incumbents in a population earn equal average fitness, 
a configuration is stable if for any sufficiently small invasion, 
it satisfies the following conditions: 
\begin{itemize}
  \item The post-entry aggregate outcome remains virtually unchanged. 
  \item No incumbent type goes extinct. 
\end{itemize} 
Under this stability concept, we first require that in any post-entry environment, 
incumbents maintain their pre-entry behavior against one another. 
Second, we require that no incumbent type in a population earns the lowest average fitness 
unless all individuals in that population earn the same average fitness. 
That is, if an incumbent type in a population earns the lowest average fitness, 
all individuals in that population earn the same average fitness.  
Logically, this condition is equivalent to stating that 
either at least one mutant type in a population earns a strictly lower average fitness than all incumbents, 
or all individuals in that population earn the same average fitness. 
We formalize multi-mutation stability based on this latter formulation.

We treat a \emph{mutation distribution} as a \emph{unit} of mutation. 
A set of mutant types fails to invade under a given distribution 
if at least one of these mutant types goes extinct in a population, even in the multi-population setting.\footnote{%
This concept is consistent with that of the multi-population ESS formulated by \citet{rCre:scegt}.}
The underlying rationale is that if certain mutants gain a fitness advantage over incumbents at the expense of other mutants, 
this advantage vanishes as soon as the latter go extinct.

Here, 
as is common in evolutionary theory, 
there are no restrictions on mutants’ preferences, 
and they may choose any possible action. 
Thus, we allow for the coexistence of mutants and incumbents in a stable configuration. 
In addition, for a given underlying game, 
we focus on evolutionarily viable outcomes rather than the emergence of a specific preference type.


\section{The Single-population Case}\label{sec:single}


In this section, the $n$ individuals for each match are drawn independently from a single population,
and cannot identify their positions in the game they play based on their preferences. 
Naturally, 
let all players have the same action set,
and suppose that the actual fitness to a player depends only on the strategies being played and not on who is playing which.
In other words, the underlying game is \emph{symmetric},
which we now proceed to define formally.

\subsection*{Symmetry in \texorpdfstring{$n$}{n}-player Environments} 

A \emph{permutation} $\tau$ of $N$ is a bijection of the set $N$ onto itself.
We say that a normal-form game $(N, (A_i), (\pi_i))$ is \emph{symmetric with respect to} $\tau$ if
$A_{\tau(i)} = A_i$ for all $i\in N$, and
$\pi_{\tau(i)}(a_1, \dots, a_n) = \pi_i(a_{\tau(1)}, \dots, a_{\tau(n)})$ for all $i\in N$ and
all $(a_1, \dots, a_n)\in \prod_{i\in N}A_i$. 
(Such a permutation $\tau$ is called a \emph{symmetry} of the game.) 
A normal-form game is \emph{symmetric} if it is symmetric with respect to every permutation of $N$.

\begin{Def}\label{def:20200526}
A normal-form game $(N, (A_i), (\pi_i))$ is said to be \emph{symmetric} if $A_i = \bar{A}$ for all $i\in N$,
and for any permutation $\tau$ of $N$, the equality
\[
\pi_{\tau(i)}(a_1, \dots, a_n) = \pi_i(a_{\tau(1)}, \dots, a_{\tau(n)})
\]
holds for all $i\in N$ and all $(a_1, \dots, a_n)\in \bar{A}^n$.\footnote{%
The condition
$\pi_{\tau(i)}(a_1, \dots, a_n) = \pi_i(a_{\tau(1)}, \dots, a_{\tau(n)})$ for all $i\in N$ is obviously equivalent to
$\pi_i(a_1, \dots, a_n) = \pi_{\tau^{-1}(i)}(a_{\tau(1)}, \dots, a_{\tau(n)})$ for all $i\in N$.
Note that \citet{pDas-eMas:eedeg} give the following definition: a game is symmetric if
$A_1 = \dots = A_n = \bar{A}$,
and $\pi_i(a_1, \dots, a_n) = \pi_{\tau(i)}(a_{\tau(1)}, \dots, a_{\tau(n)})$
for all $i\in N$,
all permutations $\tau$ of $N$,
and all $(a_1, \dots, a_n)\in \bar{A}^n$.
Their definition is not equivalent to ours except for the case of two-player games.}
\end{Def}

We shall write $(N, \bar{A}, (\pi_i))$ to denote the symmetric $n$-player underlying game with the common action set $\bar{A}$.
In every game round,
players choose strategies from $\bar{A}$ to maximize their own subjective preferences
depending only on the strategies chosen by the other players and not on the identities of the players choosing them.
This reflects that individuals have no awareness of the player positions.
Therefore,
the subjective utility that a preference type $\theta$ receives by choosing a strategy $a_0$ against $n - 1$ opponents' strategies $a_1$, \dots,~$a_{n-1}$
is specially denoted by $\theta(a_0\mid a_1, \dots, a_{n-1})$ with the property that for any permutation $\tau$ of $\{1, \dots, n - 1\}$,
the equality
\begin{equation}\label{eq:20200717}
\theta(a_0\mid a_1, \dots, a_{n-1}) = \theta(a_0\mid a_{\tau(1)}, \dots, a_{\tau(n - 1)})
\end{equation}
holds for all $a_0$, $a_1$, \dots,~$a_{n-1}\in \bar{A}$, 
which means that the subjective utility is invariant under permutations of the opponents' strategies. 
Let $\Theta_s$ be the set of von Neumann--Morgenstern utility functions defined on $\bar{A}^n$ satisfying~\eqref{eq:20200717},
and let $\mathcal{M}(\Theta_s)$ be the set of finite-support distributions over $\Theta_s$. 
The probability distribution of preference types in the population is represented by a type distribution $\mu\in \mathcal{M}(\Theta_s)$. 
Let $\Gamma(\mu)$ denote the game in which $n$ types are drawn independently according to $\mu$,
and then each player chooses a strategy from the common set $\bar{A}$ that is a subjective best reply to the strategies chosen by their opponents. 
Combining this with the underlying game $(N, \bar{A}, (\pi_i))$,
the pair $[(N, \bar{A}, (\pi_i)), \Gamma(\mu)]$ is referred to as an \emph{environment}.

Given $\mu\in \mathcal{M}(\Theta_s)$,
we denote the finite support of $\mu$ by $\spt \mu$.
Let $\bdsb{\theta} = (\theta_1, \dots, \theta_n)$ be a matched $n$-tuple where each player $i$ has the type $\theta_i\in \spt\mu$. 
Then the normal-form game induced by their preferences is $(N, (A_i), (u^{\bdsb{\theta}}_i))$,
where each $A_i = \bar{A}$ and 
each payoff function $u^{\bdsb{\theta}}_i\colon \bar{A}^n\to \mathbb{R}$ is defined by
$u^{\bdsb{\theta}}_i(a_1, \dots, a_n) = \theta_i(a_i \mid \bdsb{a}_{-i})$. 
For this induced game,
let $b_i(\theta_1, \dots, \theta_i, \dots, \theta_n)\in \Delta \bar{A}$ be the strategy used by type $\theta_i$ of player~$i$ 
against $n - 1$ types $\theta_1$, \dots,~$\theta_{i-1}$, $\theta_{i+1}$, \dots,~$\theta_n$.
Since player positions are irrelevant to players in such a game,\footnote{%
Let $\tau^{-1}(i) = j$. 
Then in the normal-form game induced by $\bdsb{\theta}' = (\theta_{\tau(1)}, \dots, \theta_{\tau(n)})$, 
player~$j$ is of type $\theta_i$ and 
his payoff function should be defined by 
$u^{\bdsb{\theta}'}_j(a_1, \dots, a_n) = \theta_{i}(a_j \mid \bdsb{a}_{-j})$.} 
the strategy used by type~$\theta_{i}$ would remain unchanged whether $\theta_{i}$ is assigned in the position $i$ or in any other position.
That is,
players' strategies would necessarily imply that
\begin{equation}\label{eq:20230208}
b_{i}(\theta_1, \dots, \theta_n) = b_{\tau^{-1}(i)}(\theta_{\tau(1)}, \dots, \theta_{\tau(n)})
\end{equation}
for every $i\in N$ and every permutation $\tau$ of $N$. 
Let $b_1$, \dots,~$b_n$ satisfy~\eqref{eq:20230208}. 
Define $b\colon (\spt\mu)^n\to (\Delta \bar{A})^n$ by $b(\bdsb{\theta}) = (b_1(\bdsb{\theta}),\dots,b_n(\bdsb{\theta}))$.
We say that the strategy function $b$ is an \emph{equilibrium} in the game $\Gamma(\mu)$ if
for all $\bdsb{\theta} = (\theta_1, \dots, \theta_n)$ in $(\spt\mu)^n$,
the strategy profile $b(\bdsb{\theta})$ is a Nash equilibrium of the normal-form game induced by $\bdsb{\theta}$,
that is, for each $i\in N$,
\[
b_i(\bdsb{\theta})\in \argmax_{\sigma\in\Delta \bar{A}} \theta_i\big( \sigma\mid b_{-i}(\bdsb{\theta}) \big).
\] 
Here, $b_{-i}(\bdsb{\theta})$ denotes the strategy profile that $\theta_i$'s opponents play. 
The set of all such equilibria in $\Gamma(\mu)$ is denoted by $B(\spt\mu)$.

Note that equation~\eqref{eq:20230208} assures us that 
players of the same type in the induced normal-form game use the same strategy.\footnote{%
Let $\theta_i = \theta_j$, and consider the permutation $\tau$ of $N$ defined by
$\tau(i) = j$, $\tau(j) = i$, and $\tau(k) = k$ for all $k\neq i$,~$j$.
Then we have $(\theta_{\tau(1)}, \dots, \theta_{\tau(n)}) = (\theta_1, \dots, \theta_n)$,
which by~\eqref{eq:20230208} implies that $b_{i}(\theta_1, \dots, \theta_n) = b_{j}(\theta_1, \dots, \theta_n)$.}
Thus to ensure that the set $B(\spt\mu)$ is nonempty, we have to check that 
for any matched $n$-tuple $(\theta_1, \dots, \theta_n)$,
there exists a Nash equilibrium $b(\theta_1, \dots, \theta_n)$ satisfying 
$b_i(\theta_1, \dots, \theta_n) = b_j(\theta_1, \dots, \theta_n)$ if $\theta_i = \theta_j$, 
which is a so-called \emph{type-homogeneous} equilibrium. 
Indeed, 
this follows from the fact that 
any finite normal-form game has a \emph{symmetry-invariant} equilibrium, 
which is due to Theorem~2 of \citet{jNash:ncg}. 
Appendix~\ref{sec:sym-inv} provides the details.

\subsection*{Configurations} 

In addition to the finite-support type distribution $\mu$ over $\Theta_s$,
the state of the population is further described by a behavioral rule $b\in B(\spt\mu)$,
which specifies a Nash Equilibrium for each match of types in $\spt \mu$.
We call the pair $(\mu, b)$ a \emph{configuration}.
Let $\theta_{i_1}$, \dots,~$\theta_{i_k}$ be types drawn independently from the population. 
We will write $\mu^{k}(\theta_{i_1}, \dots, \theta_{i_k})$ in place of 
the probability $\prod_{j\in \{i_1, \dots, i_k\}} \mu(\theta_j)$ that $\theta_{i_1}$, \dots,~$\theta_{i_k}$ are matched.\footnote{%
Here we use the assumption that the population contains an infinite number of individuals.} 
By the law of large numbers,
the \emph{average fitness} of a type $\theta\in \spt\mu$ with respect to $(\mu,b)$ is 
\[
\varPi_{\theta}(\mu;b) = \sum_{\bdsb{\theta}'_{-1}\in (\spt\mu)^{n-1}} \mu^{n-1}(\bdsb{\theta}'_{-1}) \pi_1\big( b(\theta, \bdsb{\theta}'_{-1}) \big),
\]
on which the evolution of preferences depends.
A configuration $(\mu,b)$ is said to be \emph{balanced} if all preference types in the population have the same average fitness, that is,
$\varPi_{\theta}(\mu;b) = \varPi_{\theta'}(\mu;b)$ for all $\theta$,~$\theta'\in \spt\mu$.
Because a configuration may be polymorphic, 
we define the \emph{aggregate outcome} of $(\mu,b)$ to be the correlated strategy $\varphi_{\mu,b}\in \Delta(\bar{A}^n)$ defined by
\[
\varphi_{\mu,b}(a_1, \dots, a_n) = \sum_{\bdsb{\theta}\in (\spt\mu)^n} \mu^{n}(\bdsb{\theta}) \prod_{i\in N} b_i(\bdsb{\theta})(a_i)
\] 
for all $(a_1, \dots, a_n)\in \bar{A}^n$, where $b_i(\bdsb{\theta})(a_i)$ denotes the probability which $b_i(\bdsb{\theta})$ assigns to $a_i$. 
According to~\eqref{eq:20230208}, 
it can be shown that the aggregate outcome $\varphi_{\mu,b}$ in the single-population case also possesses symmetry.

\begin{theoremEnd}[no link to theorem, restate]{prop}
Let $(\mu, b)$ be a configuration in $[(N, \bar{A}, (\pi_i)), \Gamma(\mu)]$. 
Then for any strategy profile $(a_1, \dots, a_n)\in \bar{A}^n$ and any permutation $\tau$ of $N$, 
\[
\varphi_{\mu,b}(a_1, \dots, a_n) = \varphi_{\mu,b}(a_{\tau(1)}, \dots, a_{\tau(n)}). 
\] 
Furthermore, $\pi_1(\varphi_{\mu,b}) = \pi_j(\varphi_{\mu,b})$ for $j = 2$, \dots,~$n$.
\end{theoremEnd}

\begin{proofEnd}
For any $(a_1, \dots, a_n)\in \bar{A}^n$ and any permutation $\tau$ of $N$,
we can write
\begin{align*}
\varphi_{\mu,b}(a_1, \dots, a_n)
&= \sum_{(\theta_1, \dots, \theta_n)\in (\spt\mu)^n} \Bigl( \prod_{i\in N}\mu(\theta_i) \Bigr)
\prod_{i\in N} b_i(\theta_1, \dots, \theta_n)(a_i)\\
&= \sum_{(\theta_1, \dots, \theta_n)\in (\spt\mu)^n} \Bigl( \prod_{i\in N}\mu(\theta_i) \Bigr)
\prod_{i\in N} b_{\tau(i)}(\theta_1, \dots, \theta_n)(a_{\tau(i)})
\end{align*}
which, by~\eqref{eq:20230208}, is equal to
$\sum_{(\theta_1, \dots, \theta_n)\in (\spt\mu)^n} \bigl( \prod_{i\in N}\mu(\theta_{\tau(i)}) \bigr)
\prod_{i\in N} b_i(\theta_{\tau(1)}, \dots, \theta_{\tau(n)})(a_{\tau(i)})$.
For any permutation $\tau$ of $N$, it is clear that
\[
\{\, (\theta_1, \dots, \theta_n) \mid (\theta_1, \dots, \theta_n)\in (\spt\mu)^n \,\} =
\{\, (\theta_{\tau(1)}, \dots, \theta_{\tau(n)}) \mid (\theta_1, \dots, \theta_n)\in (\spt\mu)^n \,\}.
\]
Thus we can write $\varphi_{\mu,b}(a_1, \dots, a_n)$ in another way as
\[
\varphi_{\mu,b}(a_1, \dots, a_n)
= \sum_{(\theta_1, \dots, \theta_n)\in (\spt\mu)^n} \Bigl( \prod_{i\in N}\mu(\theta_i) \Bigr)
\prod_{i\in N} b_i(\theta_1, \dots, \theta_n)(a_{\tau(i)})
= \varphi_{\mu,b}(a_{\tau(1)}, \dots, a_{\tau(n)}),
\]
as required.
In addition, for any given $j$, let $\tau$ be a permutation of $N$ with $\tau(1) = j$.
Then we also have
\begin{align*}
\pi_j(\varphi_{\mu,b})
&= \sum_{(a_1, \dots, a_n)\in \bar{A}^n} \varphi_{\mu,b}(a_1, \dots, a_n) \pi_j(a_1, \dots, a_n)\\
&= \sum_{(a_1, \dots, a_n)\in \bar{A}^n} \varphi_{\mu,b}(a_{\tau(1)}, \dots, a_{\tau(n)}) \pi_{\tau^{-1}(j)}(a_{\tau(1)}, \dots, a_{\tau(n)}) = \pi_1(\varphi_{\mu,b}),
\end{align*}
where the facts that
$\pi_j(a_1, \dots, a_n) = \pi_{\tau^{-1}(j)}(a_{\tau(1)}, \dots, a_{\tau(n)})$ and
\[
\{\, (a_1, \dots, a_n) \mid (a_1, \dots, a_n)\in \bar{A}^n \,\} =
\{\, (a_{\tau(1)}, \dots, a_{\tau(n)}) \mid (a_1, \dots, a_n)\in \bar{A}^n \,\}
\]
for any permutation $\tau$ of $N$ have been used.
\end{proofEnd}

More specifically, 
if the configuration is monomorphic, 
the symmetric strategy profile $(\sigma, \dots, \sigma)$ chosen by all players of the same type 
is also regarded as the aggregate outcome.

\subsection*{Evolutionary Stability} 

Given a type distribution $\mu\in \mathcal{M}(\Theta_s)$, which represents the incumbent population,
the types in $\spt\mu$ are called \emph{incumbents}.
To capture evolutionary stability against multiple mutations,
let different new types $\witilde{\theta}_1$, \dots,~$\witilde{\theta}_r$, called \emph{mutants},
simultaneously enter the population. 
In this paper, all possible types of mutants coming from $\cplm{(\spt\mu)}$, the complement of $\spt\mu$,
are allowed to compete. 
Let $\upsilon$ be the distribution over the $r$ mutant types, 
and suppose that a fraction $\varepsilon$ of the population is replaced by mutants. 
Then the post-entry population can be represented by the type distribution 
$\witilde{\mu} = (1 - \varepsilon)\mu + \varepsilon\upsilon$. 
Obviously, $\spt\witilde{\mu} = \spt\mu \cup \spt\upsilon$, 
where $\spt\upsilon = \{\witilde{\theta}_1, \dots, \witilde{\theta}_r\}$ and $\spt\mu \cap \spt\upsilon = \varnothing$. 
We call such $\upsilon$ an \emph{$r$th-order mutation distribution}, 
and we write $\left| \spt\upsilon \right| = r$ for the set $\spt\upsilon$ having $r$ mutant types. 
The set $\{\witilde{\theta}_1, \dots, \witilde{\theta}_r\}$ is also called an \emph{$r$th-order mutation set}. 
Note that for any type $\theta\in \spt\witilde{\mu}$, 
the post-entry average fitness of $\theta$ is continuous in $\varepsilon$.

Following the entry of a mutant group, 
multiple post-entry equilibria may emerge, 
some of which may be unfamiliar to the players. 
When a match involves new types, 
predicting which equilibrium will be selected becomes impractical. 
Thus, we impose no restrictions on players' equilibrium behavior in such scenarios. 
Conversely, 
when all players in a match are incumbents, 
we assume they have no incentive to deviate from their established behavioral patterns, 
thereby maintaining their pre-entry behavior among themselves.\footnote{%
This is consistent with our implicit assumption that the post-entry behavior would reach a state of equilibrium quickly;
see footnote~\ref{footnote:1018}.}

\begin{Def}\label{def:focal}
Let $(\mu,b)$ and $(\witilde{\mu}, \witilde{b})$ be two configurations with $\spt\mu \subset \spt\witilde{\mu}$.
The equilibrium $\witilde{b}$ in the game $\Gamma(\witilde{\mu})$ is said to be \emph{focal} with respect to $b$
if $\witilde{b}(\bdsb{\theta}) = b(\bdsb{\theta})$ for all $\bdsb{\theta}\in (\spt\mu)^n$.
Let $B(\spt\witilde{\mu}; b)\subseteq B(\spt\witilde{\mu})$ denote the set of focal equilibria with respect to $b$. 
\end{Def}

\begin{Rem}
For any given mutant group,
the set $B(\spt\witilde{\mu};b)$ must be nonempty.
Clearly, for an arbitrary focal equilibrium $\witilde{b}$,
the post-entry aggregate outcome has the desired property that 
$\varphi_{\witilde{\mu},\witilde{b}}$ converges to $\varphi_{\mu,b}$ as $\varepsilon$ tends to zero,
which means that rare mutants cannot significantly affect the aggregate outcome.
\end{Rem}

In line with the widely accepted idea that mutation is an undirected random event,\footnote{%
\citet{Monroe-etal:mbrnsat} find that mutations occur less often in functionally constrained regions of the genome in the plant Arabidopsis thaliana, 
challenging the prevailing paradigm that mutation is a directionless force in evolution.} 
it is reasonable to assume that no focal equilibrium will be chosen with certainty. 
Our multi-mutation stability concept requires that regardless of which focal equilibrium players choose, 
the entry of rare mutants cannot result in an incumbent type earning the lowest average fitness, 
unless all individuals earn equally. 
Logically, this means that after any entry of rare mutants, 
either a mutant type earns a strictly lower average fitness than every incumbent, 
or all individuals achieve the same average fitness.

\begin{Def}\label{def:20200704}
In $[(N, \bar{A}, (\pi_i)), \Gamma(\mu)]$,
a configuration $(\mu,b)$ is \emph{$r$th-order stable}, 
or \emph{stable of order~$r$}, 
if for any $r$th-order mutation distribution $\upsilon\in \mathcal{M}(\Theta_s)$ 
and for any $\witilde{b}\in B(\spt\witilde{\mu};b)$, 
there exists some $\bar{\varepsilon}\in (0,1)$ such that 
either~\ref{Cs:wiped-out} or~\ref{Cs:coexist} is satisfied for all $\varepsilon\in (0,\bar{\varepsilon})$. 

\begin{enumerate}[label=(\roman*)]
\item There exists $\witilde{\theta}_j\in \spt\upsilon$ for which
$\varPi_{\witilde{\theta}_j}(\witilde{\mu};\witilde{b})< \varPi_{\theta}(\witilde{\mu};\witilde{b})$
for every $\theta\in \spt\mu$.\label{Cs:wiped-out}
\item $\varPi_{\theta}(\witilde{\mu};\witilde{b}) = \varPi_{\wihat{\theta}}(\witilde{\mu};\witilde{b})$
for every $\theta$,~$\wihat{\theta}\in \spt\witilde{\mu}$.\label{Cs:coexist}
\end{enumerate}
A configuration $(\mu,b)$ is simply said to be \emph{stable} if it is first-order stable. 
A configuration $(\mu,b)$ is said to be \emph{infinite-order stable} if it is stable of any large orders.
\end{Def}

We also call a symmetric strategy profile $\bdsb{\sigma}\in (\Delta \bar{A})^n$ 
\emph{$r$th-order stable}, or \emph{stable of order~$r$}, 
if it is the aggregate outcome of an $r$th-order stable configuration.
Condition~\ref{Cs:wiped-out} in the above definition requires that 
the set of mutant types fails to invade under a given distribution. 
Condition~\ref{Cs:coexist} achieves the coexistence of mutants and incumbents.
This definition implies that a stable configuration must be balanced, that is,
all incumbents in a stable configuration earn the same average fitness;
see Lemma~\ref{prop:20150215} or Corollary~\ref{prop:20160330}.

As the population shares of mutants tend to zero, 
any fitness derived from matches involving a higher number of mutant opponents becomes negligible 
(see also the argument after equation~\eqref{eq:20241128}). 
Consequently, 
we can define multi-mutation stability alternatively through 
sequential comparisons of players’ fitness values when facing the same composition types of opponents. 
This alternative formulation is highly instrumental in proving Theorems~\ref{prop:20150311} and~\ref{prop:20240312}; 
Appendix~\ref{sec:alt-def} provides the details.

The following simple example shows how a polymorphic group of mutants in the single-population case can invade a stable incumbent population. 
The results of this example, 
shown by explicitly constructing preference types, 
can also be derived from the theorems which we shall prove later. 
Theorem~\ref{prop:20160106} will provide a complete characterization of stable outcomes for symmetric $2 \times 2$ games.

\begin{example}\label{exam:20151001}
Suppose that the underlying game, which represents the fitness assignment, is as follows.
Each player has the same action set $\{ a_H, a_L \}$,
and we let $\pi$ denote the material payoff function.

\begin{table}[!h]
\centering
\renewcommand{\arraystretch}{1.2}
    \begin{tabular}{r|c|c|}
      \multicolumn{1}{c}{} & \multicolumn{1}{c}{$a_H$} & \multicolumn{1}{c}{$a_L$}\\ \cline{2-3}
      $a_H$ & $0$, $0$ & $2$, $2$ \\  \cline{2-3}
      $a_L$ & $2$, $2$ & $0$, $0$ \\  \cline{2-3}
    \end{tabular}
\end{table}

\noindent
Consider a monomorphic configuration $(\mu,b)$ consisting of type $\theta$ for which 
\[
\theta(a_H\mid a_H) = \theta(a_L\mid a_L) > \theta(a_H\mid a_L) = \theta(a_L\mid a_H), 
\] 
and $b_i(\theta, \theta) = (0.5, 0.5)$ for $i = 1$,~$2$.
We first show that the configuration $(\mu, b)$ is stable.
Let $\witilde{\theta}$ be an entrant with its population share $\varepsilon$,
and let $\witilde{b}$ be an arbitrary focal equilibrium (with respect to $b$).
Then the post-entry average fitnesses of $\theta$ and $\witilde{\theta}$ are, respectively,
\[ 
\varPi_{\theta}(\witilde{\mu};\witilde{b}) =
(1-\varepsilon) \pi_1\big( (0.5, 0.5), (0.5, 0.5) \big) + \varepsilon \pi_1\big( \witilde{b}(\theta, \witilde{\theta}) \big)
\]
and
\[
\varPi_{\witilde{\theta}}(\witilde{\mu};\witilde{b}) =
(1-\varepsilon) \pi_1\big( \witilde{b}(\witilde{\theta}, \theta) \big) + 
\varepsilon \pi_1\big( \witilde{b}(\witilde{\theta}, \witilde{\theta}) \big).
\]
If $\witilde{b}_1(\witilde{\theta}, \theta)\neq (0.5, 0.5)$,
then by incumbents' preferences,
we can deduce that $\pi_1\big( \witilde{b}(\witilde{\theta}, \theta) \big)< 1$,
and hence that for $\varepsilon$ small enough, we have
$\varPi_{\theta}(\witilde{\mu};\witilde{b})> \varPi_{\witilde{\theta}}(\witilde{\mu};\witilde{b})$.
If $\witilde{b}_1(\witilde{\theta}, \theta) = (0.5, 0.5)$,
it is also easy to check that
\[
\varPi_{\theta}(\witilde{\mu};\witilde{b}) - \varPi_{\witilde{\theta}}(\witilde{\mu};\witilde{b}) =
\varepsilon \big[ 1 - \pi_1\big( \witilde{b}(\witilde{\theta}, \witilde{\theta}) \big) \big]\geq 0,
\] 
regardless of the strategy mutants play against each other. 
This shows that $(\mu, b)$ is stable, and so is the strategy pair $((0.5, 0.5), (0.5, 0.5))$.

Nevertheless, 
the configuration $(\mu, b)$ is not second-order stable. 
To see this, let 
$\upsilon$ be a distribution of two indifferent types $\witilde{\theta}_1$ and $\witilde{\theta}_2$, 
and let $\witilde{\mu} = (1 - \varepsilon)\mu + \varepsilon\upsilon$ be the post-entry type distribution. 
Suppose that the chosen focal equilibrium $\witilde{b}$ satisfies 
\begin{itemize}
\item $\witilde{b}(\witilde{\theta}_1, \witilde{\theta}_2) = (a_H, a_L)$, and 
\item $\witilde{b}(\witilde{\theta}_i, \theta) = \witilde{b}(\witilde{\theta}_i, \witilde{\theta}_i) = ((0.5, 0.5), (0.5, 0.5))$ for all $i$.
\end{itemize} 
(See also the remark below.) 
Then we obtain $\varPi_{\theta}(\witilde{\mu};\witilde{b}) = 1$ and
$\varPi_{\witilde{\theta}_i}(\witilde{\mu};\witilde{b}) = 1 + \varepsilon \upsilon(\witilde{\theta}_j)$ for all $i$,~$j$ with $i\neq j$.
Thus $(\mu, b)$ is not a second-order stable configuration.
In fact, by Theorem~\ref{prop:20160106}\ref{2b2:mi}, 
there is no stable aggregate outcome of order greater than $1$ in this game.\footnote{%
If the two-population matching setting is applied to this symmetric two-player underlying game (as will be discussed in Section~\ref{sec:multi}),
then by Theorems~\ref{prop:20150130} and~\ref{prop:20141028},
one can see that all stable strategy pairs are $(a_H, a_L)$ and $(a_L, a_H)$, both of which are indeed infinite-order stable.} 
\end{example}

\begin{Rem} 
Since we follow the idea that mutation is an undirected random event, 
all possible mutant types and all possible focal equilibria are allowed to compete. 
Technically, the indifferent types, represented by constant utility functions,
enable us easily to build up an argument about the mutants' particular actions. 
Of course, such an argument is not always needed. 
For example, 
the two indifferent types $\witilde{\theta}_1$ and $\witilde{\theta}_2$ in Example~\ref{exam:20151001} 
can be replaced by $\witilde{\theta}'_1$ and $\witilde{\theta}'_2$, respectively, where 
$\witilde{\theta}'_i(a_H\mid a_H) = \witilde{\theta}'_i(a_L\mid a_L) = 0$ and 
$\witilde{\theta}'_i(a_H\mid a_L) = \witilde{\theta}'_i(a_L\mid a_H) = i$ for $i = 1$,~$2$. 
\end{Rem}

It is intuitively clear that if an incumbent population is invaded by a mutation set,
then it is more likely to be successfully invaded by another which contains more mutant types 
as long as the action distribution resulting from a larger mutation set is identical to the one resulting from the smaller mutation set, 
which we prove formally below. 
This means that a multi-mutation stable configuration can also preserve the stability of lower orders. 
Therefore we can say that a configuration is stable of all orders if it is stable of any large orders.

\begin{theoremEnd}[no link to theorem, restate]{lem}\label{prop:20150924}
In $[(N, \bar{A}, (\pi_i)), \Gamma(\mu)]$,
let $(\mu, b)$ be a $k$th-order stable configuration with $k> 1$. 
Then it is $r$th-order stable for any positive integer $r< k$.
\end{theoremEnd}

\begin{proofEnd}
Suppose that $(\mu,b)$ is not an $r$th-order stable configuration for some positive integer $r$.
This means that
there exist an $r$th-order mutation distribution $\upsilon$ and a focal equilibrium $\witilde{b}\in B(\spt\witilde{\mu};b)$
such that for every $\bar{\varepsilon}\in (0,1)$,
the following conditions are satisfied for some $\varepsilon\in (0,\bar{\varepsilon})$: 
(1) for every $\witilde{\theta}_j\in \spt\upsilon$,
there exists $\theta\in \spt\mu$ earning the average fitness $\varPi_{\theta}(\witilde{\mu};\witilde{b})$
not higher than $\varPi_{\witilde{\theta}_j}(\witilde{\mu};\witilde{b})$;
(2) the inequality $\varPi_{\theta}(\witilde{\mu};\witilde{b})\neq \varPi_{\wihat{\theta}}(\witilde{\mu};\witilde{b})$
holds for some $\theta$,~$\wihat{\theta}\in \spt\witilde{\mu}$.

Let $k$ be an arbitrary integer with $k> r$.
To prove that the configuration $(\mu,b)$ is not $k$th-order stable, 
consider a $k$th-order mutation distribution $\upsilon'$, 
and let $\witilde{\mu}' = (1 - \varepsilon)\mu + \varepsilon\upsilon'$ be the post-entry type distribution. 
Suppose that $\spt\upsilon = \{\witilde{\theta}_1, \dots, \witilde{\theta}_r\}$ and 
$\spt\upsilon' = \spt\upsilon\cup \{\witilde{\theta}_{r+1}, \dots, \witilde{\theta}_k\}$, 
and let $\upsilon'$ also satisfy $\upsilon'(\witilde{\theta}_j) = \upsilon(\witilde{\theta}_j)$ for $j = 1$, \dots,~$r-1$ 
and $\sum_{j=r}^{k} \upsilon'(\witilde{\theta}_j) = \upsilon(\witilde{\theta}_r)$. 
In addition, suppose that 
$\witilde{\theta}_{r+1}$, \dots,~$\witilde{\theta}_k$ have preferences that induce the same behavior as the mutant type $\witilde{\theta}_r$.
Then the post-entry focal equilibrium $\witilde{b}'\in B(\spt\witilde{\mu}';b)$ can be chosen such that
$\varPi_{\wihat{\theta}}(\witilde{\mu}';\witilde{b}') = \varPi_{\wihat{\theta}}(\witilde{\mu};\witilde{b})$
for all $\wihat{\theta}\in \spt\witilde{\mu}$
and $\varPi_{\witilde{\theta}_j}(\witilde{\mu}';\witilde{b}') = \varPi_{\witilde{\theta}_r}(\witilde{\mu};\witilde{b})$
for $j = r+1$, \dots,~$k$.
According to the above non-stability conditions,
this clearly leads to the conclusion that $(\mu,b)$ is not $k$th-order stable. 
\end{proofEnd}

As demonstrated earlier, 
a stable configuration can be destabilized by an increasing number of mutant types. 
This raises the question of whether any stable configuration remains vulnerable to destabilization 
under an arbitrarily large number of mutant types. 
We will show that a stable configuration can resist invasion by any number of mutant types 
only if it exhibits a sufficiently high level of efficiency. 
Before exploring the relationship between stability and efficiency, 
we introduce several efficiency concepts, some of which are unique to the single-population case.

\subsection*{Material Efficiency in Symmetric Games} 

When focusing on single-population matching settings, 
we say that a symmetric strategy profile is efficient
if its fitness value is highest among all symmetric strategy profiles of the symmetric game. 
Note that throughout this paper, 
all the concepts of efficiency are defined with respect to the material payoff function rather than the players' subjective utility functions.

\begin{Def}\label{def:20240102}
In a symmetric game $(N, \bar{A}, (\pi_i))$,
a strategy profile $(\sigma^*, \dots, \sigma^*)\in (\Delta \bar{A})^n$, or a strategy $\sigma^*\in \Delta \bar{A}$,
is \emph{efficient} if $\pi_1(\sigma^*, \dots, \sigma^*)\geq \pi_1(\sigma, \dots, \sigma)$ for all $\sigma\in \Delta \bar{A}$.
\end{Def}

In the existing literature on the indirect evolutionary approach,
it is well known that only efficient outcomes can be stable in single-population matching settings with observable preferences.\footnote{%
See, for example, \citet{aPos:tsespsg}, \citet{eDek-jEly-oYil:ep}, and \citet{svW:eneup}.}
Besides the symmetric efficiency concept,
we will employ the concept of Pareto efficiency, applied to the noncooperative and cooperative payoff regions,
to characterize multi-mutation stability.

\begin{Def}\label{def:20150903}
For a compact subset $\mathcal{S}$ of $\mathbb{R}^n$, the set
\[
P(\mathcal{S}) = \{\, v\in \mathcal{S} \mid \text{there is no $w\in \mathcal{S}$ with $w\neq v$ and $w_i\geq v_i$ for $i = 1$, \dots,~$n$} \,\}
\]
denotes the \emph{Pareto frontier} of $\mathcal{S}$.
\end{Def}

Pareto efficiency is defined as an allocation of resources from which 
no one can be made better off without making someone else worse off. 
Let $(N, (A_i), (\pi_i))$ be a finite normal-form game
(which may be either symmetric or asymmetric). 
We call $P(\mathcal{F}_{\nc})$ the \emph{noncooperative Pareto frontier}, 
and $P(\mathcal{F}_{\co})$ the \emph{cooperative Pareto frontier}. 
For any strategy profile $\bdsb{\sigma}\in \prod_{i\in N}\Delta A_i$,
if $\pi(\bdsb{\sigma})\in P(\mathcal{F}_{\co})$, then clearly $\pi(\bdsb{\sigma})\in P(\mathcal{F}_{\nc})$.
The converse, however, is not true;
it is easy to see that
a payoff profile $\pi(\bdsb{\sigma})$ lying on $P(\mathcal{F}_{\nc})$ may not lie on $P(\mathcal{F}_{\co})$ 
(as in Example~\ref{exam:20201215}).
From the point of view of Pareto frontier,
the efficient strategy $\sigma^*\in \Delta \bar{A}$ defined for a symmetric game $(N, \bar{A}, (\pi_i))$
is such that
$\pi(\sigma^*, \dots, \sigma^*)\in P\big( \{\, \pi(\sigma, \dots, \sigma) \mid \sigma\in \Delta \bar{A} \,\} \big)$.
It follows that in $(N, \bar{A}, (\pi_i))$,
a strategy $\sigma\in \Delta \bar{A}$ which satisfies $\pi(\sigma, \dots, \sigma)\in P(\mathcal{F}_{\nc})$
(or $\pi(\sigma, \dots, \sigma)\in P(\mathcal{F}_{\co})$) is of course efficient,
but the converse does not hold.\footnote{%
For instance, 
the strategy profile $((0.5, 0.5), (0.5, 0.5))$ in Example~\ref{exam:20151001} is efficient,
but it is strongly Pareto dominated by $(a_H, a_L)$ (or $(a_L, a_H)$). 
This means that $\pi\big( (0.5, 0.5), (0.5, 0.5) \big)$ does not lie on $P(\mathcal{F}_{\nc})$ 
(and thus not on $P(\mathcal{F}_{\co})$).}
We must carefully distinguish these efficiency concepts from one another.

For any strategy profile in a normal-form game,
we can ensure that its payoff profile lies on the cooperative Pareto frontier
if it maximizes the sum of all players' payoffs over all pure strategies. 
This is because 
for any correlated strategy $\varphi\in \Delta(\prod_{i\in N}A_i)$,
we can write 
$\sum_{i\in N} \pi_i(\varphi) = \sum_{\bdsb{a}\in \prod_{i\in N}A_i} \bigl( \varphi(\bdsb{a}) \sum_{i\in N} \pi_i(\bdsb{a}) \bigr)$,
and hence
\[
\max \{\, \sum_{i\in N} \pi_i(\bdsb{a}) \mid \bdsb{a}\in \prod_{i\in N} A_i \,\} = 
\max \{\, \sum_{i\in N} \pi_i(\varphi) \mid \varphi\in \Delta(\prod_{i\in N}A_i) \,\}.
\]

\begin{theoremEnd}[]{lem}\label{prop:20240802}
Let $(N, (A_i), (\pi_i))$ be a finite normal-form game.
Suppose that $\bdsb{\sigma}\in \prod_{i\in N}\Delta A_i$ satisfies
$\sum_{i\in N} \pi_i(\bdsb{\sigma})\geq \sum_{i\in N} \pi_i(\bdsb{a})$
for all $\bdsb{a}\in \prod_{i\in N} A_i$.
Then $\pi(\bdsb{\sigma})\in P(\mathcal{F}_{\co})$.
\end{theoremEnd}

For a symmetric strategy profile in a symmetric game, 
we can show conversely that 
if its payoff profile lies on the cooperative Pareto frontier, 
then it maximizes the sum of all players' payoffs.\footnote{%
\citet{yHel-eMoh:cdp} define stability against polymorphic groups of mutants.
A strategy profile is efficient in the sense of \citet{yHel-eMoh:cdp} if it maximizes the sum of fitness payoffs.
Their result of Corollary~1 is consistent with our result of Theorem~\ref{prop:20150411}\ref{Snc:o3}.} 
This result will crucially affect several unique properties of the single-population case.

\begin{theoremEnd}[no link to theorem, restate]{lem}\label{prop:20200908}
Let $(N, \bar{A}, (\pi_i))$ be a symmetric game. 
If $\sigma^*\in \Delta \bar{A}$ satisfies $\pi(\sigma^*, \dots, \sigma^*)\in P(\mathcal{F}_{\co})$, 
then 
$\sum_{i\in N} \pi_i(\sigma^*, \dots, \sigma^*)\geq \sum_{i\in N} \pi_i(\bdsb{\sigma})$ 
for all $\bdsb{\sigma}\in (\Delta \bar{A})^n$. 
\end{theoremEnd}

\begin{proofEnd}
Suppose that there exists $\bdsb{\sigma}\in (\Delta \bar{A})^n$ such that 
$\sum_{i\in N} \pi_i(\bdsb{\sigma})> \sum_{i\in N} \pi_i(\sigma^*, \dots, \sigma^*)$. 
Since we can write 
$\sum_{i\in N} \pi_i(\bdsb{\sigma}) = 
\sum_{\bdsb{a}\in \bar{A}^n} \bigl( \varphi_{\bdsb{\sigma}}(\bdsb{a}) \sum_{i\in N} \pi_i(\bdsb{a}) \bigr)$, 
it follows that there is a pure strategy profile $(a_1, \dots, a_n)\in \bar{A}^n$ satisfying 
$\sum_{i\in N} \pi_i(a_1, \dots, a_n)> \sum_{i\in N} \pi_i(\sigma^*, \dots, \sigma^*)$. 
To prove that $\pi(\sigma^*, \dots, \sigma^*)$ does not lie on $P(\mathcal{F}_{\co})$, 
let $\varphi\in \Delta(\bar{A}^n)$ be defined by 
\[ 
\varphi(a_1, a_2, \dots, a_n) = \varphi(a_n, a_1, \dots, a_{n-1}) = \dots = \varphi(a_2, a_3, \dots, a_1) = \frac{1}{n}, 
\] 
and taking the value $0$ otherwise. 
Since $(N, \bar{A}, (\pi_i))$ is a symmetric game, we have 
\[ 
\pi_i(\varphi) = \frac{1}{n} [ \pi_i(a_1, a_2, \dots, a_n) + \pi_i(a_n, a_1, \dots, a_{n-1}) + \dots + \pi_i(a_2, a_3, \dots, a_1) ] 
= \frac{1}{n} \sum_{j\in N} \pi_j(a_1, \dots, a_n) 
\] 
for all $i\in N$. 
This implies that 
$\pi_i(\varphi)> \pi_i(\sigma^*, \dots, \sigma^*)$ for all $i\in N$, 
as desired.
\end{proofEnd}

\subsection*{Stability Results} 

Regarding the relationship between stability and efficiency, 
we will clearly demonstrate that 
efficiency levels tend to rise with higher orders of stability.
The intuition is as follows.
If the efficiency level of an outcome is not sufficiently high, 
then enough heterogeneous mutant types with the ``secret handshake'' flavor can correlate their plays to gain a fitness advantage.\footnote{%
\citet{aRob:eegdnsh} demonstrates that any inefficient ESS could be destabilized by the ``secret handshake'' mutant,
which refers to the mutants playing the inefficient outcome against the incumbents and attaining a more efficient outcome against themselves.} 
As the diversity of invading mutants increases, 
synergistic interactions among them expand their collective advantage, 
requiring the incumbent population to achieve higher performance efficiency to avoid being displaced.

Before deriving necessary conditions for multi-mutation stability, 
We present a lemma that reveals a notable property: 
in a stable configuration, 
all incumbents receive a common fitness value in every interaction among themselves, 
despite having different preference types. 
Therefore,
as we will see in Corollary~\ref{prop:20160330},
any stable configuration is balanced, 
meaning that the average fitness of any incumbent is equal to this common fitness value.

\begin{theoremEnd}[no link to theorem, restate]{lem}\label{prop:20150215}
Let $(\mu,b)$ be a stable configuration in $[(N, \bar{A}, (\pi_i)), \Gamma(\mu)]$.
Then $\pi\big( b(\theta_1, \dots, \theta_n) \big) = \pi\big( b(\theta'_1, \dots, \theta'_n) \big)$
for any $\theta_1$, \dots,~$\theta_n\in \spt\mu$ and any $\theta'_1$, \dots,~$\theta'_n\in \spt\mu$.
\end{theoremEnd}

\begin{proofEnd}
To prove the lemma, it suffices to show that
for an arbitrary integer $s$ with $1\leq s\leq n$, and
for any two multisets
$\{ \theta'_1, \dots, \theta'_s \}$ and $\{ \theta''_1, \dots, \theta''_s \}$
consisting of types in $\spt\mu$, the equality
\[
\pi_i\big( b(\theta'_1, \dots, \theta'_s, \theta_{s+1}, \dots, \theta_n) \big) =
\pi_i\big( b(\theta''_1, \dots, \theta''_s, \theta_{s+1}, \dots, \theta_n) \big)
\]
holds for all $\theta_{s+1}$, \dots,~$\theta_n\in \spt\mu$ and for $i = 1$, \dots,~$n$.

For the case $s = 1$,
we first suppose that there are two different types
$\theta'_1$,~$\theta''_1\in \spt\mu$ such that
$\pi_1\big(b(\theta'_1, \theta_2, \dots, \theta_n)\big)> \pi_1\big(b(\theta''_1, \theta_2, \dots, \theta_n)\big)$
for some types $\theta_2$, \dots,~$\theta_n\in \spt\mu$.
Consider an indifferent type $\witilde{\theta}$ entering the population.
Let the chosen focal equilibrium $\witilde{b}$ satisfy:
$\witilde{b}(\witilde{\theta}, \theta_2, \dots, \theta_n) = b(\theta'_1, \theta_2, \dots, \theta_n)$
if $\witilde{\theta}$ is matched against the incumbents $\theta_2$, \dots,~$\theta_n$;
otherwise,
$\witilde{\theta}$ mimics the play of $\theta''_1$ and all the incumbents keep playing their pre-entry plays as when matched with $\theta''_1$.
Then for a sufficiently small population share of $\witilde{\theta}$,
we have $\varPi_{\witilde{\theta}}(\witilde{\mu};\witilde{b})> \varPi_{\theta''_1}(\witilde{\mu};\witilde{b})$,
which means that the configuration $(\mu,b)$ is not stable.
Thus for any $\theta'_1$,~$\theta''_1\in \spt\mu$, the equality
\begin{equation}\label{eq:0331}
\pi_1\big( b(\theta'_1, \theta_2, \dots, \theta_n) \big) = \pi_1\big( b(\theta''_1, \theta_2, \dots, \theta_n) \big)
\end{equation}
holds for all $\theta_2$, \dots,~$\theta_n\in \spt\mu$.

Now suppose that there are two different incumbents $\theta'_1$, $\theta''_1$,
and incumbents $\theta_2$, $\theta_3$, \dots,~$\theta_n$ such that, without loss of generality,
$\pi_2\big( b(\theta'_1, \theta_2, \theta_3, \dots, \theta_n) \big)>
\pi_2\big( b(\theta''_1, \theta_2, \theta_3, \dots, \theta_n) \big)$.
Then by~\eqref{eq:0331}, we have
$\pi_i\big( b(\theta'_1, \theta'_1, \theta_3, \dots, \theta_n) \big)>
\pi_i\big( b(\theta''_1, \theta''_1, \theta_3, \dots, \theta_n) \big)$
for $i = 2$, and thus for $i = 1$,
this inequality is also true.
Consider an indifferent type $\witilde{\theta}$ entering the population.
Let the chosen focal equilibrium $\witilde{b}$ satisfy:
$\witilde{b}(\witilde{\theta}, \theta''_1, \theta_3, \dots, \theta_n) = b(\theta''_1, \theta''_1, \theta_3, \dots, \theta_n)$
if $\witilde{\theta}$ is matched against the $n - 1$ incumbents $\theta''_1$ and $\theta_3$, \dots,~$\theta_n$;
otherwise, $\witilde{\theta}$ mimics the play of $\theta'_1$
and all the incumbents keep playing their pre-entry plays as when matched with $\theta'_1$.
Then by~\eqref{eq:0331} again, we obtain
$\varPi_{\witilde{\theta}}(\witilde{\mu};\witilde{b})> \varPi_{\theta''_1}(\witilde{\mu};\witilde{b})$
for any population share of $\witilde{\theta}$,
and hence the configuration $(\mu,b)$ is not stable as desired.
This proves the case $s = 1$.

Proceeding inductively, suppose that the case $s = k$ is true for some $1\leq k\leq n - 1$.
Let $\{ \theta'_1, \dots, \theta'_{k+1} \}$ and $\{ \theta''_1, \dots, \theta''_{k+1} \}$
be two multisets consisting of types in $\spt\mu$.
Then for any $\theta_{k+2}$, \dots,~$\theta_n\in \spt\mu$, we get
\begin{align*}
\pi\big( b(\theta'_1, \dots, \theta'_k, \theta'_{k+1}, \theta_{k+2}, \dots, \theta_n) \big)
&= \pi\big( b(\theta'_1, \dots, \theta'_k, \theta''_{k+1}, \theta_{k+2}, \dots, \theta_n) \big)\\
&= \pi\big( b(\theta''_1, \dots, \theta''_k, \theta''_{k+1}, \theta_{k+2}, \dots, \theta_n) \big).
\end{align*}
This proves the lemma.
\end{proofEnd}

\begin{theoremEnd}[no link to theorem, restate]{thm}\label{prop:20150411}
Suppose that $(\mu,b)$ is a configuration in $[(N, \bar{A}, (\pi_i)), \Gamma(\mu)]$.
Let $\theta_1$, \dots,~$\theta_n\in \spt\mu$ and let $\sigma^*\in \Delta \bar{A}$ be efficient in $(N, \bar{A}, (\pi_i))$.
\begin{enumerate}[label=\upshape (\roman*)]
\item If $(\mu,b)$ is stable, then $\pi\big( b(\theta_1, \dots, \theta_n) \big) = \pi(\sigma^*, \dots, \sigma^*)$.\label{Snc:o1}
\item If $(\mu,b)$ is stable of order~$n$, then $\pi\big( b(\theta_1, \dots, \theta_n) \big) \in P(\mathcal{F}_{\nc})$.\label{Snc:o2}
\item If $(\mu,b)$ is stable of order~$n + 1$, then $\pi\big( b(\theta_1, \dots, \theta_n) \big) \in P(\mathcal{F}_{\co})$.\label{Snc:o3}
\end{enumerate}
\end{theoremEnd}

\begin{proofEnd}
To prove~\ref{Snc:o1},
it suffices, by Lemma~\ref{prop:20150215},
to show that $b(\theta, \dots, \theta)$ is efficient for any $\theta\in \spt\mu$ if $(\mu,b)$ is stable.
Suppose that there exists $\theta'\in \spt\mu$ for which $b(\theta', \dots, \theta')$ is not efficient.
Consider an indifferent type $\witilde{\theta}$ with its population share $\varepsilon$ entering the population 
in which the mutants mimic the play of $\theta'$,
except that they all play the efficient strategy $\sigma^*$ when matched against one another. 
Then there is a focal equilibrium $\witilde{b}$ such that for any $\varepsilon$, 
\[
\varPi_{\witilde{\theta}}(\witilde{\mu};\witilde{b}) - \varPi_{\theta'}(\witilde{\mu};\witilde{b}) = 
\varepsilon^{n-1} 
\Bigl( \pi_1(\sigma^*, \dots, \sigma^*) - \pi_1\big(b(\theta', \dots, \theta')\big) \Bigr)> 0, 
\] 
and thus the configuration $(\mu,b)$ is not stable.

\smallskip

For proving~\ref{Snc:o2},
suppose that there are $\theta'_1$, \dots,~$\theta'_n\in \spt\mu$ for which
$\pi\big( b(\theta'_1, \dots, \theta'_n) \big)\notin P(\mathcal{F}_{\nc})$.
Then there exists $\sigma\in (\Delta \bar{A})^n$ such that
$\pi_i(\sigma)\geq \pi_i\big( b(\theta'_1, \dots, \theta'_n) \big)$ for all $i$ with strict inequality for at least one $i$. 
Consider an $n$th-order mutation distribution $\upsilon$, 
and let the post-entry type distribution be $\witilde{\mu} = (1 - \varepsilon)\mu + \varepsilon\upsilon$. 
Suppose that $\spt\upsilon$ consists of $n$ indifferent types $\witilde{\theta}_1$, \dots,~$\witilde{\theta}_n$, 
and that the strategy profile $\sigma$ is played when $\witilde{\theta}_1$, \dots,~$\witilde{\theta}_n$ are matched together;
in all other interactions,
each $\witilde{\theta}_j$ mimics the play of $\theta'_j$ and
all the incumbents keep playing their pre-entry plays as among themselves.
Then for each $j$, 
the difference between the post-entry average fitnesses of $\witilde{\theta}_j$ and $\theta'_j$ is 
\[
\varPi_{\witilde{\theta}_j}(\witilde{\mu};\witilde{b}) - \varPi_{\theta'_j}(\witilde{\mu};\witilde{b}) =
\bigl( (n-1)! \prod_{i\neq j} \varepsilon \upsilon(\witilde{\theta}_i) \bigr)
\Bigl( \pi_j(\sigma) - \pi_j\big( b(\theta'_1, \dots, \theta'_n) \big) \Bigr). 
\] 
Since the strategy profile $\sigma$ Pareto dominates $b(\theta'_1, \dots, \theta'_n)$, 
this implies that the configuration $(\mu,b)$ is not $n$th-order stable.

\smallskip

For~\ref{Snc:o3}, we argue by contradiction.
Let $(\mu,b)$ be $(n + 1)$th-order stable, 
and suppose that there exist $\theta'_1$, \dots,~$\theta'_n\in \spt\mu$ for which
$\pi\big( b(\theta'_1, \dots, \theta'_n) \big)\notin P(\mathcal{F}_{\co})$.
By Lemma~\ref{prop:20150924} and property~\ref{Snc:o1},
we get $\pi\big( b(\theta'_1, \dots, \theta'_n) \big) = \pi(\sigma^*, \dots, \sigma^*)$,
and hence $\pi(\sigma^*, \dots, \sigma^*)\notin P(\mathcal{F}_{\co})$.
Thus, by Lemma~\ref{prop:20240802},
there exist $a_1$, \dots,~$a_n\in \bar{A}$ such that
$\sum_{i\in N} \pi_i(a_1, \dots, a_n) > n\pi_1(\sigma^*, \dots, \sigma^*)$.
Now consider an $(n+1)$th-order mutation distribution $\upsilon$ with equal probabilities 
assigned to indifferent types $\witilde{\theta}_1$, \dots,~$\witilde{\theta}_{n+1}$ in $\spt\upsilon$. 
Let $\witilde{\mu} = (1 - \varepsilon)\mu + \varepsilon\upsilon$, 
and suppose that the chosen focal equilibrium $\witilde{b}$ satisfies:
when $n$ distinct mutant types are matched,
\[
\witilde{b}(\witilde{\theta}_1, \witilde{\theta}_2, \dots, \witilde{\theta}_n) =
\witilde{b}(\witilde{\theta}_2, \witilde{\theta}_3, \dots, \witilde{\theta}_{n+1}) = \dots =
\witilde{b}(\witilde{\theta}_{n+1}, \witilde{\theta}_1, \dots, \witilde{\theta}_{n-1}) = (a_1, a_2 \dots, a_n);
\]
in the remaining matches,
each $\witilde{\theta}_j$ mimics the play of a certain incumbent type,
and all the incumbents keep playing their pre-entry plays as among themselves.
Then applying property~\ref{Snc:o1} again,
we conclude that for any $\witilde{\theta}_j\in \spt\upsilon$ and any $\theta\in \spt\mu$,
the difference between their average fitnesses is
\[
\varPi_{\witilde{\theta}_j}(\witilde{\mu};\witilde{b}) - \varPi_{\theta}(\witilde{\mu};\witilde{b}) =
(n-1)! \, \Bigl( \frac{\varepsilon}{n+1} \Bigr)^{n-1}
\Bigl( \sum_{i\in N} \pi_i(a_1, \dots, a_n) - n\pi_1(\sigma^*, \dots, \sigma^*) \Bigr)> 0, 
\] 
which contradicts the hypothesis that $(\mu, b)$ is an $(n + 1)$th-order stable configuration.
\end{proofEnd}

From Theorem~\ref{prop:20150411}\ref{Snc:o1},
it follows directly that the average fitness of any incumbent in a stable configuration is equal to the efficiency value.
Moreover, it is easy to see that the aggregate outcome of a stable configuration also has the efficiency value.

\begin{theoremEnd}[no link to theorem, restate]{cor}\label{prop:20160330}
Let $\sigma^*$ be an efficient strategy in $(N, \bar{A}, (\pi_i))$,
and let $(\mu,b)$ be a stable configuration with the aggregate outcome $\varphi_{\mu,b}$ in $[(N, \bar{A}, (\pi_i)), \Gamma(\mu)]$.
Then $(\mu, b)$ is balanced, and the relations
\[
\varPi_{\theta}(\mu;b) = \pi_1(\sigma^*, \dots, \sigma^*) = \pi_i(\varphi_{\mu,b})
\]
hold for all $\theta\in \spt\mu$ and all $i\in N$. 
\end{theoremEnd}

\begin{proofEnd}
Let $(\mu,b)$ be a stable configuration. Then for each $i\in N$, we have
\begin{align*}
\pi_i(\varphi_{\mu,b})
&= \sum_{(a_1, \dots, a_n)\in \bar{A}^n} \varphi_{\mu,b}(a_1, \dots, a_n) \pi_i(a_1, \dots, a_n)\\
&= \sum_{\bdsb{\theta}\in (\spt\mu)^n} \mu^n(\bdsb{\theta})
\sum_{(a_1, \dots, a_n)\in \bar{A}^n}
\Bigl( \prod_{j\in N} b_j(\bdsb{\theta})(a_j) \Bigr) \pi_i(a_1, \dots, a_n)\\
&= \sum_{\bdsb{\theta}\in (\spt\mu)^n} \mu^n(\bdsb{\theta}) \pi_i\big( b(\bdsb{\theta}) \big) = \pi_1(\sigma^*, \dots, \sigma^*),
\end{align*}
which follows from Theorem~\ref{prop:20150411}\ref{Snc:o1}.
The rest of the proof is straightforward.
\end{proofEnd}

In Example~\ref{exam:20151001}, it is shown that 
a polymorphic mutant group can invade a stable incumbent population.
However, 
the invasion capacity of mutants does not necessarily expand indefinitely as the number of mutant types increases. 
In fact, a configuration stable against any number of mutant types exists, provided that it achieves a certain order of stability. 
We show here that an $(n + 1)$th-order stable configuration under single-population matching is necessarily an infinite-order stable one.

\begin{theoremEnd}[no link to theorem, restate]{thm}\label{prop:20150311}
In $[(N, \bar{A}, (\pi_i)), \Gamma(\mu)]$,
a configuration $(\mu,b)$ is stable of order~$n + 1$ if and only if it is stable of any large orders.
\end{theoremEnd}

\begin{proofEnd}
One direction follows immediately from Lemma~\ref{prop:20150924}: 
if $(\mu,b)$ is not stable of order~$n + 1$, then it is not stable of order greater than $n + 1$. 
For the converse, 
suppose that $(\mu,b)$ is stable of order~$n + 1$. 
Then by Lemma~\ref{prop:20150924} and
by parts~\ref{Snc:o1} and~\ref{Snc:o3} of Theorem~\ref{prop:20150411},
we know that for all $\theta_1$, \dots,~$\theta_n\in \spt\mu$,
\[
\pi\big( b(\theta_1, \dots, \theta_n) \big) = (\pi^e, \dots, \pi^e)\in P(\mathcal{F}_{\co})
\]
where $\pi^e$ is defined to be $\max_{\sigma\in \Delta \bar{A}} \pi_1(\sigma, \dots, \sigma)$, 
and so by Lemma~\ref{prop:20200908} we see that 
\begin{equation}\label{eq:20260708} 
n\pi^{e}\geq \sum_{i=1}^{n} \pi_i(\sigma_1, \dots, \sigma_n) 
\end{equation} 
for all $(\sigma_1, \dots, \sigma_n)\in (\Delta \bar{A})^n$. 
For any positive integer $r$ greater than $n + 1$, 
consider an $r$th-order mutation distribution $\upsilon$, 
and let $\witilde{\mu} = (1 - \varepsilon)\mu + \varepsilon\upsilon$. 
We have to show that $(\mu, b)$ is multi-mutation stable against invasion by mutant types in $\spt\upsilon$ playing any focal equilibrium.
In order to find out which type earns a higher average fitness,
we will make sequential comparisons of players' fitness values earned when matched against the same composition types of opponents.

We first examine the case of a matched $n$-tuple consisting of one mutant and $n-1$ incumbents.
Suppose that there exist $\witilde{\theta}_1\in \spt\upsilon$ and $\theta_2$, \dots,~$\theta_n\in \spt\mu$ such that
$\pi_1\big( \witilde{b}(\witilde{\theta}_1, \theta_2, \dots, \theta_n) \big)> \pi^e$ for some $\witilde{b}\in B(\spt\witilde{\mu};b)$. 
This implies that if an indifferent type 
that mimics the play of $\witilde{\theta}_1$ against $\theta_2$, \dots,~$\theta_n$
and mimics the play of a certain incumbent type in all other interactions 
appears alone in the incumbent population, 
then this new type could earn a higher average fitness than all the incumbents as long as its population share is sufficiently small, 
a contradiction.
Thus for any $\witilde{b}\in B(\spt\witilde{\mu};b)$,
we have $\pi_1\big( \witilde{b}(\witilde{\theta}_1, \theta_2, \dots, \theta_n) \big)\leq \pi^e$
for all $\witilde{\theta}_1\in \spt\upsilon$ and all $\theta_2$, \dots,~$\theta_n\in \spt\mu$. 
If there exist $\witilde{\theta}_1\in \spt\upsilon$ and $\theta_2$, \dots,~$\theta_n\in \spt\mu$ such that
$\pi_1\big( \witilde{b}(\witilde{\theta}_1, \theta_2, \dots, \theta_n) \big)< \pi^e$ for some $\witilde{b}\in B(\spt\witilde{\mu};b)$,
then for $\varepsilon$ small enough,
the mutant type $\witilde{\theta}_1$ earns a strictly lower average fitness than all the incumbents, 
as desired.
Therefore for any $\witilde{b}\in B(\spt\witilde{\mu};b)$,
we only have to consider $\pi_1\big( \witilde{b}(\witilde{\theta}_1, \theta_2, \dots, \theta_n) \big) = \pi^e$
for all $\witilde{\theta}_1\in \spt\upsilon$ and all $\theta_2$, \dots,~$\theta_n\in \spt\mu$.
Since $\pi_1\big( b(\theta_1, \dots, \theta_n) \big) = \pi^e$
for all $\theta_1$, \dots,~$\theta_n\in \spt\mu$,
we certainly have
$F^{\witilde{\mu}}_{\witilde{\theta}_j}(0) = F^{\witilde{\mu}}_{\theta}(0)$ 
for all $\witilde{\theta}_j\in \spt\upsilon$ and all $\theta\in \spt\mu$.

Regarding the fitness values of the incumbents in the above case, 
it follows from~\eqref{eq:20260708} that 
for any $\witilde{b}\in B(\spt\witilde{\mu};b)$,
we have
$\sum_{j=2}^n \pi_j\big( \witilde{b}(\witilde{\theta}_1, \theta_2, \dots, \theta_n) \big)\leq (n-1)\pi^e$
for all $\witilde{\theta}_1\in \spt\upsilon$ and all $\theta_2$, \dots,~$\theta_n\in \spt\mu$.
Without loss of generality, 
suppose that there exist 
$\witilde{\theta}_1\in \spt\upsilon$ and $\theta_2$, \dots,~$\theta_n\in \spt\mu$ such that 
$\pi_2\big( \witilde{b}(\witilde{\theta}_1, \theta_2, \dots, \theta_n) \big)< \pi^e$ for some $\witilde{b}\in B(\spt\witilde{\mu};b)$. 
From this, 
it follows that if an indifferent type 
that mimics the play of $\witilde{\theta}_1$ against $\theta_2$, \dots,~$\theta_n$
and mimics the play of a certain incumbent type in all other interactions 
appears alone in the incumbent population, 
then this new type could earn the average fitness $\pi^e$, which is higher than that of $\theta_2$, 
a contradiction.
Thus for any $\witilde{b}\in B(\spt\witilde{\mu};b)$,
and for any $\witilde{\theta}_1\in \spt\upsilon$ and any $\theta_2$, \dots,~$\theta_n\in \spt\mu$,
it happens that
$\pi_j\big( \witilde{b}(\witilde{\theta}_1, \theta_2, \dots, \theta_n) \big) = \pi^e$ for $j = 2$, \dots,~$n$, 
and therefore 
$F^{\witilde{\mu}}_{\theta}(1) = F^{\witilde{\mu}}_{\theta'}(1)$ for all $\theta$,~$\theta'\in \spt\mu$.

Now for every integer $k$ with $1\leq k\leq n-2$,
suppose inductively that for each $1\leq m\leq k$, 
and for each $\witilde{b}\in B(\spt\witilde{\mu};b)$ chosen by 
any $\witilde{\theta}_1$, \dots,~$\witilde{\theta}_m\in \spt\upsilon$ and any $\theta_{m+1}$, \dots,~$\theta_n\in \spt\mu$,
\[
\sum_{j=1}^m \pi_j\big( \witilde{b}(\witilde{\theta}_1, \dots, \witilde{\theta}_m, \theta_{m+1}, \dots, \theta_n) \big) = m\pi^e,
\]
and the equality
$\pi_j\big( \witilde{b}(\witilde{\theta}_1, \dots, \witilde{\theta}_m, \theta_{m+1}, \dots, \theta_n) \big) = \pi^e$
holds for $j = m+1$, \dots,~$n$,
which implies that 
$F^{\witilde{\mu}}_{\theta}(m) = F^{\witilde{\mu}}_{\theta'}(m)$ for all $\theta$,~$\theta'\in \spt\mu$.
In addition, for each $1\leq m\leq k$ and each $\witilde{b}\in B(\spt\witilde{\mu};b)$, 
we suppose that
\begin{equation}\label{eq:weighted-sum}
F^{\witilde{\mu}}_{\witilde{\theta}_j}(m - 1) = F^{\witilde{\mu}}_{\theta}(m - 1)
\end{equation}
for all $\witilde{\theta}_j\in \spt\upsilon$ and all $\theta\in \spt\mu$.

We then study the case where $k + 1$ members of a matched $n$-tuple are mutants.
Suppose that there exist $\witilde{\theta}_1$, \dots,~$\witilde{\theta}_{k+1}\in \spt\upsilon$
and $\theta_{k+2}$, \dots,~$\theta_n\in \spt\mu$ such that
$\sum_{j=1}^{k+1} \pi_j\big( \witilde{b}(\witilde{\theta}_1, \dots, \witilde{\theta}_{k+1}, \theta_{k+2}, \dots, \theta_n) \big)>
(k+1)\pi^e$
for some $\witilde{b}\in B(\spt\witilde{\mu};b)$. 
This leads us to consider 
$k + 2$ indifferent types $\witilde{\theta}^0_1$, \dots,~$\witilde{\theta}^0_{k+2}$
which play a focal equilibrium $\witilde{b}^0$ satisfying the conditions:
\begin{align*}
\witilde{b}^0(\witilde{\theta}^0_1, \witilde{\theta}^0_2, \dots, \witilde{\theta}^0_{k+1}, \theta_{k+2}, \dots, \theta_n)
&= \witilde{b}^0(\witilde{\theta}^0_2, \witilde{\theta}^0_3, \dots, \witilde{\theta}^0_{k+2}, \theta_{k+2}, \dots, \theta_n)\\
&= \cdots = \witilde{b}^0(\witilde{\theta}^0_{k+2}, \witilde{\theta}^0_1, \dots, \witilde{\theta}^0_k, \theta_{k+2}, \dots, \theta_n)\\
&= \witilde{b}(\witilde{\theta}_1, \witilde{\theta}_2, \dots, \witilde{\theta}_{k+1}, \theta_{k+2}, \dots, \theta_n);
\end{align*}
in all other interactions, 
each of the $k + 2$ types mimics the play of a certain incumbent type. 
If equal fractions of these $k + 2$ types enter the population, 
then for a sufficiently small fraction of mutants, 
it can be verified that each of these new types would earn a higher average fitness than all the incumbents, a contradiction. 
Thus for any $\witilde{b}\in B(\spt\witilde{\mu};b)$, we have 
$\sum_{j=1}^{k+1} \pi_j\big( \witilde{b}(\witilde{\theta}_1, \dots, \witilde{\theta}_{k+1}, \theta_{k+2}, \dots, \theta_n) \big)\leq (k+1)\pi^e$
for all $\witilde{\theta}_1$, \dots,~$\witilde{\theta}_{k+1}\in \spt\upsilon$ and all $\theta_{k+2}$, \dots,~$\theta_n\in \spt\mu$.

Observe that for any $\theta\in \spt\mu$ and any $\witilde{b}\in B(\spt\witilde{\mu};b)$, 
the induction hypothesis implies 
\[ 
\sum_{\witilde{\theta}_j\in \spt\upsilon} \varepsilon \upsilon(\witilde{\theta}_j) 
\bigl( F^{\witilde{\mu}}_{\witilde{\theta}_j}(k) - F^{\witilde{\mu}}_{\theta}(k) \bigr) 
= \sum_{\witilde{\theta}_j\in \spt\upsilon} \varepsilon \upsilon(\witilde{\theta}_j) 
\sum_{[\bdsb{\witilde{\theta}}_k, \bdsb{\theta}_{r(k)}]\in \mathcal{G}_{(k, r(k))}} 
C_{[\bdsb{\witilde{\theta}}_k, \bdsb{\theta}_{r(k)}]}^{\witilde{\mu}} 
\Bigl( \pi_1 \big( \witilde{b}(\witilde{\theta}_j, \bdsb{\witilde{\theta}}_k, \bdsb{\theta}_{r(k)}) \big) - 
\pi^e \Bigr). 
\] 
Furthermore, this is equal to 
\begin{equation}\label{eq:20260714} 
\sum_{\bdsb{\wihat{\theta}}\in (\spt\upsilon)^{k+1}\times (\spt\mu)^{n-1-k}}
\frac{(n-1)!}{(k+1)! (n-1-k)!} \times
\witilde{\mu}^{n}(\bdsb{\wihat{\theta}}) \times
\sum_{j=1}^{k+1} \Bigl( \pi_j\big( \witilde{b}(\bdsb{\wihat{\theta}}) \big) - \pi^e \Bigr), 
\end{equation}
which is thus less than or equal to $0$. 
If it happens that
$F^{\witilde{\mu}}_{\witilde{\theta}_j}(k) \neq F^{\witilde{\mu}}_{\theta}(k)$ 
for some $\witilde{\theta}_j\in \spt\upsilon$ and some $\witilde{b}\in B(\spt\witilde{\mu};b)$, 
then applying~\eqref{eq:weighted-sum},
as in the proof of Theorem~\ref{prop:20241111},
we conclude that for $\varepsilon$ small enough,
there is at least one mutant type in $\spt\upsilon$ earning a strictly lower average fitness than all the incumbents, as desired.
Therefore for any $\witilde{b}\in B(\spt\witilde{\mu};b)$, 
we only have to consider
$F^{\witilde{\mu}}_{\witilde{\theta}_j}(k) = F^{\witilde{\mu}}_{\theta}(k)$
for all $\witilde{\theta}_j\in \spt\upsilon$ and all $\theta\in \spt\mu$, 
which also implies that 
\begin{equation}\label{eq:mu-sum}
\sum_{j=1}^{k+1} \pi_j\big( \witilde{b}(\witilde{\theta}_1, \dots, \witilde{\theta}_{k+1}, \theta_{k+2}, \dots, \theta_n) \big) = (k+1)\pi^e
\end{equation}
for all $\witilde{\theta}_1$, \dots,~$\witilde{\theta}_{k+1}\in \spt\upsilon$ and all $\theta_{k+2}$, \dots,~$\theta_n\in \spt\mu$.

Again, 
for any $\witilde{b}\in B(\spt\witilde{\mu};b)$, 
it follows from~\eqref{eq:20260708} that 
$\sum_{j=k+2}^n \pi_j\big( \witilde{b}(\witilde{\theta}_1, \dots, \witilde{\theta}_{k+1}, \theta_{k+2}, \dots, \theta_n) \big)\leq (n-k-1)\pi^e$ 
for all $\witilde{\theta}_1$, \dots,~$\witilde{\theta}_{k+1}\in \spt\upsilon$ and all $\theta_{k+2}$, \dots,~$\theta_n\in \spt\mu$.
Accordingly we may suppose without loss of generality that there exist 
$\witilde{\theta}_1$, \dots,~$\witilde{\theta}_{k+1}\in \spt\upsilon$ and $\theta_{k+2}$, \dots,~$\theta_n\in \spt\mu$ such that 
$\pi_{k+2}\big( \witilde{b}(\witilde{\theta}_1, \dots, \witilde{\theta}_{k+1}, \theta_{k+2}, \dots, \theta_n) \big)< \pi^e$ 
for some $\witilde{b}\in B(\spt\witilde{\mu};b)$. 
This also leads us to consider $k + 2$ new types 
that behave like $\witilde{\theta}^0_1$, \dots,~$\witilde{\theta}^0_{k+2}$ mentioned above. 
Then we can verify by~\eqref{eq:mu-sum} that 
the population might be invaded by these $k + 2$ types in equal fractions,
each of whom would earn the average fitness $\pi^e$, higher than that of $\theta_{k+2}$,
a contradiction.
Thus for any $\witilde{b}\in B(\spt\witilde{\mu};b)$, 
and for any $\witilde{\theta}_1$, \dots,~$\witilde{\theta}_{k+1}\in \spt\upsilon$ 
and any $\theta_{k+2}$, \dots,~$\theta_n\in \spt\mu$,
we must have
\[
\pi_j\big( \witilde{b}(\witilde{\theta}_1, \dots, \witilde{\theta}_{k+1}, \theta_{k+2}, \dots, \theta_n) \big) = \pi^e
\]
for $j = k+2$, \dots,~$n$, 
and therefore 
$F^{\witilde{\mu}}_{\theta}(k + 1) = F^{\witilde{\mu}}_{\theta'}(k + 1)$ for all $\theta$,~$\theta'\in \spt\mu$.

So far,
for any $\witilde{b}\in B(\spt\witilde{\mu};b)$,
we have shown that for each $\theta\in \spt\mu$ and each $\bdsb{\wihat{\theta}}_{-1}\in (\spt \witilde{\mu})^{n-1}$,
the fitness value $\pi_1\big( \witilde{b}(\theta, \bdsb{\wihat{\theta}}_{-1}) \big)$ can only be considered equal to $\pi^e$,
which implies that for all $\theta$,~$\theta'\in \spt\mu$,
it suffices to consider 
$F^{\witilde{\mu}}_{\theta}(m) = F^{\witilde{\mu}}_{\theta'}(m)$ for $0\leq m\leq n - 1$.
In addition,
for all $\witilde{\theta}_j\in \spt\upsilon$ and all $\theta\in \spt\mu$,
we only have to consider
$F^{\witilde{\mu}}_{\witilde{\theta}_j}(m) = F^{\witilde{\mu}}_{\theta}(m)$
for $0\leq m\leq n - 2$.

To complete the proof, 
let us compare $F^{\witilde{\mu}}_{\witilde{\theta}_j}(n - 1)$ with $F^{\witilde{\mu}}_{\theta}(n - 1)$, 
where $\witilde{\theta}_j$ is a mutant type in $\spt\upsilon$, 
and $F^{\witilde{\mu}}_{\theta}(n - 1)$ is a constant for all incumbent types. 
Given any $\theta\in \spt\mu$ and any $\witilde{b}\in B(\spt\witilde{\mu};b)$, 
we can apply~\eqref{eq:20260714} with $k = n - 1$ to deduce that 
\[
\sum_{\witilde{\theta}_j\in \spt\upsilon} \varepsilon \upsilon(\witilde{\theta}_j) 
\bigl( F^{\witilde{\mu}}_{\witilde{\theta}_j}(n - 1) - F^{\witilde{\mu}}_{\theta}(n - 1) \bigr) =
\sum_{\bdsb{\witilde{\theta}}\in (\spt\upsilon)^n} \frac{\witilde{\mu}^{n}(\bdsb{\witilde{\theta}})}{n}
\sum_{i\in N}
\Bigl( \pi_i\big( \witilde{b}(\bdsb{\witilde{\theta}}) \big) - \pi^e \Bigr),
\] 
which by~\eqref{eq:20260708} is less than or equal to $0$. 
Therefore, we conclude that
$F^{\witilde{\mu}}_{\witilde{\theta}_j}(n - 1) = F^{\witilde{\mu}}_{\theta}(n - 1)$ 
for all $\witilde{\theta}_j\in \spt\upsilon$ and all $\theta\in \spt\mu$, 
or there exists $\witilde{\theta}_j\in \spt\upsilon$ such that 
$F^{\witilde{\mu}}_{\witilde{\theta}_j}(n - 1)< F^{\witilde{\mu}}_{\theta}(n - 1)$ 
for all $\theta\in \spt\mu$, as desired 
(see the remark after Theorem~\ref{prop:20241111}). 
\end{proofEnd}

Mathematically, 
the $(n + 1)$th-order stability establishes a boundary condition. 
In economic terms, 
local resistance to minor invasions automatically extrapolates to immunity against coalitional threats of any scale, 
obviating the need to evaluate infinitely complex composite strategies.

\begin{Rem} 
Notice that the proof of Theorem~\ref{prop:20150311} also shows that 
the hypothesis of $(n + 1)$th-order stability can be replaced by the weaker condition that 
$(\mu,b)$ is $n$th-order stable and 
$\pi(\sigma^*, \dots, \sigma^*)\in P(\mathcal{F}_{\co})$ for an efficient strategy $\sigma^*$. 
Hence, by Corollary~\ref{prop:20160330}, it follows that 
\emph{in $[(N, \bar{A}, (\pi_i)), \Gamma(\mu)]$, 
a configuration $(\mu,b)$ is infinite-order stable if and only if 
it is an $n$th-order stable configuration with the aggregate outcome $\varphi_{\mu,b}$ satisfying 
$\pi(\varphi_{\mu,b})\in P(\mathcal{F}_{\co})$}. 
\end{Rem}

Next, we turn our attention to sufficient conditions for multi-mutation stability. 
We first establish the results for the familiar two-player case before generalizing them to $n$-player games. 
In Theorem~\ref{prop:20160105} below, 
we demonstrate that whether a symmetric strict Nash equilibrium in a two-player game is multi-mutation stable 
depends crucially on the height of its efficiency level. 
The intuition for parts~\ref{Ssc:o1} and~\ref{Ssc:o2} is straightforward.\footnote{%
The result in part~\ref{Ssc:o1} of Theorem~\ref{prop:20160105} is due to \citet{eDek-jEly-oYil:ep}.} 
Consider a monomorphic population of types for which the symmetric strict Nash equilibrium strategy is strictly dominant, 
and suppose that a sufficiently small fraction of the population is replaced by mutants. 
A mutant type will go extinct if it plays any other strategy against incumbents. 
Furthermore,
when mutants play against each other, 
a sufficiently high level of efficiency guarantees that 
no mutant can earn a higher average fitness without worsening the fitness of other mutants. 
Finally, in Theorem~\ref{prop:20160105}\ref{Ssc:o3},
the maximum sum of material payoffs, 
a property established in Lemma~\ref{prop:20200908} that applies specifically to symmetric games, 
plays a pivotal role in preventing invasion by polymorphic mutant groups, 
hinting at sufficient conditions for multi-mutation stability in $n$-player games.

\begin{theoremEnd}[no link to theorem, restate]{thm}\label{prop:20160105}
Let the strategy pair $(a^*,a^*)$ be a strict Nash equilibrium of a symmetric two-player game $(\{1, 2\}, \bar{A}, (\pi_i))$.
\begin{enumerate}[label=\upshape (\roman*)]
\item If $a^*$ is an efficient strategy, then $(a^*,a^*)$ is stable.\label{Ssc:o1}
\item If $\pi(a^*,a^*)\in P(\mathcal{F}_{\nc})$, then $(a^*,a^*)$ is second-order stable.\label{Ssc:o2}
\item If $\pi(a^*,a^*)\in P(\mathcal{F}_{\co})$, then $(a^*,a^*)$ is infinite-order stable.\label{Ssc:o3}
\end{enumerate}
\end{theoremEnd}

\begin{proofEnd}
Let $(a^*,a^*)$ be a strict Nash equilibrium of $(\{1, 2\}, \bar{A}, (\pi_i))$.
Suppose that $(\mu,b)$ is a monomorphic configuration in which $a^*$ a strictly dominant strategy for $\theta$, 
so that $(a^*, a^*)$ is the aggregate outcome of $(\mu, b)$. 
For any given positive integer $r$, 
consider an $r$th-order mutation distribution $\upsilon$. 
Let $\spt\upsilon = \{\witilde{\theta}_1, \dots, \witilde{\theta}_r\}$, 
and let $\witilde{\mu} = (1 - \varepsilon)\mu + \varepsilon\upsilon$. 
Then for any $\witilde{b}\in B(\spt\witilde{\mu};b)$, 
the difference 
$\varPi_{\theta}(\witilde{\mu};\witilde{b}) - \varPi_{\witilde{\theta}_i}(\witilde{\mu};\witilde{b})$
between the post-entry average fitnesses of $\theta$ and $\witilde{\theta}_i\in \spt\upsilon$ is equal to 
\[
(1 - \varepsilon)
\Bigl( \pi_1(a^*,a^*) - \pi_1\big( \witilde{b}_1(\witilde{\theta}_i, \theta), a^* \big) \Bigr) +
\sum_{j=1}^r \varepsilon \upsilon(\witilde{\theta}_j)
\Bigl( \pi_1\big( a^*, \witilde{b}_2(\theta, \witilde{\theta}_j) \big) -
\pi_1\big( \witilde{b}(\witilde{\theta}_i, \witilde{\theta}_j) \big) \Bigr).
\]
If $\witilde{b}_1(\witilde{\theta}_i, \theta)\neq a^*$ for some $\witilde{\theta}_i\in \spt\upsilon$,
then since $(a^*,a^*)$ is a strict Nash equilibrium,
we must have
$\varPi_{\theta}(\witilde{\mu};\witilde{b}) > \varPi_{\witilde{\theta}_i}(\witilde{\mu};\witilde{b})$ 
for all sufficiently small $\varepsilon$, as desired.
Thus we only have to consider
$\witilde{b}_1(\witilde{\theta}_i, \theta) = a^*$ for all $\witilde{\theta}_i\in \spt\upsilon$, 
and hence 
\[
\varPi_{\theta}(\witilde{\mu};\witilde{b}) - \varPi_{\witilde{\theta}_i}(\witilde{\mu};\witilde{b}) =
\sum_{j=1}^r \varepsilon \upsilon(\witilde{\theta}_j)
\Bigl( \pi_1( a^*, a^*) -
\pi_1\big( \witilde{b}(\witilde{\theta}_i, \witilde{\theta}_j) \big) \Bigr).
\]
In the case $\left| \spt\upsilon \right| = 1$, part~\ref{Ssc:o1} follows directly from the condition that $a^*$ is an efficient strategy. 

\smallskip

For~\ref{Ssc:o2},
let $\left| \spt\upsilon \right| = 2$, and let $\pi(a^*,a^*)\in P(\mathcal{F}_{\nc})$.
Then $(a^*,a^*)$ is efficient,
and hence for $i = 1$,~$2$,
\[
\varPi_{\theta}(\witilde{\mu};\witilde{b}) - \varPi_{\witilde{\theta}_i}(\witilde{\mu};\witilde{b})
\geq \varepsilon \upsilon(\witilde{\theta}_j) 
\Bigl( \pi_1(a^*, a^*) - \pi_1\big(\witilde{b}(\witilde{\theta}_i, \witilde{\theta}_j)\big) \Bigr),
\]
where $\witilde{\theta}_j\in \spt\upsilon$, distinct from $\witilde{\theta}_i$.
Since $\pi(a^*,a^*)\in P(\mathcal{F}_{\nc})$, 
we conclude that for any $\varepsilon$,
either
$\varPi_{\theta}(\witilde{\mu};\witilde{b}) = \varPi_{\witilde{\theta}_i}(\witilde{\mu};\witilde{b})$
for all $i = 1$,~$2$,
or
$\varPi_{\theta}(\witilde{\mu};\witilde{b}) > \varPi_{\witilde{\theta}_i}(\witilde{\mu};\witilde{b})$
for some $i = 1$,~$2$.
Thus $(\mu,b)$ is a second-order stable configuration.

\smallskip

For part~\ref{Ssc:o3}, suppose that $\spt\upsilon$ has an arbitrary number of mutant types, 
and let $\pi(a^*,a^*)\in P(\mathcal{F}_{\co})$. 
Notice that we can write 
$\sum_{i=1}^r \varepsilon \upsilon(\witilde{\theta}_i) 
\sum_{j=1}^r \varepsilon \upsilon(\witilde{\theta}_j) \pi_1\big(\witilde{b}(\witilde{\theta}_i, \witilde{\theta}_j)\big)$ 
as 
\[ 
\sum_{i=1}^r \big(\varepsilon \upsilon(\witilde{\theta}_i)\big)^2 \pi_1\big(\witilde{b}(\witilde{\theta}_i, \witilde{\theta}_i)\big)
+ \sum_{i=1}^{r-1} \sum_{j=i+1}^r \varepsilon \upsilon(\witilde{\theta}_i) \varepsilon \upsilon(\witilde{\theta}_j)
\Bigl( \pi_1\big(\witilde{b}(\witilde{\theta}_i, \witilde{\theta}_j)\big) + \pi_2\big(\witilde{b}(\witilde{\theta}_i, \witilde{\theta}_j)\big) \Bigr), 
\] 
which is less than or equal to 
$\sum_{i=1}^r \varepsilon \upsilon(\witilde{\theta}_i) \sum_{j=1}^r \varepsilon \upsilon(\witilde{\theta}_j) \pi_1(a^*, a^*)$, 
by applying Lemma~\ref{prop:20200908} with $\pi(a^*, a^*)\in P(\mathcal{F}_{\co})$. 
It follows that 
$\sum_{j=1}^r \varepsilon \upsilon(\witilde{\theta}_j) 
\bigl( \pi_1(a^*, a^*) - \pi_1\big( \witilde{b}(\witilde{\theta}_i, \witilde{\theta}_j) \big) \bigr) = 0$ 
for all $\witilde{\theta}_i\in \spt\upsilon$, 
or that 
$\sum_{j=1}^r \varepsilon \upsilon(\witilde{\theta}_j) 
\bigl( \pi_1(a^*, a^*) - \pi_1\big( \witilde{b}(\witilde{\theta}_i, \witilde{\theta}_j) \big) \bigr)> 0$ 
for some $\witilde{\theta}_i\in \spt\upsilon$. 
Hence for any $\varepsilon$,
we have either
$\varPi_{\theta}(\witilde{\mu};\witilde{b}) = \varPi_{\witilde{\theta}_i}(\witilde{\mu};\witilde{b})$
for all $\witilde{\theta}_i\in \spt\upsilon$, or
$\varPi_{\theta}(\witilde{\mu};\witilde{b}) > \varPi_{\witilde{\theta}_i}(\witilde{\mu};\witilde{b})$
for some $\witilde{\theta}_i\in \spt\upsilon$, 
thereby proving the infinite-order stability of $(a^*, a^*)$.
(One can also prove this claim
by applying the remark after Theorem~\ref{prop:20150311} to $n = 2$,
combined with the result of~\ref{Ssc:o2}.)
\end{proofEnd}

We now generalize Theorem~\ref{prop:20160105} to symmetric $n$-player games. 
While a strict Nash equilibrium effectively prevents unilateral deviations, 
the complex interaction structure of $n$-player games requires accounting for coalitional deviations as well. 
To generalize the strict Nash equilibrium concept to accommodate such coalitional deviations, 
we first adopt the \emph{strictly strong Nash equilibrium} concept, 
which requires the defining payoff inequality to hold strictly for all coalition members.

\begin{Def}\label{def:20210120}
Let $(N, (A_i), (\pi_i))$ be a finite normal-form game.
A strategy profile $\bdsb{x}$ is a \emph{strictly strong Nash equilibrium} if
for any nonempty $J\subseteq N$, 
and for any $\bdsb{\sigma}_J\in \prod_{j\in J}\Delta A_j$ with $\bdsb{\sigma}_J\neq \bdsb{x}_J$,
there exists some $k\in J$ such that $\pi_k(\bdsb{x})> \pi_k(\bdsb{\sigma}_J, \bdsb{x}_{-J})$.\footnote{%
The notation $\bdsb{\sigma}_J\neq \bdsb{x}_J$ means that $\sigma_j\neq x_j$ for some $j\in J$ (rather than for all $j\in J$).
The strategy profile $(\bdsb{\sigma}_J, \bdsb{x}_{-J})$ results from $\bdsb{x}$ by replacing $x_j$ with $\sigma_j$ for each $j\in J$.}
\end{Def}

This equilibrium concept requires that
at least one deviating player suffers a loss from any coalitional deviation. 
Consequently, 
the payoff profile of a strictly strong Nash equilibrium must lie on the noncooperative Pareto frontier. 
In symmetric games, where players choosing identical actions receive identical payoffs, 
a symmetrically strictly strong Nash equilibrium $(a^*, \dots, a^*)$ is inherently efficient, 
and the efficient strategy $a^*$ must be unique for the game.

Now, 
consider a population initially consisting of types for which the strictly strong Nash equilibrium strategy $a^*$ is strictly dominant, 
and suppose mutants of a single type enter this population. 
Any deviation by these mutants from the equilibrium profile $(a^*, \dots, a^*)$ 
will result in strictly lower payoffs for the deviating mutants, 
who share the same type and play the same action. 
Therefore, 
if a symmetric $n$-player game possesses a symmetrically strictly strong Nash equilibrium, 
this equilibrium constitutes the unique stable strategy profile in the game 
(a result formalized in Proposition~\ref{prop:20201130}). 
Given that its payoff profile lies on the noncooperative Pareto frontier, 
one might expect it to yield a second-order stable outcome, 
as stated in Theorem~\ref{prop:20160105}\ref{Ssc:o2}. 
However, 
as the following example shows, 
this conjecture does not hold for multi-player games.

\begin{example}\label{exam:20201215}
Let the following three-player game represent the fitness assignment.
Here,
the set of actions available to each player is $\{a_1, a_2\}$.

\begin{table}[!h]
\centering
\renewcommand{\arraystretch}{1.2}
    \begin{tabular}{r|c|c|}
      \multicolumn{1}{c}{} & \multicolumn{1}{c}{$a_1$} & \multicolumn{1}{c}{$a_2$}\\ \cline{2-3}
      $a_1$ & $10$, $10$, $10$                               & $0$, $0$, $\phantom{3}0$  \\  \cline{2-3}
      $a_2$ & $\phantom{1}0$, $\phantom{1}0$, $\phantom{1}0$ & $1$, $1$, $30$ \\  \cline{2-3}
      \multicolumn{1}{c}{} & \multicolumn{2}{c}{$a_1$}
    \end{tabular} \qquad
    \begin{tabular}{r|c|c|}
      \multicolumn{1}{c}{} & \multicolumn{1}{c}{$a_1$} & \multicolumn{1}{c}{$a_2$}\\ \cline{2-3}
      $a_1$ & $0$, $\phantom{3}0$, $0$ & $30$, $1$, $1$ \\  \cline{2-3}
      $a_2$ & $1$, $30$, $1$           & $\phantom{3}0$, $0$, $0$ \\  \cline{2-3}
      \multicolumn{1}{c}{} & \multicolumn{2}{c}{$a_2$}
    \end{tabular}
\end{table}

\noindent
One can check that $(a_1, a_1, a_1)$ is a strictly strong Nash equilibrium, 
and hence its payoff profile $\pi(a_1, a_1, a_1)\in P(\mathcal{F}_{\nc})$. 
(By Lemma~\ref{prop:20200908}, we know that $\pi(a_1, a_1, a_1)\notin P(\mathcal{F}_{\co})$.)
From the previous discussion, the strategy profile $(a_1, a_1, a_1)$ is stable 
(see also Proposition~\ref{prop:20201130}). 
We now prove that 
this strictly strong Nash equilibrium $(a_1, a_1, a_1)$, endowed with $\pi(a_1, a_1, a_1)\in P(\mathcal{F}_{\nc})$, 
is not second-order stable.

Let $(\mu, b)$ be a configuration with the aggregate outcome $(a_1, a_1, a_1)$.
Then for all incumbents $\theta$, $\theta'$, and $\theta''$,
we have $b(\theta, \theta', \theta'') = (a_1, a_1, a_1)$.
Consider equal fractions of two indifferent types $\witilde{\theta}_1$ and $\witilde{\theta}_2$ entering the population. 
Suppose that the chosen focal equilibrium $\witilde{b}$ satisfies the conditions: 
for any $i$,~$j\in \{1, 2\}$ with $i\neq j$,
we have $\witilde{b}(\witilde{\theta}_i, \witilde{\theta}_j, \witilde{\theta}_j) = (a_1, a_2, a_2)$,
and furthermore
\[
\witilde{b}(\witilde{\theta}_i, \theta, \theta') =
\witilde{b}(\witilde{\theta}_i, \witilde{\theta}_i, \theta) =
\witilde{b}(\witilde{\theta}_i, \witilde{\theta}_j, \theta) =
\witilde{b}(\witilde{\theta}_i, \witilde{\theta}_i, \witilde{\theta}_i) = (a_1, a_1, a_1)
\]
for arbitrary incumbents $\theta$ and $\theta'$.
Comparing the post-entry average fitness of an incumbent type $\theta$ with that of the two mutant types $\witilde{\theta}_1$ and $\witilde{\theta}_2$,
we find that
\[
\varPi_{\witilde{\theta}_1}(\witilde{\mu};\witilde{b}) = \varPi_{\witilde{\theta}_2}(\witilde{\mu};\witilde{b}) =
\varPi_{\theta}(\witilde{\mu};\witilde{b}) + 0.5 \varepsilon^2. 
\] 
This means that
incumbents who induce the aggregate outcome $(a_1, a_1, a_1)$ will be displaced by such two mutant types.
Thus $(a_1, a_1, a_1)$ is not a second-order stable outcome.
\end{example}

Instead, 
a symmetrically strictly strong Nash equilibrium of a symmetric three-player game is second-order stable 
if it maximizes the sum of material payoffs 
(which, by Lemmas~\ref{prop:20240802} and~\ref{prop:20200908}, 
is equivalent to its payoff profile lying on the cooperative Pareto frontier). 
However, unlike the result of Theorem~\ref{prop:20160105}\ref{Ssc:o3},
this stronger condition cannot ensure higher-order stability in symmetric three-player games. 
Furthermore, in games with more than three players,
such an equilibrium may fail to be second-order stable, 
although it maximizes the sum of material payoffs. 
A detailed formal analysis of these findings is presented in Appendix~\ref{sec:counterexamples}.

So far, we have observed that a group of mutants may act collectively as if they form coalitions. 
Examples~\ref{exam:20201215}, \ref{exam:20210123}, and~\ref{exam:20210124} 
illustrate various possibilities in which mutants' losses incurred from specific deviations can be compensated for through other matches, 
thereby increasing their average fitnesses. 
As the number of mutant types or the number of players in a game increases, 
the possible structures of coalitions become more diverse. 
This diversity provides mutant types with more opportunities to offset losses caused by certain deviations. 
Consequently, 
these findings motivate us to define a new equilibrium concept 
that prevents mutants from obtaining sufficient compensation across different coalitional groups.

\begin{Def}\label{def:k-unionNE}
Let $(N, \bar{A}, (\pi_i))$ be a symmetric $n$-player game,
and let $k$ be an integer with $1\leq k\leq n$.
For a sub-profile $\bdsb{\sigma}_J\in (\Delta \bar{A})^{|J|}$,
let $D(\bdsb{\sigma}_J)$ denote the set of distinct strategies in $\bdsb{\sigma}_J$.
A strategy profile $\bdsb{x}$ is a \emph{strict $k$-type union equilibrium} if
for any nonempty $J\subseteq N$, 
and for any $\bdsb{\sigma}_J\in (\Delta \bar{A})^{|J|}$
with $\left| D(\bdsb{\sigma}_J) \right|\leq k$ and $\bdsb{\sigma}_J\neq \bdsb{x}_J$,
\[
\sum_{j\in J}\pi_j(\bdsb{x})> \sum_{j\in J}\pi_j(\bdsb{\sigma}_J, \bdsb{x}_{-J}).
\]
\end{Def}

A strict $k$-type union equilibrium is a strategy profile such that
for any nonempty subset $J$ of $N$,
and for any players in $J$ choosing $\bdsb{\sigma}_J$ with the restriction $\left| D(\bdsb{\sigma}_J) \right|\leq k$,
the sum of the payoffs of these players in $J$ will be strictly lower if any of them deviates.
The condition $\left| D(\bdsb{\sigma}_J) \right|\leq k$ 
only concerns the number of distinct strategies in $\bdsb{\sigma}_J$; 
it applies specially to symmetric games,
where $n$ players have the same action set and player positions are ignored.
Roughly speaking, 
this equilibrium concept implies that 
losses caused by certain deviations cannot be compensated for by specific coalitional groups, 
and thus guarantees a particular order stability.

\begin{theoremEnd}[no link to theorem, restate]{thm}\label{prop:20240312}
Let $(a^*, \dots, a^*)$ be a strict $k$-type union equilibrium of a symmetric $n$-player game $(N, \bar{A}, (\pi_i))$.
Then it is stable of order~$k$.
\end{theoremEnd}

\begin{proofEnd}
Let $(a^*, \dots, a^*)$ be a strict $k$-type union equilibrium.
Suppose that $(\mu, b)$ is a monomorphic configuration in which $a^*$ is a strictly dominant strategy for $\theta$, 
so that $(a^*, \dots, a^*)$ is the aggregate outcome of $(\mu, b)$. 
Consider an arbitrary $k$th-order mutation distribution $\upsilon$, 
and let $\witilde{\mu} = (1 - \varepsilon)\mu + \varepsilon\upsilon$ be the post-entry type distribution. 
Then we have to prove that $(\mu, b)$ is stable against the mutants in $\spt\upsilon$ playing any focal equilibrium.
We do this by induction on the number of mutant players who appear among a player's opponents.

First, let all opponents of a player be incumbents.
If a mutant player chooses a strategy different from $a^*$ against the incumbent opponents who always play $a^*$,
then since $(a^*, \dots, a^*)$ is a strict $k$-type union equilibrium,
this mutant type must incur a loss,
and so for $\varepsilon$ small enough, it will go extinct, as desired.
Therefore, we shall only consider the case in which mutants always choose $a^*$ against all their opponents being incumbents.
This implies at once that
$F^{\witilde{\mu}}_{\witilde{\theta}_j}(0) = F^{\witilde{\mu}}_{\theta}(0)$
for all $\witilde{\theta}_j\in \spt\upsilon$.

Now for each $m$ with $1\leq m \leq n-1$,
suppose inductively that for any $1\leq s\leq m$ and any $\witilde{b}\in B(\spt\witilde{\mu};b)$, we have
$\witilde{b}_i(\witilde{\theta}_1, \dots, \witilde{\theta}_s, \theta, \dots, \theta) = a^*$
for all $i\in N$ and all $\witilde{\theta}_1$, \dots,~$\witilde{\theta}_s\in \spt\upsilon$.
From this, we also have
\begin{equation}\label{eq:weighted-sum-0317}
F^{\witilde{\mu}}_{\witilde{\theta}_j}(s - 1) = F^{\witilde{\mu}}_{\theta}(s - 1)
\end{equation}
for all $\witilde{\theta}_j\in \spt\upsilon$.
Next, we treat the case where a matched $n$-tuple contains $m+1$ mutants.
We can write 
$\sum_{\witilde{\theta}_j\in \spt\upsilon} \varepsilon \upsilon(\witilde{\theta}_j) 
\bigl( F^{\witilde{\mu}}_{\witilde{\theta}_j}(m) - F^{\witilde{\mu}}_{\theta}(m) \bigr)$ 
as 
\[
\sum_{\witilde{\theta}_j\in \spt\upsilon} \varepsilon \upsilon(\witilde{\theta}_j) 
\sum_{[\bdsb{\witilde{\theta}}_m, \bdsb{\theta}_{r(m)}]\in \mathcal{G}_{(m, r(m))}} 
C_{[\bdsb{\witilde{\theta}}_m, \bdsb{\theta}_{r(m)}]}^{\witilde{\mu}} 
\Bigl( 
\pi_1 \big( \witilde{b}(\witilde{\theta}_j, \bdsb{\witilde{\theta}}_m, \bdsb{\theta}_{r(m)}) \big) - 
\pi_1 \big( \witilde{b}(\theta, \bdsb{\witilde{\theta}}_m, \bdsb{\theta}_{r(m)}) \big) \Bigr), 
\] 
and by the induction hypothesis,
this is equal to
\[
\sum_{\bdsb{\wihat{\theta}}\in (\spt\upsilon)^{m+1}\times \{ \theta \}^{n-1-m}}
\frac{(n-1)!}{(m+1)! (n-1-m)!} \times
\witilde{\mu}^{n}(\bdsb{\wihat{\theta}}) \times
\sum_{j=1}^{m+1} \Bigl( \pi_j\big( \witilde{b}(\bdsb{\wihat{\theta}}) \big) - \pi_j(a^*, \dots, a^*) \Bigr).
\]
For each $\bdsb{\wihat{\theta}}\in (\spt\upsilon)^{m+1}\times \{ \theta \}^{n-1-m}$,
notice that
$\big| D\big( ( \witilde{b}_j(\bdsb{\wihat{\theta}}) )_{j = 1, \dots, m+1} \big) \big|\leq k$
since $\left| \spt\upsilon \right| = k$.
If there is a mutant in such a match $\bdsb{\wihat{\theta}}$ choosing a strategy different from $a^*$,
then since $(a^*, \dots, a^*)$ is a strict $k$-type union equilibrium,
we must have 
$\sum_{\witilde{\theta}_j\in \spt\upsilon} \varepsilon \upsilon(\witilde{\theta}_j) 
\bigl( F^{\witilde{\mu}}_{\witilde{\theta}_j}(m) - F^{\witilde{\mu}}_{\theta}(m) \bigr)< 0$. 
Furthermore,
applying~\eqref{eq:weighted-sum-0317},
we conclude that there exists at least one mutant type in $\spt\upsilon$
earning a strictly lower average fitness than the incumbent $\theta$ as long as $\varepsilon$ is sufficiently small,
as desired.
Thus, for $\bdsb{\wihat{\theta}}\in (\spt\upsilon)^{m+1}\times \{ \theta \}^{n-1-m}$,
we only have to consider all mutants in $\bdsb{\wihat{\theta}}$ choosing the strategy $a^*$.
It follows that
$F^{\witilde{\mu}}_{\witilde{\theta}_j}(m) = F^{\witilde{\mu}}_{\theta}(m)$
for all $\witilde{\theta}_j\in \spt\upsilon$.
This completes the induction,
and so the strategy profile $(a^*, \dots, a^*)$ is $k$th-order stable.
\end{proofEnd}

When the parameter $k$ in Definition~\ref{def:k-unionNE} equals the total number of players $n$, 
the strict $n$-type union equilibrium concept 
encompasses all possible coalitional groups, where each player may either deviate to an alternative strategy or remain unchanged. 
In an environment with an $n$-player underlying game,
the number of distinct strategies chosen by a mutant subgroup cannot exceed $n$, 
even if an unlimited number of mutant types enter the population. 
Based on this observation, 
we expect this equilibrium condition to ensure not only $n$th-order stability but also infinite-order stability. 
Alternatively, 
this can be viewed from another perspective: 
a strict $n$-type union equilibrium inherently maximizes the sum of material payoffs. 
Consequently, 
by Lemma~\ref{prop:20240802}, 
its payoff profile must lie on the cooperative Pareto frontier. 
It then follows from Theorem~\ref{prop:20240312} and the remark after Theorem~\ref{prop:20150311} that 
this equilibrium outcome is indeed infinite-order stable.

\begin{theoremEnd}[]{cor}\label{prop:20210130}
Let $(a^*, \dots, a^*)$ be a strict $n$-type union equilibrium of a symmetric $n$-player game $(N, \bar{A}, (\pi_i))$.
Then it is stable of all orders.
\end{theoremEnd}

\subsubsection*{Symmetric $2\times 2$ Games} 

Finally, we give a characterization of multi-mutation stable outcomes for the class of symmetric $2\times 2$ games,
which are most common in the evolutionary game theory literature.
For the single-mutation case,
\citet{eDek-jEly-oYil:ep} show that
existence of an efficient pure strategy is necessary and sufficient for a stable outcome to exist in symmetric $2\times 2$ games,
with the exception of a set of measure zero.
In the previous examples,
we have seen that games with stable strategy profiles may not have higher-order stable ones.
However,
we shall see by Theorem~\ref{prop:20160106}\ref{2b2:pu} that
any efficient pure symmetric strategy profile in a symmetric $2\times 2$ game is not only stable, but also stable of all orders.
Moreover,
a symmetric $2\times 2$ game without any efficient pure strategy definitely has no infinite-order stable outcomes. 
(This is because when $2A< B + C$, the pair $\pi(\sigma^*, \sigma^*)$ defined in Lemma~\ref{prop:20160104} 
does not lie on $P(\mathcal{F}_{\co})$, to which we apply parts~\ref{Snc:o1} and~\ref{Snc:o3} of Theorem~\ref{prop:20150411}.)
Thus, in symmetric $2\times 2$ games,
the existence of an efficient pure strategy is completely equivalent to the existence of an infinite-order stable outcome.

For characterizing multi-mutation stable outcomes in symmetric $2\times 2$ games,
let the game with the material payoff function $\pi$ be defined by the following bimatrix game $G$.
\begin{table}[!h]
\centering
\renewcommand{\arraystretch}{1.2}
    \begin{tabular}{r|c|c|}
      \multicolumn{1}{c}{} & \multicolumn{1}{c}{$a_1$} & \multicolumn{1}{c}{$a_2$}\\ \cline{2-3}
      $a_1$ & $A$, $A$ & $B$, $C$ \\  \cline{2-3}
      $a_2$ & $C$, $B$ & $D$, $D$ \\  \cline{2-3}
    \end{tabular}
\end{table}

\noindent
Without loss of generality, suppose that $A\geq D$.
We shall need the following lemma, and it can be shown by a direct calculation.

\begin{theoremEnd}[]{lem}\label{prop:20160104}
When $2A\geq B + C$, the strategy $a_1$ is efficient with the fitness value $A$.
When $2A< B + C$, the efficient strategy $\sigma^*$ is such that $\sigma^*(a_1) = \frac{B + C - 2D}{2(B + C - A -D)}$
with the fitness value $\frac{(B + C - 2A)^2}{4(B + C - A -D)} + A$,
which is equal to $\frac{(B + C - 2D)^2}{4(B + C - A -D)} + D$.
\end{theoremEnd}

\begin{Rem}
More precisely,
the pure strategy $a_1$ is the unique efficient strategy if $2A\geq B + C$ and $A> D$.
If $2A> B + C$ and $A = D$, all efficient strategies are the two pure strategies $a_1$ and $a_2$;
if $2A = B + C$ and $A = D$, all strategies are efficient.
Finally,
if $2A< B + C$,
then the mixed strategy $\sigma^*$ is the unique efficient strategy.
\end{Rem}

\begin{theoremEnd}[no link to theorem, restate]{thm}\label{prop:20160106}
Under the symmetric $2\times 2$ game $G$ described above, we have the following results.
\begin{enumerate}[label=\upshape (\roman*)]
\item If $(a_1, a_1)$ is efficient, then it is infinite-order stable.\label{2b2:pu}
\item If $(a_1, a_1)$ is not efficient, then the unique efficient strategy profile $(\sigma^*, \sigma^*)$ is stable if and only if $B = C> A$.
      Furthermore, there is no stable aggregate outcome of order greater than $1$ whenever $B = C> A$.\label{2b2:mi}
\end{enumerate}
\end{theoremEnd}

\begin{proofEnd}
To prove~\ref{2b2:pu}, let $(a_1, a_1)$ be efficient.
Then, by Lemma~\ref{prop:20160104}, we must have $2A\geq B + C$,
and hence by Lemma~\ref{prop:20240802} $\pi(a_1, a_1)\in P(\mathcal{F}_{\co})$.
If $A> C$, then $(a_1, a_1)$ is a strict Nash equilibrium,
and so by Theorem~\ref{prop:20160105}\ref{Ssc:o3} we see that $(a_1, a_1)$ is infinite-order stable.
Next let $A\leq C$, which implies $A\geq B$. 
We shall show that there exists an infinite-order stable configuration with the aggregate outcome $(a_1, a_1)$.

Let $(\mu,b)$ be a monomorphic configuration in which the incumbent type $\theta\in \spt\mu$ is defined by
$\theta(a_i, a_j) = 1$ if $(a_i, a_j) = (a_2, a_2)$ and $\theta(a_i, a_j) = 0$ otherwise,
and the incumbents choose the strategy profile $b(\theta, \theta) = (a_1, a_1)$.
For any positive integer $r$, 
consider an $r$th-order mutation distribution $\upsilon$. 
Let $\spt\upsilon = \{\witilde{\theta}_1, \dots, \witilde{\theta}_r\}$, 
and let $\witilde{\mu} = (1 - \varepsilon)\mu + \varepsilon\upsilon$. 
Then for any $\witilde{b}\in B(\spt\witilde{\mu};b)$,
the post-entry average fitnesses of the incumbent type $\theta$ and $\witilde{\theta}_i\in \spt\upsilon$ are, respectively,
\[
\varPi_{\theta}(\witilde{\mu};\witilde{b}) = (1-\varepsilon)A +
\sum_{j=1}^r \varepsilon\upsilon(\witilde{\theta}_j) \pi_1\big(\witilde{b}(\theta, \witilde{\theta}_j)\big)
\]
and
\[
\varPi_{\witilde{\theta}_i}(\witilde{\mu};\witilde{b}) =
(1-\varepsilon) \pi_1\big(\witilde{b}(\witilde{\theta}_i, \theta)\big) +
\sum_{j=1}^r \varepsilon\upsilon(\witilde{\theta}_j) \pi_1\big(\witilde{b}(\witilde{\theta}_i, \witilde{\theta}_j)\big).
\]
Let us consider the possible cases.
\vspace{2.5pt}

\noindent
\textbf{Case 1}:
$B = A\leq C$. We have $A = B = C$ since $2A\geq B + C$.
Suppose that $A = D$. Then clearly $\varPi_{\theta}(\witilde{\mu};\witilde{b}) = \varPi_{\witilde{\theta}_i}(\witilde{\mu};\witilde{b})$
for all $\witilde{\theta}_i\in \spt\upsilon$.
Let $A> D$. 
If there exists $\witilde{\theta}_i\in \spt\upsilon$ choosing $\witilde{b}_1(\witilde{\theta}_i, \theta)$ not equal to $a_1$, 
then $\witilde{b}_2(\witilde{\theta}_i, \theta) = a_2$, and hence
$\varPi_{\theta}(\witilde{\mu};\witilde{b}) > \varPi_{\witilde{\theta}_i}(\witilde{\mu};\witilde{b})$
as long as $\varepsilon$ is sufficiently small.
If $\witilde{b}_1(\witilde{\theta}_i, \theta) = a_1$ for all $\witilde{\theta}_i\in \spt\upsilon$, 
then since $A = C$, we have $\varPi_{\theta}(\witilde{\mu};\witilde{b}) = A$ 
regardless of the strategy the incumbent type $\theta$ chooses against the mutants, 
and thus $\varPi_{\theta}(\witilde{\mu};\witilde{b})\geq \varPi_{\witilde{\theta}_i}(\witilde{\mu};\witilde{b})$
for all $\witilde{\theta}_i\in \spt\upsilon$.
\vspace{2.5pt}

\noindent
\textbf{Case 2}:
$B< A\leq C$.
If $\witilde{b}_1(\witilde{\theta}_i, \theta)$ chosen by some $\witilde{\theta}_i\in \spt\upsilon$ is a mixed strategy,
then $\witilde{b}_2(\witilde{\theta}_i, \theta) = a_2$,
and hence for $\varepsilon$ small enough,
we obtain
$\varPi_{\theta}(\witilde{\mu};\witilde{b}) > \varPi_{\witilde{\theta}_i}(\witilde{\mu};\witilde{b})$.
Thus, for any mutant type $\witilde{\theta}_i$,
we let $\witilde{b}_1(\witilde{\theta}_i, \theta) = a_1$ or $a_2$.
If $\witilde{b}_1(\witilde{\theta}_i, \theta) = a_1$ for some $\witilde{\theta}_i\in \spt\upsilon$, 
then the incumbent type $\theta$ is indifferent between $a_1$ and $a_2$.
Suppose that $\witilde{b}_2(\witilde{\theta}_i, \theta) \neq a_1$.
Then
$\varPi_{\theta}(\witilde{\mu};\witilde{b}) > \varPi_{\witilde{\theta}_i}(\witilde{\mu};\witilde{b})$
as long as $\varepsilon$ is sufficiently small.
Therefore,
we only have to consider $\witilde{b}_2(\witilde{\theta}_i, \theta) = a_1$
whenever $\witilde{b}_1(\witilde{\theta}_i, \theta) = a_1$.
In addition,
if $\witilde{b}_1(\witilde{\theta}_i, \theta) = a_2$ for some $\witilde{\theta}_i\in \spt\upsilon$,
it must be that $\witilde{b}_2(\witilde{\theta}_i, \theta) = a_2$.

Now let $A> D$,
and suppose that there exists $\witilde{\theta}_i\in \spt\upsilon$ choosing $a_2$ against $\theta$.
Then since $\witilde{b}_2(\witilde{\theta}_i, \theta) = a_2$ whenever $\witilde{b}_1(\witilde{\theta}_i, \theta) = a_2$,
we obtain $\pi_1\big(\witilde{b}(\witilde{\theta}_i, \theta)\big) = D$.
It follows that
$\varPi_{\theta}(\witilde{\mu};\witilde{b}) > \varPi_{\witilde{\theta}_i}(\witilde{\mu};\witilde{b})$
as long as $\varepsilon$ is sufficiently small.
Therefore, when $A> D$,
we only consider $\witilde{b}_1(\witilde{\theta}_i, \theta) = a_1$ for all $\witilde{\theta}_i\in \spt\upsilon$.
Since $\witilde{b}_2(\witilde{\theta}_i, \theta) = a_1$ whenever $\witilde{b}_1(\witilde{\theta}_i, \theta) = a_1$,
the difference between the post-entry average fitnesses of the incumbent type $\theta$ and $\witilde{\theta}_i\in \spt\upsilon$ is
\begin{equation}\label{eq:Pco}
\varPi_{\theta}(\witilde{\mu};\witilde{b}) - \varPi_{\witilde{\theta}_i}(\witilde{\mu};\witilde{b})
= \sum_{j=1}^r \varepsilon\upsilon(\witilde{\theta}_j)
\Bigl( A - \pi_1\big(\witilde{b}(\witilde{\theta}_i, \witilde{\theta}_j)\big) \Bigr).
\end{equation}
As in the proof of part~\ref{Ssc:o3} of Theorem~\ref{prop:20160105},
the conditions $A> D$ and $2A\geq B + C$ must imply that
\[
\sum_{i=1}^r \varepsilon\upsilon(\witilde{\theta}_i)
\Bigl( \varPi_{\theta}(\witilde{\mu};\witilde{b}) - \varPi_{\witilde{\theta}_i}(\witilde{\mu};\witilde{b}) \Bigr)\geq 0,
\]
and thus for any $\varepsilon$,
either
$\varPi_{\theta}(\witilde{\mu};\witilde{b}) = \varPi_{\witilde{\theta}_i}(\witilde{\mu};\witilde{b})$
for all $\witilde{\theta}_i\in \spt\upsilon$, or
$\varPi_{\theta}(\witilde{\mu};\witilde{b}) > \varPi_{\witilde{\theta}_i}(\witilde{\mu};\witilde{b})$
for some $\witilde{\theta}_i\in \spt\upsilon$.
Finally, suppose that $A = D$.
Then since we only have to consider
$\witilde{b}(\witilde{\theta}_i, \theta) = (a_1, a_1)$ or $(a_2, a_2)$ for any $\witilde{\theta}_i\in \spt\upsilon$,
the pair $\pi\big(\witilde{b}(\witilde{\theta}_i, \theta)\big)$ of fitness payoffs must be $(A, A)$,
no matter which strategy pair they choose.
Again, we obtain~\eqref{eq:Pco},
and so we can conclude that the mutants coexist with the incumbents,
or the mutation set $\spt\upsilon$ fails to invade.
This completes the proof of part~\ref{2b2:pu}.

\smallskip

For part~\ref{2b2:mi}, suppose that $(a_1, a_1)$ is not efficient.
Then, by Lemma~\ref{prop:20160104}, the inequality $2A< B + C$ holds and
the efficient strategy $\sigma^*$ is such that $\sigma^*(a_1) = \frac{B + C - 2D}{2(B + C - A -D)}$, which will be denoted by $\alpha^*$.
Let $(\mu,b)$ be a stable configuration with the aggregate outcome $(\sigma^*, \sigma^*)$, 
and suppose that $B\neq C$. 
Given $\theta$,~$\theta'\in \spt\mu$, 
we let $b(\theta, \theta') = (\sigma, \sigma')$. 
Then by part~\ref{Snc:o1} of Theorem~\ref{prop:20150411}, we have $\pi(\sigma, \sigma') = \pi(\sigma^*, \sigma^*)$. 
Using the assumption $B\neq C$, 
it is easy to verify that the equality $\pi_1(\sigma, \sigma') = \pi_2(\sigma, \sigma')$ implies that $\sigma = \sigma'$. 
Since the efficient strategy here is unique,
we conclude that $b(\theta, \theta') = (\sigma^*, \sigma^*)$ for all $\theta$,~$\theta'\in \spt\mu$.
Consider an indifferent type $\witilde{\theta}^0$ entering the population with its population share $\varepsilon$.
Then for any $\witilde{b}\in B(\spt\witilde{\mu};b)$, the average fitness of $\theta\in \spt\mu$ is
\[ 
\varPi_{\theta}(\witilde{\mu};\witilde{b}) = (1-\varepsilon) \pi_1(\sigma^*,\sigma^*) + \varepsilon \pi_1\big( \witilde{b}(\theta, \witilde{\theta}^0) \big);
\]
the average fitness of the mutant type $\witilde{\theta}^0$ is
\[
\varPi_{\witilde{\theta}^0}(\witilde{\mu};\witilde{b}) =
(1-\varepsilon) \sum_{\theta\in \spt\mu}\mu(\theta) \pi_1\big( \witilde{b}(\witilde{\theta}^0, \theta) \big) +
\varepsilon \pi_1\big(\witilde{b}(\witilde{\theta}^0, \witilde{\theta}^0)\big).
\]
Suppose that $\witilde{b}_1(\witilde{\theta}^0, \theta) = \sigma^*$ for all $\theta\in \spt\mu$.
Then each $\theta\in \spt\mu$ is indifferent between $a_1$ and $a_2$ when matched against $\witilde{\theta}^0$,
and we let $\witilde{b}_2(\witilde{\theta}^0, \theta) = (q_{\theta}, 1 - q_{\theta})$,
where $q_{\theta}\in [0, 1]$. 
It follows that 
\[
\pi_1\big( \witilde{b}(\witilde{\theta}^0, \theta) \big) =
\alpha^* q_{\theta} A + \alpha^* (1 - q_{\theta}) B + (1 - \alpha^*) q_{\theta} C + (1 - \alpha^*)(1 - q_{\theta})D.
\]
Notice that the stability of $(\mu,b)$ requires
$\pi_1(\sigma^*,\sigma^*)\geq \sum_{\theta\in \spt\mu}\mu(\theta) \pi_1\big( \witilde{b}(\witilde{\theta}^0, \theta) \big)$.
If $q_{\theta} = 1$ for all $\theta\in \spt\mu$,
then the inequality $B\geq C$ can be derived (using $\pi_1(\sigma^*,\sigma^*) = \frac{(B + C - 2A)^2}{4(B + C - A -D)} + A$).
If $q_{\theta} = 0$ for all $\theta\in \spt\mu$,
then the inequality $C\geq B$ can be derived (using $\pi_1(\sigma^*,\sigma^*) = \frac{(B + C - 2D)^2}{4(B + C - A -D)} + D$). 
Thus $B = C$, 
which contradicts the assumption $B\neq C$. 
Therefore, the efficient strategy profile $(\sigma^*, \sigma^*)$ being stable implies that $B = C> A$.

Conversely, suppose that $B = C> A$.
Define a preference type $\theta^*$ by $\theta^*(a_1, a_1) = 1 - \alpha^*$, $\theta^*(a_2, a_2) = \alpha^*$, and $\theta^*(a_i, a_j) = 0$ whenever $a_i\neq a_j$.
Let $(\mu,b)$ be a monomorphic configuration, where the population consists of $\theta^*$ for which $b(\theta^*, \theta^*) = (\sigma^*, \sigma^*)$.
Consider a mutant type $\witilde{\theta}$ entering the population with a population share $\varepsilon$.
Then for any $\witilde{b}\in B(\spt\witilde{\mu};b)$, 
the difference between the post-entry average fitnesses of $\theta^*$ and $\witilde{\theta}$ is
\[
\varPi_{\theta^*}(\witilde{\mu};\witilde{b}) - \varPi_{\witilde{\theta}}(\witilde{\mu};\witilde{b})
= (1-\varepsilon)
\Bigl( \pi_1(\sigma^*,\sigma^*) - \pi_1\big( \witilde{b}(\witilde{\theta}, \theta^*) \big) \Bigr) +
\varepsilon
\Bigl( \pi_1\big( \witilde{b}(\theta^*, \witilde{\theta}) \big) - \pi_1\big(\witilde{b}(\witilde{\theta}, \witilde{\theta})\big) \Bigr).
\]
Suppose that $\witilde{b}_1(\witilde{\theta}, \theta^*) = (p, 1 - p)$ for some $p\in [0, 1]$.
Then $\witilde{b}_2(\witilde{\theta}, \theta^*) = a_1$ if $p> \alpha^*$;
$\witilde{b}_2(\witilde{\theta}, \theta^*) = a_2$ if $p< \alpha^*$;
$\theta^*$ is indifferent between $a_1$ and $a_2$ if $p = \alpha^*$.
By $B = C$, the fitness value $\pi_1(\sigma^*,\sigma^*)$ satisfies
\begin{align*}
\pi_1(\sigma^*,\sigma^*) &= \alpha^* A + (1 - \alpha^*) B\\
&= \alpha^* C + (1 - \alpha^*) D.
\end{align*}
When $p> \alpha^*$, then since $B = C> A$, we have
\[
\pi_1\big( \witilde{b}(\witilde{\theta}, \theta^*) \big) = pA + (1 - p)C
< \alpha^* A + (1 - \alpha^*) B = \pi_1(\sigma^*,\sigma^*).
\]
Similarly, when $p< \alpha^*$,
we also have $\pi_1\big( \witilde{b}(\witilde{\theta}, \theta^*) \big)< \pi_1(\sigma^*,\sigma^*)$ since $B = C> D$.
Thus, in both cases,
$\varPi_{\theta^*}(\witilde{\mu};\witilde{b})> \varPi_{\witilde{\theta}}(\witilde{\mu};\witilde{b})$
as long as $\varepsilon$ is sufficiently small.
Finally, let $p = \alpha^*$.
Then, by applying $B = C$ again,
it is not hard to verify that
\[
\pi_1\big( \witilde{b}(\witilde{\theta}, \theta^*) \big) = \pi_1\big( \witilde{b}(\theta^*, \witilde{\theta}) \big) = \pi_1(\sigma^*,\sigma^*),
\]
no matter which strategy the incumbent $\theta^*$ chooses.
Clearly, the efficiency of $(\sigma^*,\sigma^*)$ would imply
$\varPi_{\theta^*}(\witilde{\mu};\witilde{b})\geq \varPi_{\witilde{\theta}}(\witilde{\mu};\witilde{b})$.
Hence, $(\sigma^*,\sigma^*)$ is a stable aggregate outcome.

In addition, 
whenever $B = C> A$, 
it is easy to see that the strategy profile $(\sigma^*,\sigma^*)$ is strongly Pareto dominated by $(a_1, a_2)$ (or $(a_2, a_1)$). 
Given any stable configuration, 
by Theorem~\ref{prop:20150411}\ref{Snc:o1}, 
the payoff pair incumbents can obtain when matched against each other must be $\pi(\sigma^*,\sigma^*)$, 
which does not lie on $P(\mathcal{F}_{\nc})$. 
Thus, by Lemma~\ref{prop:20150924} and Theorem~\ref{prop:20150411}\ref{Snc:o2},
if $B = C> A$,
there is no stable aggregate outcome of order greater than $1$.
\end{proofEnd}

In the case where $2A< B + C$,
it is easy to verify that the efficient strategy profile $(\sigma^*,\sigma^*)$ is a Nash equilibrium of the symmetric game $G$
if and only if $B = C$.
Therefore,
by Theorem~\ref{prop:20160106}\ref{2b2:mi},
we can say that whenever $2A< B + C$,
the strategy profile $(\sigma^*,\sigma^*)$ is stable if and only if it is a Nash equilibrium of this game.\footnote{%
Example~\ref{exam:20151001} can serve to demonstrate the properties in case $2A< B + C$.}


\section{The Multi-population Case}\label{sec:multi}


Unlike the single-population case,
many relevant strategic interactions take place among individuals from separate populations.
In such a situation,
player positions are identified,
and players in different positions may have different action sets.
Thus players' actual fitnesses would be given by an asymmetric underlying game.\footnote{%
Asymmetric interactions in the indirect evolutionary approach can be captured by more than one setting
(see \citet{ysTu-wtJuang:epmp}),
as in the standard evolutionary game theory (see \citet{pTay:essttp} and \citet{rCre:scegt} for multi-population stability criteria,
and see \citet{rSel:nessaac} for embedding the asymmetric interactions in a larger symmetric game),
where evolution acts upon strategies.}
Here, we study evolutionarily stable preferences against multiple mutations in a multi-population matching setting.
Suppose that there are $n$ populations, each of which consists of an infinite number of individuals.
In every match, $n$ players are drawn independently, one at random from each population, to play the game.
The essential point here, unlike in Section~\ref{sec:single}, is that a preference type never interacts with itself;
a player will only meet opponents coming from the other populations.

Let $(N, (A_i), (\pi_i))$ be the underlying game (with $n$ players' material payoff functions).
This game can be either symmetric or asymmetric.
For the set $\Theta$ of von Neumann--Morgenstern utility functions on $\prod_{i\in N}A_i$,
we denote by $\mathcal{M}(\Theta^n)$
the set of joint distributions of $n$ independent random variables defined on the same sample space $\Theta$ with finite supports.
Suppose that the probability distribution of preference types of individuals in all $n$ populations is $\mu\in \mathcal{M}(\Theta^n)$,
and let $\spt\mu$ denote the support of $\mu$.
Then for a matched $n$-tuple $\bdsb{\theta} = (\theta_1, \dots, \theta_n)\in \spt\mu$,
its probability is $\mu(\bdsb{\theta}) = \prod_{i\in N} \mu_i(\theta_i)$,
where $\mu_i$ is the marginal distribution over all types in the $i$th population.
In addition,
for $k\in N$, the conditional probability
$\mu(\bdsb{\theta}_{-k}|\theta_k)$ is equal to $\prod_{i\neq k} \mu_i(\theta_i)$,
and hence we write $\mu_{-k}(\bdsb{\theta}_{-k})$ for $\mu(\bdsb{\theta}_{-k}|\theta_k)$.
Of course,
$\spt\mu = \prod_{i\in N}\spt\mu_i$,
and similarly we shall write $\spt\mu_{-k}$ for $\prod_{i\neq k} \spt\mu_i$.
Let, as in Section~\ref{sec:single},
$\Gamma(\mu)$ denote the game
which allows us to analyze strategic interactions among players~$1$, \dots,~$n$ having types in $\spt\mu_1$, \dots,~$\spt\mu_n$, respectively.
Combining this with the underlying game,
we refer to the pair $[(N, (A_i), (\pi_i)), \Gamma(\mu)]$ as an \emph{environment}.

\subsection*{Configurations} 

A \emph{strategy} of player~$i$ in $\Gamma(\mu)$ is a function $b_i\colon \spt\mu\to \Delta A_i$.
For $(\theta_i, \bdsb{\theta}_{-i})\in \spt\mu$,
player~$i$ of type $\theta_i$ chooses the strategy $b_i(\theta_i, \bdsb{\theta}_{-i})$ against opponents of types $\bdsb{\theta}_{-i}$
in the game induced by their preferences.
The function $b\colon \spt\mu\to\prod_{i\in N}\Delta A_i$ defined by $b(\bdsb{\theta}) = (b_1(\bdsb{\theta}),\dots,b_n(\bdsb{\theta}))$
is an \emph{equilibrium} in $\Gamma(\mu)$ if for all $\bdsb{\theta}\in\spt\mu$,
the strategy profile $b(\bdsb{\theta})$ is a Nash equilibrium of the normal-form game induced by $\bdsb{\theta}$, that is, for each $i\in N$,
\[
b_i(\bdsb{\theta})\in \argmax_{\sigma_i\in\Delta A_i} \theta_i\big( \sigma_i, b_{-i}(\bdsb{\theta}) \big).
\]
Let $B(\spt\mu)$ be the set of all such equilibria in $\Gamma(\mu)$.
%
Given $\mu\in \mathcal{M}(\Theta^n)$, let $b\in B(\spt\mu)$ be chosen.
We call the pair $(\mu,b)$ a \emph{configuration},
which describes the state of the $n$ populations.
The \emph{aggregate outcome} of $(\mu,b)$ is the correlated strategy $\varphi_{\mu,b}\in \Delta(\prod_{i\in N}A_i)$ defined by
\[
\varphi_{\mu,b}(a_1, \dots, a_n) = \sum_{\bdsb{\theta}\in \spt\mu} \mu(\bdsb{\theta}) \prod_{i\in N} b_i(\bdsb{\theta})(a_i)
\]
for all $(a_1, \dots, a_n)\in \prod_{i\in N}A_i$.
More specifically, 
if the configuration is monomorphic,\footnote{%
Here, for a monomorphic configuration, all individuals in the same population have the same preference type.} 
the strategy profile $\bdsb{\sigma}\in \prod_{i\in N}\Delta A_i$ 
chosen by all matched $n$-tuples is also regarded as the aggregate outcome. 
For $i\in N$, the \emph{average fitness} of a type $\theta_i\in \spt\mu_i$ with respect to $(\mu,b)$ is 
\[
\varPi_{\theta_i}(\mu;b) = \sum_{\bdsb{\theta}'_{-i}\in \spt\mu_{-i}} \mu_{-i}(\bdsb{\theta}'_{-i})\pi_i\big( b(\theta_i,\bdsb{\theta}'_{-i}) \big).
\]
A configuration $(\mu,b)$ is said to be \emph{balanced} if for each $i\in N$,
the equality $\varPi_{\theta_i}(\mu;b) = \varPi_{\theta'_i}(\mu;b)$ holds for all $\theta_i$,~$\theta'_i\in \spt\mu_i$.

\subsection*{Evolutionary Stability} 

Let $\mu\in \mathcal{M}(\Theta^n)$ be the incumbent preference distribution.
For $i\in N$,
the types in $\spt\mu_i$ are the so-called \emph{incumbents}. 
To study multi-mutation stability in a multi-population matching setting, 
consider a type distribution $\upsilon\in \mathcal{M}(\Theta^n)$ satisfying 
$\spt\mu_i \cap \spt\upsilon_i = \varnothing$ for all $i\in N$. 
Let $J_{\upsilon}$ defined by 
\[ 
J_{\upsilon} = \{\, i\in N \mid \spt\upsilon_i \neq \varnothing \,\} 
\] 
be a nonempty set. 
For every $j\in J_{\upsilon}$, 
suppose that the new types in $\spt\upsilon_j$, called \emph{mutants}, 
enter the $j$th population simultaneously, 
and that a fraction $\varepsilon_j$ of the $j$th population is replaced by these mutants. 
Define 
$\supnorm{(\left| \spt\upsilon_i \right|)} = \max_{i\in J_{\upsilon}} \left| \spt\upsilon_i \right|$. 
We call such $\upsilon$ an \emph{$m$th-order mutation distribution} if $\supnorm{(\left| \spt\upsilon_i \right|)} = m$. 
For each $i\in N$, 
the post-entry type distribution in the $i$th population is 
\[ 
\witilde{\mu}_i = 
\begin{cases}
    (1 - \varepsilon_i)\mu_i + \varepsilon_i\upsilon_i & \text{if $i\in J_{\upsilon}$,}\\
    \mu_i & \text{if $i\notin J_{\upsilon}$.}
\end{cases}
\] 
For these marginal distributions $\witilde{\mu}_i$, 
$i = 1$, \dots,~$n$, 
we denote their joint distribution by $\witilde{\mu}$. 
We write $\bdsb{\varepsilon}_{\upsilon}$ for $(\varepsilon_i)_{i\in J_{\upsilon}}$, 
and further define $\supnorm{\bdsb{\varepsilon}_{\upsilon}} = \max_{i\in J_{\upsilon}} \varepsilon_i$. 
Then obviously 
$\supnorm{\bdsb{\varepsilon}_{\upsilon}}\rightarrow 0$ if and only if $\varepsilon_i\rightarrow 0$ for all $i\in J_{\upsilon}$. 
In addition, 
we sometimes call $\prod_{i\in J_{\upsilon}} \spt\upsilon_i$ 
an \emph{$m$th-order mutation set} if $\supnorm{(\left| \spt\upsilon_i \right|)} = m$.

For evolutionary stability,
we require that incumbents in a post-entry environment continue to play their pre-entry actions 
against incumbent opponents from the other populations. 
However,
whenever a match includes non-incumbent players,
all equilibria of the game induced by their respective preferences must be taken into account.

\begin{Def}
Let $(\mu,b)$ and $(\witilde{\mu}, \witilde{b})$ be two configurations satisfying $\spt\mu \subset \spt\witilde{\mu}$.
The equilibrium $\witilde{b}$ in the game $\Gamma(\witilde{\mu})$ is called \emph{focal} with respect to $b$
if $\witilde{b}(\bdsb{\theta}) = b(\bdsb{\theta})$ for all $\bdsb{\theta}\in \spt\mu$.
The set of all focal equilibria with respect to $b$ is denoted by $B(\spt\witilde{\mu}; b)$,
and is called the \emph{focal set}.\footnote{%
In the multi-population case, the focal set is always nonempty after any entry of mutants,
and the post-entry aggregate outcome can be made arbitrarily close to the pre-entry one by taking the population shares of mutants small enough,
as discussed in the remark after Definition~\ref{def:focal}.}
\end{Def}

As in the single-population case, 
our multi-mutation stability concept introduced in Section~\ref{sec:basic} is used here. 
The definition of a multi-mutation stable configuration for the multi-population case is as follows.


\begin{Def}\label{def:20210427}
In $[(N, (A_i), (\pi_i)), \Gamma(\mu)]$,
a configuration $(\mu,b)$ is \emph{$m$th-order stable}, or \emph{stable of order~$m$}, 
if for any $m$th-order mutation distribution $\upsilon\in \mathcal{M}(\Theta^n)$ 
and for any $\witilde{b}\in B(\spt\witilde{\mu};b)$,
there exists some $\bar{\varepsilon}\in (0,1)$ such that
either~\ref{Cm:wiped-out} or~\ref{Cm:coexist} is satisfied 
for all $\bdsb{\varepsilon}_{\upsilon}$ with $\supnorm{\bdsb{\varepsilon}_{\upsilon}}\in (0,\bar{\varepsilon})$.

\begin{enumerate}[label=\upshape (\roman*)]
\item For some $j\in J_{\upsilon}$ and for some $\witilde{\theta}_j^s\in \spt\upsilon_j$, 
$\varPi_{\witilde{\theta}_j^s}(\witilde{\mu};\witilde{b})< \varPi_{\theta_j}(\witilde{\mu};\witilde{b})$ for every $\theta_j\in \spt\mu_j$.\label{Cm:wiped-out}
\item $\varPi_{\theta_i}(\witilde{\mu};\witilde{b}) = \varPi_{\wihat{\theta}_i}(\witilde{\mu};\witilde{b})$
for every $i\in N$ and for every $\theta_i$,~$\wihat{\theta}_i\in \spt \witilde{\mu}_i$.\label{Cm:coexist}
\end{enumerate}
A configuration $(\mu,b)$ is said to be \emph{stable} if it is first-order stable. 
A configuration $(\mu,b)$ is said to be \emph{infinite-order stable} if it is stable of any large orders.
\end{Def}

We also call a strategy profile $\bdsb{\sigma}\in\prod_{i\in N}\Delta A_i$ \emph{$m$th-order stable}, 
or \emph{stable of order~$m$}, 
if it is the aggregate outcome of an $m$th-order stable configuration. 
Proposition~\ref{prop:20160329} will show that any stable configuration in the multi-population case is balanced.

For the standard setting of evolutionary game theory, where players are programmed to play some strategies,
there are two popular ways of extending the definition of an evolutionarily stable strategy (ESS)
from a single-population setting to a multi-population setting.
One is introduced by \citet{pTay:essttp}, and the other is introduced by \citet{rCre:scegt};
see also \citet{jWei:egt} and \citet{wSan:pged}.\footnote{%
The multi-population stability criterion suggested by \citet{pTay:essttp} is based on
average fitness values aggregated over all player positions.
\citet{rCre:scegt} introduced a seemingly weaker criterion for multi-population evolutionary stability:
the stability is ensured if entrants earn strictly less than the residents in at least one population.
Indeed, these two criteria are equivalent in standard evolutionary game theory.}
Our multi-population stability concept of preference evolution is consistent with that of the two-species ESS formulated by \citet{rCre:scegt}.
It can be shown that our stability criterion is weaker than that due to \citet{pTay:essttp}; see \citet{ysTu-wtJuang:epmp}.

In a multi-population matching setting, 
the underlying game can be either symmetric or asymmetric.
Even if this game is symmetric,
the evolutionary outcome induced by multi-population matching can diverge significantly from that of single-population matching.
Below, we present an example to illustrate this divergence. 
This example also verifies that a stable outcome may be destabilized by a higher-order mutation set in the multi-population case.

\begin{example}\label{exam:20150930}
Let the following symmetric two-player game $G$ represent the fitness assignment,
and suppose that all players have the same action set $\{a_{T}, a_{M}, a_{B}\}$.

\begin{table}[!h]
\centering
\renewcommand{\arraystretch}{1.2}
    \begin{tabular}{r|c|c|c|}
      \multicolumn{1}{c}{} & \multicolumn{1}{c}{$a_{T}$} & \multicolumn{1}{c}{$a_{M}$}
                           & \multicolumn{1}{c}{$a_{B}$}\\ \cline{2-4}
      $a_{T}$ & $4$, $4$ & $0$, $0$ & $0$, $0$ \\  \cline{2-4}
      $a_{M}$ & $0$, $0$ & $0$, $0$ & $3$, $9$ \\  \cline{2-4}
      $a_{B}$ & $0$, $0$ & $9$, $3$ & $0$, $0$ \\  \cline{2-4}
    \end{tabular}
\end{table}

\noindent
For this symmetric game $G$,
observe that $a_{T}$ is the unique efficient strategy,
and that
$P(\mathcal{F}_{\nc}) = \{ (4, 4), (3, 9), (9, 3) \}$ and
$P(\mathcal{F}_{\co})\bigcap \mathcal{F}_{\nc} = \{ (3, 9), (9, 3) \}$.
The three Pareto-efficient payoff pairs $(4, 4)$, $(3, 9)$, and $(9, 3)$ correspond exactly to the three strategy pairs
$(a_{T}, a_{T})$, $(a_{M}, a_{B})$, and $(a_{B}, a_{M})$, respectively.
Moreover, the three strategy pairs are all strictly strong Nash equilibria of $G$.
With these properties,
the stable outcomes can be examined more easily.

If the environment is considered from the point of view of a single-population matching setting (discussed in Section~\ref{sec:single}),
then applying Theorems~\ref{prop:20150411} and~\ref{prop:20160105} (and Proposition~\ref{prop:20201130}), 
we see that the strategy pair $(a_T, a_T)$ is the unique stable outcome in $G$,
which is also stable of order~$2$, but not stable of the higher orders.
On the other hand,
if we consider players drawn from two different populations,
then the strategy pair $(a_T, a_T)$ is not the only stable outcome in $G$.
More precisely, if we apply Theorems~\ref{prop:20150130} and~\ref{prop:20141028},
we can conclude that
all the other stable strategy pairs are $(a_{M}, a_{B})$ and $(a_{B}, a_{M})$,
and both of them are infinite-order stable.

Here we directly verify that under two-population matching,
the strategy pair $(a_{T}, a_{T})$ is stable of order~$1$,
but not stable of order~$2$.
First suppose that $a_{T}$ is strictly dominant for the incumbents in each of the two populations.
Consider two mutant types entering the first and second populations, respectively,
with sufficiently small population shares.
Since $(a_{T}, a_{T})$ is a strict Nash equilibrium of $G$,
any mutant type will go extinct if mutants of this type choose any other strategy against the incumbents.
In addition, when two mutants from different populations are matched,
Pareto efficiency implies that
there is no way of improving some mutant's fitness without worsening the fitness of the other mutant.
Thus, the strategy pair $(a_{T}, a_{T})$ is a stable aggregate outcome; see also Theorem~\ref{prop:20141028}.

To see that $(a_{T}, a_{T})$ is not stable of order~$2$,
let $(\mu,b)$ be an arbitrary configuration with the aggregate outcome $(a_{T}, a_{T})$.
Then we have $b(\bdsb{\theta}) = (a_{T}, a_{T})$ for all $\bdsb{\theta}\in \spt\mu$.
For each $i\in \{1, 2\}$,
consider two indifferent types $\witilde{\theta}_i^1$ and $\witilde{\theta}_i^2$
which enter the $i$th population simultaneously and which choose their strategies such that
(i) 
$\witilde{b}(\witilde{\theta}_1^1, \witilde{\theta}_2^1) = (a_{M}, a_{B})$,
$\witilde{b}(\witilde{\theta}_1^1, \witilde{\theta}_2^2) = (a_{B}, a_{M})$,
$\witilde{b}(\witilde{\theta}_1^2, \witilde{\theta}_2^1) = (a_{B}, a_{M})$,
and $\witilde{b}(\witilde{\theta}_1^2, \witilde{\theta}_2^2) = (a_{M}, a_{B})$; 
(ii) when a mutant $\witilde{\theta}_i^s$ is matched against an incumbent $\theta_{-i}\in \spt\mu_{-i}$, 
we have $\witilde{b}(\witilde{\theta}_i^s, \theta_{-i}) = (a_{T}, a_{T})$.\footnote{%
For two-player games in the multi-population case, 
the subscript $-i$ on $\theta_{-i}$ and $\mu_{-i}$ denotes the population index $j$ with $j\in \{1, 2\}\setminus \{i\}$, 
and the pair $(\witilde{\theta}_i^s, \theta_{-i})$ 
refers to the ordered pair consisting of two preference types arranged in ascending order according to their population indices.} 
Then the average fitness of any incumbent $\theta_i$ is $\varPi_{\theta_i}(\witilde{\mu};\witilde{b}) = 4$.
However, if we let $\witilde{\theta}_i^s$'s population share equal $\varepsilon/2$ for all $i$,~$s\in \{1,2\}$,
then the average fitness of any mutant $\witilde{\theta}_i^s$ is
\[
\varPi_{\witilde{\theta}_i^s}(\witilde{\mu};\witilde{b}) =
(1 - \varepsilon)\cdot 4 + \frac{\varepsilon}{2}\cdot (3 + 9) = 4 + 2\varepsilon.
\]
Thus $\varPi_{\witilde{\theta}_i^s}(\witilde{\mu};\witilde{b}) > \varPi_{\theta_i}(\witilde{\mu};\witilde{b})$ for all $i$,~$s\in \{1,2\}$,
which means that the strategy pair $(a_{T}, a_{T})$ is not stable of order~$2$.
\end{example}

Unsurprisingly, 
a multi-mutation stable configuration must imply the stability of lower orders in the multi-population case. 
The proof proceeds analogously to that of Lemma~\ref{prop:20150924}.

\begin{theoremEnd}[]{lem}\label{prop:20150901}
In $[(N, (A_i), (\pi_i)), \Gamma(\mu)]$,
if $(\mu,b)$ is a $k$th-order stable configuration with $k> 1$, then it is stable of order~$m$ for any $m< k$.
\end{theoremEnd}

\subsection*{Stability Results} 

For a stable configuration in the multi-population model,
\citet{ysTu-wtJuang:epmp} demonstrate that
the payoff profile for any matched $n$-tuple of incumbents must lie on the noncooperative Pareto frontier.
Here, under mild assumptions,
the Pareto-efficient payoff profile induced by a sufficiently high-order stable configuration
can be represented as a point on the cooperative Pareto frontier.
Our assumptions require the underlying game's payoff matrices to have strictly rational entries 
and players to choose mixed strategies with rational probabilities. 
The assumptions we use are that the payoff matrices for the underlying game have exclusively rational entries,
and that the probabilities in the mixed strategies chosen by players are all rational numbers.
We denote by $\rcvx{\mathcal{F}}_{\nc}$ the set of payoff profiles achievable via such \emph{rational} mixed strategies.

\begin{textAtEnd} 
Before proving Theorem~\ref{prop:20150130}\ref{Mnc:o2}, 
we need the lemmas below. 
\end{textAtEnd}

\begin{theoremEnd}[all end]{lem}\label{prop:20210604}
Let $A$ be an $m\times n$ matrix over $\mathbb{Q}$, and let $b\in \mathbb{Q}^m$.
Then the set $\{\, x\in \mathbb{Q}^n \mid Ax = b \,\}$ of rational solutions to the system $Ax = b$
is dense in the set $\{\, x\in \mathbb{R}^n \mid Ax = b \,\}$ of real solutions.
\end{theoremEnd}

\begin{proofEnd}
Observe that for any one of the two solution sets,
it is nonempty if and only if
the rank of the matrix $A$ equals the rank of the augmented matrix $(A, b)$.
Let $x_0\in \mathbb{Q}^n$ satisfy $Ax_0 = b$,
and let $v_1$, \dots,~$v_r\in \mathbb{Q}^n$ form a basis of the space
$\{\, x\in \mathbb{Q}^n \mid Ax = 0 \,\}$.
Then
\[
\{\, x\in \mathbb{Q}^n \mid Ax = b \,\} =
\{\, x_0 + \lambda_1 v_1 + \dots + \lambda_r v_r \mid \lambda_1, \dots, \lambda_r\in \mathbb{Q} \,\}.
\]
Since $v_1$, \dots,~$v_r$ also form a basis of the space
$\{\, x\in \mathbb{R}^n \mid Ax = 0 \,\}$,
we can write the set of real solutions as
\[
\{\, x\in \mathbb{R}^n \mid Ax = b \,\} =
\{\, x_0 + \lambda_1 v_1 + \dots + \lambda_r v_r \mid \lambda_1, \dots, \lambda_r\in \mathbb{R} \,\}.
\]
Since $\mathbb{Q}$ is dense in $\mathbb{R}$,
the conclusion of the lemma follows directly.
\end{proofEnd}

\begin{theoremEnd}[all end]{lem}\label{prop:20210606}
Let $r$ be a positive integer
and let $(s_1, \dots, s_m)$ be an $m$-tuple in $\{ 1, \dots, r \}^m$.
Then for each $j\in \{0, 1, \dots, r - 1\}$, there are $r^{m-1}$ such $m$-tuples for which
$s_1 + \dots +s_m \equiv j \bmod{r}$.
\end{theoremEnd}

\begin{proofEnd}
For $j = 0$, $1$, \dots,~$r - 1$, let
\[
X_j = \{\, (s_1, \dots, s_m)\in \{ 1, \dots, r \}^m \mid s_1 + \dots +s_m \equiv j \bmod{r} \,\},
\]
and let $f\colon X_0\to X_j$ be defined by $(s_1, s_2, \dots, s_m)\mapsto (\tau(s_1), s_2, \dots, s_m)$,
where $\tau(s_1)$ is an integer between $1$ and $r$ such that $s_1 + j \equiv \tau(s_1) \bmod{r}$.
Then $f$ is a bijection from $X_0$ onto $X_j$.
Furthermore, since $\{ 1, \dots, r \}^m$ is the disjoint union of the sets $X_j$,
$j = 0$, $1$, \dots,~$r - 1$,
the result follows.
\end{proofEnd}

\begin{theoremEnd}[no link to theorem, restate]{thm}\label{prop:20150130}
Let $(\mu,b)$ be a configuration in $[(N, (A_i), (\pi_i)), \Gamma(\mu)]$.
\begin{enumerate}[label=\upshape (\roman*)]
\item If $(\mu,b)$ is stable, then $\pi\big( b(\bdsb{\theta}) \big)\in P(\mathcal{F}_{\nc})$ for every $\bdsb{\theta}\in \spt\mu$.\label{Mnc:o1}
\item Let $\pi(\prod_{i\in N}A_i)\subset \mathbb{Q}^n$ and let $(\mu,b)$ be infinite-order stable.
Then for $\bdsb{\theta}\in \spt\mu$,
the payoff profile $\pi\big( b(\bdsb{\theta}) \big)\in P(\mathcal{F}_{\co})$
when $\pi\big( b(\bdsb{\theta}) \big)\in \rcvx{\mathcal{F}}_{\nc}$.\label{Mnc:o2}
\end{enumerate}
\end{theoremEnd}

\begin{proofEnd}
To prove~\ref{Mnc:o1}, suppose that there exists $\bdsb{\bar{\theta}} = (\bar{\theta}_1, \dots, \bar{\theta}_n)\in\spt\mu$ for which
$\pi\big( b(\bdsb{\bar{\theta}}) \big)\notin P(\mathcal{F}_{\nc})$.
This means that there exists $\bdsb{\sigma}\in\prod_{i\in N}\Delta A_i$ such that
$\pi_i(\bdsb{\sigma})\geq \pi_i\big(b(\bdsb{\bar{\theta}})\big)$ for all $i\in N$
with strict inequality for at least one $i$.
For each $i$th population, consider an indifferent type $\witilde{\theta}_i$ entering the population,
and write $\bdsb{\witilde{\theta}}$ for $(\witilde{\theta}_1, \dots, \witilde{\theta}_n)$.
Let the chosen equilibrium $\witilde{b}\in B(\spt\witilde{\mu};b)$ satisfy: 
(1) $\witilde{b}(\bdsb{\witilde{\theta}}) = \bdsb{\sigma}$;
(2) for any proper subset $J\subset N$ and for any $\bdsb{\theta}_{-J}\in \prod_{i\notin J} \spt\mu_i$,
we have $\witilde{b}(\bdsb{\witilde{\theta}}_J, \bdsb{\theta}_{-J}) = b(\bdsb{\bar{\theta}}_J, \bdsb{\theta}_{-J})$.
Then for each $i\in N$,
the difference between the post-entry average fitnesses of $\witilde{\theta}_i$ and the incumbent type $\bar{\theta}_i$ is
\[
\varPi_{\witilde{\theta}_i}(\witilde{\mu};\witilde{b}) - \varPi_{\bar{\theta}_i}(\witilde{\mu};\witilde{b}) =
\witilde{\mu}_{-i}(\bdsb{\witilde{\theta}}_{-i})
\Bigl( \pi_i(\bdsb{\sigma}) - \pi_i\big(b(\bdsb{\bar{\theta}})\big) \Bigr).
\]
Since $\bdsb{\sigma}$ Pareto dominates $b(\bdsb{\bar{\theta}})$,
we conclude that $(\mu,b)$ is not a stable configuration.

\smallskip

For part~\ref{Mnc:o2},
suppose that there exists $\bdsb{\bar{\theta}} = (\bar{\theta}_1, \dots, \bar{\theta}_n)\in \spt\mu$ for which 
$\pi\big( b(\bdsb{\bar{\theta}}) \big)\in \rcvx{\mathcal{F}}_{\nc}\setminus P(\mathcal{F}_{\co})$.
Then there is a payoff point $\bdsb{w}\in \mathcal{F}_{\co}$ such that 
$w_i> \pi_i\big(b(\bdsb{\bar{\theta}})\big)$ for all $i\in T$ and
$w_i = \pi_i\big(b(\bdsb{\bar{\theta}})\big)$ for all $i\in N\setminus T$ where $T$ is a nonempty subset of $N$. 
Since $\bdsb{w}$ is a convex combination of the pure-payoff profiles,
there exist some $\bdsb{a}^1$, \dots,~$\bdsb{a}^k\in \prod_{i\in N}A_i$ and
positive real numbers $\lambda_1$, \dots,~$\lambda_k$ with $\sum_{j=1}^{k} \lambda_j = 1$ such that
$\bdsb{w} = \sum_{j=1}^{k} \lambda_j\pi(\bdsb{a}^j)$.
Consider the equations
$\pi_i(\bdsb{a}^1) x_1 + \dots + \pi_i(\bdsb{a}^k) x_k = \pi_i\big(b(\bdsb{\bar{\theta}})\big)$, $i\in N\setminus T$,
and the equation $x_1 + \dots + x_k = 1$ together,
which form a system of linear equations for $(x_1, \dots, x_k)$ with coefficients in the field $\mathbb{Q}\subset \mathbb{R}$.
Then the $k$-tuple $(\lambda_1, \dots, \lambda_k)$ of real numbers mentioned above is a solution of this system of linear equations.
By Lemma~\ref{prop:20210604},
the real solution $(\lambda_1, \dots, \lambda_k)$ can be approximated arbitrarily closely by rational solutions to this system.
Thus we can find another payoff point $\bdsb{u}$, written
\[
\bdsb{u} = \frac{r_1}{r}\pi(\bdsb{a}^1) + \dots + \frac{r_k}{r}\pi(\bdsb{a}^k)
\]
where $\sum_{j=1}^k r_j = r$ and each $r_j\in \mathbb{Z}^{+}$,
such that
$u_i> \pi_i\big(b(\bdsb{\bar{\theta}})\big)$ for all $i\in T$ and
$u_i = \pi_i\big(b(\bdsb{\bar{\theta}})\big)$ for all $i\in N\setminus T$.

For each $i$th population, 
consider a mutation distribution $\upsilon_i$ with equal probabilities 
assigned to indifferent types 
$\witilde{\theta}_i^1$, \dots,~$\witilde{\theta}_i^r$ which form the set $\spt\upsilon_i$. 
We define $\bdsb{\alpha}^1$, \dots,~$\bdsb{\alpha}^r\in \prod_{i\in N}A_i$, respectively, to be
$\bdsb{\alpha}^1 = \dots = \bdsb{\alpha}^{r_1} = \bdsb{a}^1$, and
\[
\bdsb{\alpha}^{r_1 + \dots + r_{j-1} + 1} = \dots = \bdsb{\alpha}^{r_1 + \dots + r_{j-1} + r_j} = \bdsb{a}^j
\]
for $j = 2$, \dots,~$k$.
Now suppose that the chosen focal equilibrium $\witilde{b}$ satisfies 
$\witilde{b}(\bdsb{\witilde{\theta}}_J, \bdsb{\theta}_{-J}) = b(\bdsb{\bar{\theta}}_J, \bdsb{\theta}_{-J})$ 
for any proper subset $J\subset N$, 
and for any $\bdsb{\witilde{\theta}}_J\in \prod_{i\in J}\spt\upsilon_i$ and any $\bdsb{\theta}_{-J}\in \prod_{i\notin J} \spt\mu_i$. 
In addition,
when mutants are matched against one another,
suppose that
\[
\witilde{b}(\witilde{\theta}_1^{s_1}, \dots, \witilde{\theta}_n^{s_n}) =
\bdsb{\alpha}^{\tau(s_1, \dots, s_n)+1}
\]
for any $s_1$, \dots,~$s_n\in \{1, \dots, r\}$,
where $\tau(s_1, \dots ,s_n)$ is an integer between $0$ and $r-1$ satisfying $s_1 + \dots + s_n \equiv \tau(s_1, \dots ,s_2) \bmod{r}$. 
Let $\witilde{\mu}_i = (1 - \varepsilon)\mu_i + \varepsilon\upsilon_i$ for all $i\in N$. 
Then for each $i\in N$ and for each $\witilde{\theta}_i^s\in \spt\upsilon_i$, 
by using Lemma~\ref{prop:20210606}, we have
\begin{align*}
\sum_{\bdsb{\witilde{\theta}}_{-i}\in \prod_{j\neq i}\spt\upsilon_j}
\witilde{\mu}_{-i}(\bdsb{\witilde{\theta}}_{-i})
\pi_i\big( \witilde{b}(\witilde{\theta}_i^s, \bdsb{\witilde{\theta}}_{-i}) \big)
&= \frac{\varepsilon^{n-1}}{r} \big( r_1\pi_i(\bdsb{a}^1) + \dots + r_k\pi_i(\bdsb{a}^k) \big)\\
&= \varepsilon^{n-1} u_i =
\sum_{\bdsb{\witilde{\theta}}_{-i}\in \prod_{j\neq i}\spt\upsilon_j}
\witilde{\mu}_{-i}(\bdsb{\witilde{\theta}}_{-i}) u_i,
\end{align*}
and thus the difference between the post-entry average fitnesses of $\witilde{\theta}_i^s$ 
and the incumbent type $\bar{\theta}_i$ is
\begin{align*}
\varPi_{\witilde{\theta}_i^s}(\witilde{\mu};\witilde{b}) - \varPi_{\bar{\theta}_i}(\witilde{\mu};\witilde{b})
&= \sum_{\bdsb{\witilde{\theta}}_{-i}\in \prod_{j\neq i}\spt\upsilon_j} \witilde{\mu}_{-i}(\bdsb{\witilde{\theta}}_{-i})
\Bigl( \pi_i\big( \witilde{b}(\witilde{\theta}_i^s, \bdsb{\witilde{\theta}}_{-i}) \big) -
\pi_i\big( \witilde{b}(\bar{\theta}_i, \bdsb{\witilde{\theta}}_{-i}) \big) \Bigr)\\
&= \sum_{\bdsb{\witilde{\theta}}_{-i}\in \prod_{j\neq i}\spt\upsilon_j} \witilde{\mu}_{-i}(\bdsb{\witilde{\theta}}_{-i})
\Bigl( u_i - \pi_i\big( b(\bdsb{\bar{\theta}}) \big) \Bigr).
\end{align*}
Since the payoff point $\bdsb{u}$ Pareto dominates $\pi\big(b(\bdsb{\bar{\theta}})\big)$,
the configuration $(\mu,b)$ is not infinite-order stable.
\end{proofEnd}

\begin{Rem}
The assumption that $\pi(\prod_{i\in N}A_i)\subset \mathbb{Q}^n$ in Theorem~\ref{prop:20150130}\ref{Mnc:o2}
is needed to ensure that
when a payoff profile $\pi(\bdsb{\bar{\sigma}})$ achievable by rational mixed strategies does not lie on $ P(\mathcal{F}_{\co})$,
we can find a convex combination of the pure-payoff profiles with rational coefficients
such that $\pi(\bdsb{\bar{\sigma}})$ is Pareto dominated by it.\footnote{%
Methodologically, 
rational probabilities enable a constructive proof by translating strategy proportions into explicit play schedules 
across a finite number of mutant types per population, 
where the denominator specifies the total count of these types.} 
In general $n$-player games, 
such a convex combination may not exist.
\citet{2017arXiv170501454T} have given an example of this situation in a three-player game,
and proved, however, that for any payoff pair in a general two-player game, 
if it does not lie on $ P(\mathcal{F}_{\co})$, 
it must be Pareto dominated by some convex combination, with rational coefficients, of the pure-payoff pairs.
Consequently, 
\emph{for any two-player underlying game in a two-population matching setting, 
every infinite-order stable configuration $(\mu,b)$ possesses the property that
$\pi\big( b(\bdsb{\theta}) \big)\in P(\mathcal{F}_{\co})$ for all $\bdsb{\theta}\in \spt\mu$}. 
This follows from an argument similar to that in the proof of Theorem~\ref{prop:20150130}\ref{Mnc:o2}.
\end{Rem}

Even though a game may have multiple Pareto-efficient allocations,
\citet{ysTu-wtJuang:epmp} show that
the payoff profiles for all matched $n$-tuple of incumbents in a stable configuration have the same Pareto-efficient form,
and thus that any stable configuration must be balanced. 
It also follows that 
the fitness vector of a stable aggregate outcome consists of the fitness values obtained in any match of the incumbents. 
These stability properties also hold in this paper.\footnote{%
Proposition~\ref{prop:20160329} is due to Theorem~3.8 of \citet{ysTu-wtJuang:epmp},
and the proofs for the results also work here.}

\begin{theoremEnd}[]{prop}\label{prop:20160329}
Let $(\mu,b)$ be a stable configuration in $[(N, (A_i), (\pi_i)), \Gamma(\mu)]$ with the aggregate outcome $\varphi_{\mu,b}$.
Then $(\mu,b)$ is balanced, and for each $i\in N$,
\[ 
\varPi_{\theta_i}(\mu;b) = \pi_i\big( b(\bdsb{\theta}) \big) = \pi_i(\varphi_{\mu,b}) 
\]
for any $\theta_i\in \spt\mu_i$ and any $\bdsb{\theta}\in \spt\mu$. 
\end{theoremEnd}

Given a stable configuration in the multi-population case, 
all incumbent types in the same population obtain the same fitness from all incumbent interactions. 
This interestingly reveals that 
the fitness an incumbent can obtain is independent of both its own type and its incumbent opponents' types.

Next, we focus on finding sufficient conditions for multi-mutation stability,
beginning with the two-population case.
Example~\ref{exam:20150930} suggests that in an arbitrary two-player game,
a strict Nash equilibrium with payoff pair lying on the noncooperative Pareto frontier may be stable,
but may not be second-order stable. 
We need a stronger condition for higher-order stability.

\begin{theoremEnd}[no link to theorem, restate]{thm}\label{prop:20141028}
Let $(a^*_1,a^*_2)$ be a strict Nash equilibrium of $(\{1, 2\}, (A_i), (\pi_i))$.
\begin{enumerate}[label=\upshape (\roman*)]
\item If $\pi(a^*_1,a^*_2)\in P(\mathcal{F}_{\nc})$, then $(a^*_1,a^*_2)$ is stable.\label{Msc:o1}
\item If $\pi(a^*_1,a^*_2)\in P(\mathcal{F}_{\co})$, then $(a^*_1,a^*_2)$ is infinite-order stable.\label{Msc:o2}
\end{enumerate}
\end{theoremEnd}

\begin{proofEnd}
Let $(a^*_1,a^*_2)$ be a strict Nash equilibrium of $(\{1, 2\}, (A_i), (\pi_i))$.
Suppose that $(\mu, b)$ is a monomorphic configuration where each $i$th population consists of $\theta_i^*$
for which $a^*_i$ is strictly dominant, 
so that $(a^*_1, a^*_2)$ is the aggregate outcome of $(\mu, b)$. 
Consider an arbitrary mutation distribution $\upsilon\in \mathcal{M}(\Theta^2)$. 
In the case when $\left| J_{\upsilon} \right|=1$, 
it is clear that mutants have no fitness advantage, since $(a^*_1,a^*_2)$ is a strict Nash equilibrium.

Next, suppose that $J_{\upsilon} = \{1, 2\}$. 
Let $\witilde{\mu}_i = (1 - \varepsilon_i)\mu_i + \varepsilon_i\upsilon_i$ for $i = 1$,~$2$, 
and let $\witilde{b}\in B(\spt\witilde{\mu};b)$ be chosen. 
Then for each $i$, 
the post-entry average fitnesses of $\theta_i^*$ 
and $\witilde{\theta}_i\in \spt\upsilon_i$ are, respectively,
\[
\varPi_{\theta_i^*}(\witilde{\mu};\witilde{b}) = (1 - \varepsilon_{-i}) \pi_i(a^*_1,a^*_2) +
\sum_{\witilde{\theta}_{-i}\in \spt\upsilon_{-i}} \varepsilon_{-i} \upsilon_{-i}(\witilde{\theta}_{-i}) \pi_i\big( a^*_i, \witilde{b}_{-i}(\theta_i^*, \witilde{\theta}_{-i}) \big)
\] 
and 
\[
\varPi_{\witilde{\theta}_i}(\witilde{\mu};\witilde{b}) = (1 - \varepsilon_{-i})
\pi_i\big( \witilde{b}_i(\witilde{\theta}_i, \theta_{-i}^*), a^*_{-i} \big) +
\sum_{\witilde{\theta}_{-i}\in \spt\upsilon_{-i}} \varepsilon_{-i} \upsilon_{-i}(\witilde{\theta}_{-i}) \pi_i\big( \witilde{b}(\witilde{\theta}_i, \witilde{\theta}_{-i}) \big). 
\]
Suppose that $\witilde{b}_i(\witilde{\theta}_i, \theta_{-i}^*) \neq a^*_i$ for some $i$ and some $\witilde{\theta}_i\in \spt\upsilon_i$.
Then since $(a^*_1, a^*_2)$ is a strict Nash equilibrium,
we have
$\varPi_{\theta_i^*}(\witilde{\mu};\witilde{b}) > \varPi_{\witilde{\theta}_i}(\witilde{\mu};\witilde{b})$
whenever $\varepsilon_{-i}$ is sufficiently small,
as desired.
Thus in the following,
we let
$\witilde{b}_i(\witilde{\theta}_i, \theta_{-i}^*) = a^*_i$ for all $i$ and all $\witilde{\theta}_i\in \spt\upsilon_i$.
Then the difference between the post-entry average fitnesses of the incumbent $\theta_i^*$
and a mutant type $\witilde{\theta}_i$ is
\[
\varPi_{\theta_i^*}(\witilde{\mu};\witilde{b}) - \varPi_{\witilde{\theta}_i}(\witilde{\mu};\witilde{b}) =
\sum_{\witilde{\theta}_{-i}\in \spt\upsilon_{-i}} \varepsilon_{-i} \upsilon_{-i}(\witilde{\theta}_{-i})
\Bigl( \pi_i(a^*_1,a^*_2) - \pi_i\big( \witilde{b}(\witilde{\theta}_i, \witilde{\theta}_{-i}) \big) \Bigr).
\]
To prove~\ref{Msc:o1}, 
consider the case where $\left| \spt\upsilon_1 \right| = \left| \spt\upsilon_2 \right| = 1$. 
Then the claim follows directly from the fact that $\pi(a^*_1,a^*_2)\in P(\mathcal{F}_{\nc})$.

\smallskip

For part~\ref{Msc:o2}, we first define a payoff point $\bdsb{u}\in \mathcal{F}_{\co}$ by
\[
\bdsb{u} = \sum_{\witilde{\theta}_{1}\in \spt\upsilon_{1}}\sum_{\witilde{\theta}_{2}\in \spt\upsilon_{2}} 
\frac{\varepsilon_{1} \upsilon_{1}(\witilde{\theta}_{1}) \varepsilon_{2} \upsilon_{2}(\witilde{\theta}_{2})}
{\sum_{\witilde{\theta}'_{1}\in \spt\upsilon_{1}}\sum_{\witilde{\theta}'_{2}\in \spt\upsilon_{2}}
\varepsilon_{1} \upsilon_{1}(\witilde{\theta}'_{1}) \varepsilon_{2} \upsilon_{2}(\witilde{\theta}'_{2})}
\pi\big( \witilde{b}(\witilde{\theta}_1, \witilde{\theta}_2) \big),
\]
which can be written in the form
\[
\bdsb{u}
= \sum_{\witilde{\theta}_{i}\in \spt\upsilon_{i}} 
\frac{\varepsilon_{i} \upsilon_{i}(\witilde{\theta}_{i})}
{\sum_{\witilde{\theta}'_{i}\in \spt\upsilon_{i}} \varepsilon_{i} \upsilon_{i}(\witilde{\theta}'_{i})}
\sum_{\witilde{\theta}_{-i}\in \spt\upsilon_{-i}} 
\frac{\varepsilon_{-i} \upsilon_{-i}(\witilde{\theta}_{-i})}
{\sum_{\witilde{\theta}'_{-i}\in \spt\upsilon_{-i}} \varepsilon_{-i} \upsilon_{-i}(\witilde{\theta}'_{-i})}
\pi\big( \witilde{b}(\witilde{\theta}_i, \witilde{\theta}_{-i}) \big),
\]
where $i = 1$ or $i = 2$.
Since $\pi(a^*_1, a^*_2)\in P(\mathcal{F}_{\co})$,
it follows that $\pi(a^*_1, a^*_2) = \bdsb{u}$ or $\pi_i(a^*_1,a^*_2)> u_i$ for some $i\in \{1, 2\}$.
From this we conclude that
\[
\pi_i(a^*_1,a^*_2) = 
\sum_{\witilde{\theta}_{-i}\in \spt\upsilon_{-i}} 
\frac{\varepsilon_{-i} \upsilon_{-i}(\witilde{\theta}_{-i})}
{\sum_{\witilde{\theta}'_{-i}\in \spt\upsilon_{-i}} \varepsilon_{-i} \upsilon_{-i}(\witilde{\theta}'_{-i})}
\pi_i\big( \witilde{b}(\witilde{\theta}_i, \witilde{\theta}_{-i}) \big)
\]
for every $i$ and every $\witilde{\theta}_i$, or
\[
\pi_i(a^*_1,a^*_2) > \sum_{\witilde{\theta}_{-i}\in \spt\upsilon_{-i}} 
\frac{\varepsilon_{-i} \upsilon_{-i}(\witilde{\theta}_{-i})}
{\sum_{\witilde{\theta}'_{-i}\in \spt\upsilon_{-i}} \varepsilon_{-i} \upsilon_{-i}(\witilde{\theta}'_{-i})}
\pi_i\big( \witilde{b}(\witilde{\theta}_i, \witilde{\theta}_{-i}) \big)
\]
for some $i$ and some $\witilde{\theta}_i$.
This obviously implies that for any $\bdsb{\varepsilon}_{\upsilon}$,
either
$\varPi_{\theta_i^*}(\witilde{\mu};\witilde{b}) = \varPi_{\witilde{\theta}_i}(\witilde{\mu};\witilde{b})$
for every $i$ and every $\witilde{\theta}_i$, or
$\varPi_{\theta_i^*}(\witilde{\mu};\witilde{b}) > \varPi_{\witilde{\theta}_i}(\witilde{\mu};\witilde{b})$
for some $i$ and some $\witilde{\theta}_i$.
Hence the configuration $(\mu, b)$ is infinite-order stable.
\end{proofEnd}

As noted in the single-population case, 
it is expected that in order to ensure stability in multi-player games,
we need a stronger condition on deviations.
However,
unlike the result of Proposition~\ref{prop:20201130},
a strictly strong Nash equilibrium of a three-player game in the multi-population case may not be stable,
although it maximizes the sum of material payoffs. 
Below, we show how mutants from different populations
can interact with one another to destabilize such an efficient equilibrium outcome.

\begin{example}\label{exam:20150101}
Let the underlying game be the following three-player game,
where the action set of player~$i$ is $\{a_{i1}, a_{i2}\}$ for $i = 1$,~$2$,~$3$.

\begin{table}[!h]
\centering
\renewcommand{\arraystretch}{1.2}
    \begin{tabular}{r|c|c|}
      \multicolumn{1}{c}{} & \multicolumn{1}{c}{$a_{21}$} & \multicolumn{1}{c}{$a_{22}$}\\ \cline{2-3}
      $a_{11}$ & $3$, $3$, $3$ & $7$, $0$, $0$  \\  \cline{2-3}
      $a_{12}$ & $0$, $0$, $7$ & $0$, $0$, $7$ \\  \cline{2-3}
      \multicolumn{1}{c}{} & \multicolumn{2}{c}{$a_{31}$}
    \end{tabular} \qquad
    \begin{tabular}{r|c|c|}
      \multicolumn{1}{c}{} & \multicolumn{1}{c}{$a_{21}$} & \multicolumn{1}{c}{$a_{22}$}\\ \cline{2-3}
      $a_{11}$ & $0$, $7$, $0$ & $7$, $0$, $0$ \\  \cline{2-3}
      $a_{12}$ & $0$, $7$, $0$ & $1$, $1$, $1$ \\  \cline{2-3}
      \multicolumn{1}{c}{} & \multicolumn{2}{c}{$a_{32}$}
    \end{tabular}
\end{table}

\noindent
It is not hard to check that $(a_{11},a_{21},a_{31})$ is a strictly strong Nash equilibrium, 
which also maximizes the sum of material payoffs. 
Even so, this efficient equilibrium is not stable.
To verify this,
let $(\mu, b)$ be a configuration with the aggregate outcome $(a_{11},a_{21},a_{31})$.
Then we have $b(\bdsb{\theta}) = (a_{11},a_{21},a_{31})$ for all $\bdsb{\theta}\in \spt\mu$.
For each $i$, consider an indifferent type $\witilde{\theta}_i$ entering the $i$th population, 
and suppose that the chosen focal equilibrium $\witilde{b}$ satisfies:
for any $\theta_i\in \spt\mu_i$, $i = 1$,~$2$,~$3$,
\begin{gather*}
\witilde{b}(\witilde{\theta}_1,\theta_2,\theta_3) =
\witilde{b}(\theta_1,\witilde{\theta}_2,\theta_3) =
\witilde{b}(\theta_1,\theta_2,\witilde{\theta}_3) =
\witilde{b}(\witilde{\theta}_1,\witilde{\theta}_2,\witilde{\theta}_3) = (a_{11},a_{21},a_{31}),\\
\witilde{b}_1(\witilde{\theta}_1,\witilde{\theta}_2,\theta_3)=a_{11},\qquad
\witilde{b}_2(\witilde{\theta}_1,\witilde{\theta}_2,\theta_3)=a_{22},\\
\witilde{b}_1(\witilde{\theta}_1,\theta_2,\witilde{\theta}_3)=a_{12},\qquad
\witilde{b}_3(\witilde{\theta}_1,\theta_2,\witilde{\theta}_3)=a_{31},\\
\witilde{b}_2(\theta_1,\witilde{\theta}_2,\witilde{\theta}_3)=a_{21},\qquad
\witilde{b}_3(\theta_1,\witilde{\theta}_2,\witilde{\theta}_3)=a_{32}.
\end{gather*}
Let $\witilde{\theta}_i$'s population share equal $\varepsilon$ for all $i$.
Then for $i = 1$,~$2$,~$3$,
the difference between the post-entry average fitnesses of $\witilde{\theta}_i$ and $\theta_i$ is
\[
\varPi_{\witilde{\theta}_i}(\witilde{\mu}; \witilde{b}) - \varPi_{\theta_i}(\witilde{\mu}; \witilde{b}) =
\varepsilon + 2\varepsilon^2> 0,
\] 
regardless of the strategy an incumbent chooses against two mutant opponents. 
Therefore,
the strategy profile $(a_{11},a_{21},a_{31})$ is not stable.
\end{example}

In this example, 
mutants can earn higher average fitnesses by forming diverse coalitions; 
losses caused by certain deviations can be sufficiently compensated for by specific coalitional groups. 
To strengthen the strictly strong Nash equilibrium condition, we require that 
any subset of players will strictly decrease the sum of their payoffs if any of them deviates.

\begin{Def}\label{def:unionNE}
Let $(N, (A_i), (\pi_i))$ be a finite normal-form game.
A strategy profile $\bdsb{x}$ is a \emph{strict union equilibrium} if
for any nonempty $J\subseteq N$, 
and for any $\bdsb{\sigma}_J\in \prod_{j\in J}\Delta A_j$ with $\bdsb{\sigma}_J\neq \bdsb{x}_J$,
\[
\sum_{j\in J}\pi_j(\bdsb{x})> \sum_{j\in J}\pi_j(\bdsb{\sigma}_J, \bdsb{x}_{-J}).
\]
\end{Def}

In fact,
a strict union equilibrium in a symmetric game is just a strict $n$-type union equilibrium.
This equilibrium concept allows for all possible strategies chosen by mutants who are matched with or without the incumbents.
It is therefore not surprising that
this equilibrium outcome is strong enough to guarantee infinite-order stability, not just stability.

\begin{theoremEnd}[no link to theorem, restate]{thm}\label{prop:20210614}
Let $\bdsb{a}^*$ be a strict union equilibrium of $(N, (A_i), (\pi_i))$.
Then $\bdsb{a}^*$ is stable of all orders.
\end{theoremEnd}

\begin{proofEnd} 
Let $\bdsb{a}^* = (a^*_1, \dots, a^*_n)$ be a strict union equilibrium of $(N, (A_i), (\pi_i))$,
and suppose that each $i$th population consists of $\theta^*_i$
for which $a^*_i$ is strictly dominant.
We denote this monomorphic configuration by $(\mu,b)$,
so the aggregate outcome of $(\mu, b)$ is $\bdsb{a}^*$
and the support of $\mu$ consists of $\bdsb{\theta}^* = (\theta^*_1, \dots, \theta^*_n)$. 
Consider an arbitrary mutation distribution $\upsilon\in \mathcal{M}(\Theta^n)$, 
and let $\witilde{\mu}_i = (1 - \varepsilon_i)\mu_i + \varepsilon_i\upsilon_i$ for all $i\in N$. 
For brevity, we shall abbreviate $J_{\upsilon}$ as $J$. 
To verify infinite-order stability,
it suffices to show that
for any focal equilibrium $\witilde{b}$,
there exists $\bar{\varepsilon}\in (0,1)$ such that the inequality 
\begin{equation}\label{eq:20260727} 
\sum_{j\in J} \sum_{\witilde{\theta}_{j}\in \spt\upsilon_{j}} \varepsilon_{j} \upsilon_{j}(\witilde{\theta}_{j})
\Bigl(
\varPi_{\theta_j^*}(\witilde{\mu};\witilde{b}) - \varPi_{\witilde{\theta}_j}(\witilde{\mu};\witilde{b})
\Bigr) \geq 0
\end{equation} 
holds for all $\bdsb{\varepsilon}_{\upsilon}$ with $\supnorm{\bdsb{\varepsilon}_{\upsilon}}\in (0,\bar{\varepsilon})$. 
(This implies that for $\supnorm{\bdsb{\varepsilon}_{\upsilon}}$ small enough, 
either 
$\varPi_{\theta_j^*}(\witilde{\mu};\witilde{b})> \varPi_{\witilde{\theta}_j}(\witilde{\mu};\witilde{b})$ 
for some $j$ and some $\witilde{\theta}_j$, or 
$\varPi_{\theta_j^*}(\witilde{\mu};\witilde{b}) = \varPi_{\witilde{\theta}_j}(\witilde{\mu};\witilde{b})$ 
for all $j$ and all $\witilde{\theta}_j$.) 
By direct calculation, 
the left-hand side of~\eqref{eq:20260727} is equal to 
\begin{multline}\label{eq:20190804}
\sum_{j\in J} \sum_{\witilde{\theta}_{j}\in \witilde{\varTheta}_j}
\Bigl( \varepsilon_{j} \upsilon_{j}(\witilde{\theta}_{j}) \prod_{i\in J_{-j}}(1 - \varepsilon_{i}) \Bigr)
\Bigl( \pi_j(\bdsb{a}^*) - \pi_j\big( \witilde{b}(\witilde{\theta}_j, \bdsb{\theta}^*_{-j}) \big) \Bigr)\\
+
\sum_{j\in J} \sum_{\witilde{\theta}_{j}\in \witilde{\varTheta}_j} \sum_{\substack{|T|=1\\ T\subseteq J_{-j}}} \sum_{\bdsb{\witilde{\theta}}_T\in \witilde{\varTheta}_T}
\Bigl( \varepsilon_{j} \upsilon_{j}(\witilde{\theta}_{j}) \witilde{\mu}_T(\bdsb{\witilde{\theta}}_T) \prod_{i\in J_{-j}\setminus T}(1 - \varepsilon_{i}) \Bigr)
\Bigl( \pi_j\big( \witilde{b}(\theta^*_j, \bdsb{\witilde{\theta}}_T, \bdsb{\theta}^*_{-j -T}) \big) -
\pi_j\big( \witilde{b}(\witilde{\theta}_j, \bdsb{\witilde{\theta}}_T, \bdsb{\theta}^*_{-j -T}) \big) \Bigr)\\
+ \dots +
\sum_{j\in J} \sum_{\witilde{\theta}_{j}\in \witilde{\varTheta}_j} \sum_{\bdsb{\witilde{\theta}}_{J_{-j}}\in \witilde{\varTheta}_{J_{-j}}}
\varepsilon_{j} \upsilon_{j}(\witilde{\theta}_{j}) \witilde{\mu}_{J_{-j}}(\bdsb{\witilde{\theta}}_{J_{-j}})
\Bigl( \pi_j\big( \witilde{b}(\theta^*_j, \bdsb{\witilde{\theta}}_{J_{-j}}, \bdsb{\theta}^*_{-J}) \big) -
\pi_j\big( \witilde{b}(\witilde{\theta}_j, \bdsb{\witilde{\theta}}_{J_{-j}}, \bdsb{\theta}^*_{-J}) \big) \Bigr),
\end{multline} 
where 
$\witilde{\varTheta}_j = \spt\upsilon_j$, 
$J_{-j} = J\setminus \{j\}$,
$\witilde{\varTheta}_T = \prod_{i\in T}\witilde{\varTheta}_i$,
and $\witilde{\mu}_T(\bdsb{\witilde{\theta}}_T) = \prod_{i\in T} \witilde{\mu}_i(\witilde{\theta}_i)$ 
for $\bdsb{\witilde{\theta}}_T = (\witilde{\theta}_i)_{i\in T}$.

For any $j\in J$, any $\witilde{\theta}_j\in \witilde{\varTheta}_j$,
any $T\subseteq J_{-j}$, and any $\bdsb{\witilde{\theta}}_T\in \witilde{\varTheta}_T$, 
we write the difference, in~\eqref{eq:20190804}, 
between the two fitnesses earned by $\theta^*_j$ and $\witilde{\theta}_j$
against $(\bdsb{\witilde{\theta}}_T, \bdsb{\theta}^*_{-j -T})$ in the form
\begin{multline*}
\pi_j\big( \witilde{b}(\theta^*_j, \bdsb{\witilde{\theta}}_T, \bdsb{\theta}^*_{-j -T}) \big) -
\pi_j\big( \witilde{b}(\witilde{\theta}_j, \bdsb{\witilde{\theta}}_T, \bdsb{\theta}^*_{-j -T}) \big)\\
=
\Bigl( \pi_j\big( \witilde{b}(\theta^*_j, \bdsb{\witilde{\theta}}_T, \bdsb{\theta}^*_{-j -T}) \big) - \pi_j(\bdsb{a}^*) \Bigr) +
\Bigl( \pi_j(\bdsb{a}^*) - \pi_j\big( \witilde{b}(\witilde{\theta}_j, \bdsb{\witilde{\theta}}_T, \bdsb{\theta}^*_{-j -T}) \big) \Bigr).
\end{multline*}
Then classify all differences 
$\pi_j(\bdsb{a}^*) - \pi_j\big( \witilde{b}(\bdsb{\witilde{\theta}}_M, \bdsb{\theta}^*_{-M}) \big)$ and 
$\pi_j\big( \witilde{b}(\bdsb{\witilde{\theta}}_M, \bdsb{\theta}^*_{-M}) \big) - \pi_j(\bdsb{a}^*)$, 
$\varnothing\neq M\subseteq J$, according to the number $\left| M \right|$ of mutants in a match.
In this way,
the sum of the weighted differences taken from~\eqref{eq:20190804} for $\bdsb{\witilde{\theta}}_M$ with $\left| M \right| =k$ is either of the form
\begin{multline}\label{eq:20190805}
\sum_{j\in J} \sum_{\witilde{\theta}_{j}\in \witilde{\varTheta}_j}
\sum_{\substack{|T|=k-1\\ T\subseteq J_{-j}}} \sum_{\bdsb{\witilde{\theta}}_T\in \witilde{\varTheta}_T}
\Bigl( \varepsilon_{j} \upsilon_{j}(\witilde{\theta}_{j}) \witilde{\mu}_T(\bdsb{\witilde{\theta}}_T) \prod_{i\in J_{-j}\setminus T}(1 - \varepsilon_{i}) \Bigr)
\Bigl( \pi_j(\bdsb{a}^*) -
\pi_j\big( \witilde{b}(\witilde{\theta}_j, \bdsb{\witilde{\theta}}_T, \bdsb{\theta}^*_{-j -T}) \big) \Bigr)\\
+
\sum_{j\in J} \sum_{\witilde{\theta}_{j}\in \witilde{\varTheta}_j}
\sum_{\substack{|T|=k\\ T\subseteq J_{-j}}} \sum_{\bdsb{\witilde{\theta}}_T\in \witilde{\varTheta}_T}
\Bigl( \varepsilon_{j} \upsilon_{j}(\witilde{\theta}_{j}) \witilde{\mu}_T(\bdsb{\witilde{\theta}}_T) \prod_{i\in J_{-j}\setminus T}(1 - \varepsilon_{i}) \Bigr)
\Bigl( \pi_j\big( \witilde{b}(\theta^*_j, \bdsb{\witilde{\theta}}_T, \bdsb{\theta}^*_{-j -T}) \big) - \pi_j(\bdsb{a}^*) \Bigr) 
\end{multline}
when $1\leq k\leq |J| - 1$,
or of the form
$\sum_{j\in J} \sum_{\bdsb{\witilde{\theta}}_J\in \witilde{\varTheta}_J} \witilde{\mu}_J(\bdsb{\witilde{\theta}}_J)
\Bigl( \pi_j(\bdsb{a}^*) - \pi_j\big( \witilde{b}(\bdsb{\witilde{\theta}}_J, \bdsb{\theta}^*_{-J}) \big) \Bigr)$
when $k = |J|$.

First, it is clear that for any focal equilibrium $\witilde{b}$, 
\[
\sum_{j\in J} \sum_{\bdsb{\witilde{\theta}}_J\in \witilde{\varTheta}_J} \witilde{\mu}_J(\bdsb{\witilde{\theta}}_J)
\Bigl( \pi_j(\bdsb{a}^*) - \pi_j\big( \witilde{b}(\bdsb{\witilde{\theta}}_J, \bdsb{\theta}^*_{-J}) \big) \Bigr) =
\sum_{\bdsb{\witilde{\theta}}_J\in \witilde{\varTheta}_J} \witilde{\mu}_J(\bdsb{\witilde{\theta}}_J) \sum_{j\in J}
\Bigl( \pi_j(\bdsb{a}^*) - \pi_j\big( \witilde{b}(\bdsb{\witilde{\theta}}_J, \bdsb{\theta}^*_{-J}) \big) \Bigr) \geq 0
\]
since $\bdsb{a}^*$ is a strict union equilibrium.
Next, for any nonempty subset $H\subset J$, 
notice that the sum of the weighted differences taken from~\eqref{eq:20190805} 
for a given mutant subgroup $\bdsb{\witilde{\theta}}_H\in \witilde{\varTheta}_H$ can be written as
\begin{multline}\label{eq:20250509}
\sum_{j\in H}
\Bigl( \witilde{\mu}_H(\bdsb{\witilde{\theta}}_H) \prod_{i\in J\setminus H}(1 - \varepsilon_{i}) \Bigr)
\Bigl( \pi_j(\bdsb{a}^*) - \pi_j\big( \witilde{b}(\bdsb{\witilde{\theta}}_H, \bdsb{\theta}^*_{-H}) \big) \Bigr)\\
+
\sum_{j\in J\setminus H} \sum_{\witilde{\theta}_{j}\in \witilde{\varTheta}_j}
\Bigl( \varepsilon_{j} \upsilon_{j}(\witilde{\theta}_{j}) \witilde{\mu}_H(\bdsb{\witilde{\theta}}_H) \prod_{i\in J_{-j}\setminus H}(1 - \varepsilon_{i}) \Bigr)
\Bigl( \pi_j\big( \witilde{b}(\bdsb{\witilde{\theta}}_H, \bdsb{\theta}^*_{-H}) \big) - \pi_j(\bdsb{a}^*) \Bigr), 
\end{multline} 
and we denote it by $S_H(\bdsb{\witilde{\theta}}_H)$.
If $\witilde{b}_j(\bdsb{\witilde{\theta}}_H, \bdsb{\theta}^*_{-H}) = a^*_j$ for all $j\in H$,
then the sum $S_H(\bdsb{\witilde{\theta}}_H) = 0$.
In addition, 
if $\witilde{b}_j(\bdsb{\witilde{\theta}}_H, \bdsb{\theta}^*_{-H}) \neq a^*_j$ for some $j\in H$,
then since $\bdsb{a}^*$ is a strict union equilibrium,
and each $\frac{\varepsilon_{j} \upsilon_{j}(\witilde{\theta}_{j})}{1 - \varepsilon_j}$ 
converges to zero as $\supnorm{\bdsb{\varepsilon}_{\upsilon}}\rightarrow 0$,
it follows that there exists
$\bar{\varepsilon}_{\bdsb{\witilde{\theta}}_H}\in (0,1)$ such that 
$S_H(\bdsb{\witilde{\theta}}_H)> 0$ 
for all $\bdsb{\varepsilon}_{\upsilon}$ with 
$\supnorm{\bdsb{\varepsilon}_{\upsilon}}\in (0,\bar{\varepsilon}_{\bdsb{\witilde{\theta}}_H})$. 
So we are done
by the fact that
\begin{multline*}
\sum_{j\in J} \sum_{\witilde{\theta}_{j}\in \witilde{\varTheta}_j} \varepsilon_{j} \upsilon_{j}(\witilde{\theta}_{j})
\Bigl(
\varPi_{\theta_j^*}(\witilde{\mu};\witilde{b}) - \varPi_{\witilde{\theta}_j}(\witilde{\mu};\witilde{b})
\Bigr)\\
=
\sum_{\varnothing\neq H\varsubsetneq J} \sum_{\bdsb{\witilde{\theta}}_H\in \witilde{\varTheta}_H} S_H(\bdsb{\witilde{\theta}}_H)
+ \sum_{j\in J} \sum_{\bdsb{\witilde{\theta}}_J\in \witilde{\varTheta}_J} \witilde{\mu}_J(\bdsb{\witilde{\theta}}_J)
\Bigl( \pi_j(\bdsb{a}^*) - \pi_j\big( \witilde{b}(\bdsb{\witilde{\theta}}_J, \bdsb{\theta}^*_{-J}) \big) \Bigr),
\end{multline*}
and that $\witilde{\varTheta}_H$ is a finite set and the number of subsets of $J$ is also finite.
\end{proofEnd}

Since a strict union equilibrium does not always exist, especially in a multi-player game,
we give a slightly weaker version of this equilibrium.\footnote{%
We are grateful to an anonymous referee for this weaker equilibrium concept.}
Also, we will prove the analogue of Theorem~\ref{prop:20141028} for multi-player games.

\begin{Def}\label{def:(n-1)-unionNE}
Let $(N, (A_i), (\pi_i))$ be a finite normal-form game.
A strategy profile $\bdsb{x}$ is a \emph{strict $(n - 1)$-union equilibrium} if
for any nonempty proper subset $J$ of $N$, 
and for any $\bdsb{\sigma}_J\in \prod_{j\in J}\Delta A_j$ with $\bdsb{\sigma}_J\neq \bdsb{x}_J$,
\[
\sum_{j\in J}\pi_j(\bdsb{x})> \sum_{j\in J}\pi_j(\bdsb{\sigma}_J, \bdsb{x}_{-J}).
\]
\end{Def}

\begin{theoremEnd}[no link to theorem, restate]{thm}\label{prop:20250307}
Let $\bdsb{a}^*$ be a strict $(n - 1)$-union equilibrium of $(N, (A_i), (\pi_i))$.
\begin{enumerate}[label=\upshape (\roman*)]
\item If $\pi(\bdsb{a}^*)\in P(\mathcal{F}_{\nc})$, then $\bdsb{a}^*$ is stable.\label{gMsc:o1}
\item If $\pi(\bdsb{a}^*)\in P(\mathcal{F}_{\co})$, then $\bdsb{a}^*$ is infinite-order stable.\label{gMsc:o2}
\end{enumerate}
\end{theoremEnd}

\begin{proofEnd}
Let $\bdsb{a}^* = (a^*_1, \dots, a^*_n)$ be a strict $(n - 1)$-union equilibrium of $(N, (A_i), (\pi_i))$. 
Suppose that each $i$th population consists of $\theta^*_i$ for which $a^*_i$ is strictly dominant, 
and denote this monomorphic configuration by $(\mu, b)$. 
Thus the aggregate outcome of $(\mu, b)$ is $\bdsb{a}^*$, 
and the support of $\mu$ consists of $\bdsb{\theta}^* = (\theta^*_1, \dots, \theta^*_n)$.
To prove~\ref{gMsc:o1}, 
let $J$ ba an arbitrary nonempty subset of $N$, 
and for each $j\in J$, consider a mutant type $\witilde{\theta}_j$ entering the $j$th population with its population share $\varepsilon_j$. 
Then for any focal equilibrium $\witilde{b}$, 
as in the proof of Theorem~\ref{prop:20210614}, 
we can write
\[
\sum_{j\in J} \varepsilon_j
\Big(
\varPi_{\theta^*_j}(\witilde{\mu};\witilde{b}) - \varPi_{\witilde{\theta}_j}(\witilde{\mu};\witilde{b})
\Big)
=
\sum_{\varnothing\neq H\varsubsetneq J} S_H
+ \sum_{j\in J}
\witilde{\mu}_J(\bdsb{\witilde{\theta}}_J)
\Bigl( \pi_j(\bdsb{a}^*) - \pi_j\big( \witilde{b}(\bdsb{\witilde{\theta}}_J, \bdsb{\theta}^*_{-J}) \big) \Bigr),
\]
where $S_H$ is defined by
\begin{multline*}
S_H =
\sum_{j\in H}
\Bigl( \witilde{\mu}_H(\bdsb{\witilde{\theta}}_H) \prod_{i\in J\setminus H}(1 - \varepsilon_i) \Bigr)
\Bigl( \pi_j(\bdsb{a}^*) - \pi_j\big( \witilde{b}(\bdsb{\witilde{\theta}}_H, \bdsb{\theta}^*_{-H}) \big) \Bigr)\\
+
\sum_{j\in J\setminus H}
\Bigl( \varepsilon_j \witilde{\mu}_H(\bdsb{\witilde{\theta}}_H) \prod_{i\in J_{-j}\setminus H}(1 - \varepsilon_i) \Bigr)
\Bigl( \pi_j\big( \witilde{b}(\bdsb{\witilde{\theta}}_H, \bdsb{\theta}^*_{-H}) \big) - \pi_j(\bdsb{a}^*) \Bigr).
\end{multline*}
First let $J$ be a proper subset of $N$. 
Since $\bdsb{a}^*$ is a strict $(n - 1)$-union equilibrium, 
the argument used in proving Theorem~\ref{prop:20210614} shows that 
there exists $\bar{\varepsilon}\in (0,1)$ such that the inequality
\[
\sum_{j\in J} \varepsilon_j
\Big(
\varPi_{\theta^*_j}(\witilde{\mu};\witilde{b}) - \varPi_{\witilde{\theta}_j}(\witilde{\mu};\witilde{b})
\Big) \geq 0
\]
holds for all $(\varepsilon_j)_{j\in J}$ with $\max_{j\in J} \varepsilon_j\in (0,\bar{\varepsilon})$, 
as desired.

Now let $J = N$.
Given a nonempty proper subset $H\subset N$, 
as we show in the proof of Theorem~\ref{prop:20210614}, 
it is easy to see that
if $\witilde{b}_j(\bdsb{\witilde{\theta}}_H, \bdsb{\theta}^*_{-H}) = a^*_j$ for all $j\in H$,
then $S_H = 0$;
otherwise,
$S_H> 0$ as long as mutants' population shares are sufficiently small. 
This means that for mutants' population shares small enough, 
we have $S_H\geq 0$, 
no matter which focal equilibrium the players choose. 
If we can find a nonempty proper subset $\bar{H}\subset N$ such that
$\witilde{b}_j(\bdsb{\witilde{\theta}}_{\bar{H}}, \bdsb{\theta}^*_{-\bar{H}}) \neq a^*_j$ for some $j\in \bar{H}$,
then since $\bdsb{a}^*$ is a strict $(n - 1)$-union equilibrium, 
we have 
$\sum_{j\in \bar{H}} \pi_j(\bdsb{a}^*)> 
\sum_{j\in \bar{H}} \pi_j\big( \witilde{b}(\bdsb{\witilde{\theta}}_{\bar{H}}, \bdsb{\theta}^*_{-\bar{H}}) \big)$. 
In addition, 
since each $\frac{\varepsilon_i}{1 - \varepsilon_i}$ 
converges to zero as mutants' population shares tend to zero, 
we can choose a barrier $\bar{\varepsilon}_{\bar{H}}\in (0,1)$ such that
\[
S_{\bar{H}}
+ \sum_{i\in N}
\witilde{\mu}(\bdsb{\witilde{\theta}})
\Bigl( \pi_i(\bdsb{a}^*) - \pi_i\big( \witilde{b}(\bdsb{\witilde{\theta}}) \big) \Bigr)> 0
\] 
for all $(\varepsilon_i)_{i\in N}$ with $\max_{i\in N} \varepsilon_i\in (0, \bar{\varepsilon}_{\bar{H}})$. 
Therefore, we can eventually choose another barrier $\bar{\varepsilon}\in (0,1)$ such that the desired inequality
\[
\sum_{i\in N} \varepsilon_i
\Big(
\varPi_{\theta^*_i}(\witilde{\mu};\witilde{b}) - \varPi_{\witilde{\theta}_i}(\witilde{\mu};\witilde{b})
\Big)
=
\sum_{\varnothing\neq H\varsubsetneq N} S_H
+ \sum_{i\in N}
\witilde{\mu}(\bdsb{\witilde{\theta}})
\Bigl( \pi_i(\bdsb{a}^*) - \pi_i\big( \witilde{b}(\bdsb{\witilde{\theta}}) \big) \Bigr)> 0
\]
holds for all $(\varepsilon_i)_{i\in N}$ with $\max_{i\in N} \varepsilon_i\in (0,\bar{\varepsilon})$.

Finally,
if for every proper subset $H$ of $N$, 
we always have $\witilde{b}_j(\bdsb{\witilde{\theta}}_H, \bdsb{\theta}^*_{-H}) = a_j^*$ for all $j\in H$,
then obviously,
the post-entry average fitnesses of $\theta^*_i$ and $\witilde{\theta}_i$ in each $i$th population are
$\varPi_{\theta^*_i}(\witilde{\mu};\witilde{b}) = \pi_i(\bdsb{a}^*)$ and
\[
\varPi_{\witilde{\theta}_i}(\witilde{\mu};\witilde{b}) =
\bigl( 1 - \witilde{\mu}_{-i}(\bdsb{\witilde{\theta}}_{-i}) \bigr)
\pi_i(\bdsb{a}^*) +
\witilde{\mu}_{-i}(\bdsb{\witilde{\theta}}_{-i})
\pi_i\big( \witilde{b}(\bdsb{\witilde{\theta}}) \big),
\]
respectively.
Since $\pi(\bdsb{a}^*)\in P(\mathcal{F}_{\nc})$,
it follows that the configuration $(\mu, b)$ is stable,
which completes the proof of part~\ref{gMsc:o1}.

\smallskip

For part~\ref{gMsc:o2}, 
consider an arbitrary mutation distribution $\upsilon\in \mathcal{M}(\Theta^n)$, 
and let $\witilde{\mu}_i = (1 - \varepsilon_i)\mu_i + \varepsilon_i\upsilon_i$ for all $i\in J_{\upsilon}$. 
It is enough to let $J_{\upsilon} = N$, 
since the proof of Theorem~\ref{prop:20210614} can be applied directly to the case when $J_{\upsilon}$ is a proper subset of $N$. 
For any nonempty proper subset $H\subset N$ and for any mutant subgroup $\bdsb{\witilde{\theta}}_H\in \witilde{\varTheta}_H$,
the same argument as in the proof of Theorem~\ref{prop:20210614} shows that
no matter which focal equilibrium the players choose,
the sum $S_H(\bdsb{\witilde{\theta}}_H)$ defined by~\eqref{eq:20250509} is always greater than or equal to $0$
as long as $\supnorm{\bdsb{\varepsilon}_{\upsilon}}$ is sufficiently small.

We first suppose that there is a nonempty proper subset $\bar{H}\subset N$ such that
$\witilde{b}_j(\bdsb{\witilde{\theta}}'_{\bar{H}}, \bdsb{\theta}^*_{-\bar{H}}) \neq a^*_j$
for some $j\in \bar{H}$ and for some $\bdsb{\witilde{\theta}}'_{\bar{H}}\in \witilde{\varTheta}_{\bar{H}}$. 
Then, since $\bdsb{a}^*$ is a strict $(n - 1)$-union equilibrium, 
we clearly have 
$\sum_{j\in \bar{H}} \pi_j(\bdsb{a}^*)> 
\sum_{j\in \bar{H}} \pi_j\big( \witilde{b}(\bdsb{\witilde{\theta}}'_{\bar{H}}, \bdsb{\theta}^*_{-\bar{H}}) \big)$. 
For any given $\bdsb{\witilde{\theta}}\in \witilde{\varTheta}$,
observe that
$\frac{\witilde{\mu}_{\bar{H}}(\bdsb{\witilde{\theta}}_{\bar{H}})}{\witilde{\mu}_{\bar{H}}(\bdsb{\witilde{\theta}}'_{\bar{H}})}$
is constant no matter what the values of $\varepsilon_j$ for all $j\in \bar{H}$ are,
which implies that the quotient
$\frac{\witilde{\mu}(\bdsb{\witilde{\theta}})}
{\witilde{\mu}_{\bar{H}}(\bdsb{\witilde{\theta}}'_{\bar{H}}) \prod_{i\notin \bar{H}}(1 - \varepsilon_i)}$ 
converges to zero as $\supnorm{\bdsb{\varepsilon}_{\upsilon}}\rightarrow 0$. 
In addition, since 
each $\frac{\varepsilon_{j} \upsilon_{j}(\witilde{\theta}_{j})}{1 - \varepsilon_j}$ also converges to zero 
as $\supnorm{\bdsb{\varepsilon}_{\upsilon}}\rightarrow 0$,
it follows that we can choose a barrier $\bar{\varepsilon}_{\bdsb{\witilde{\theta}}'_{\bar{H}}}\in (0,1)$ such that
\[
S_{\bar{H}}(\bdsb{\witilde{\theta}}'_{\bar{H}})
+
\sum_{i\in N} \sum_{\bdsb{\witilde{\theta}}\in \witilde{\varTheta}} \witilde{\mu}(\bdsb{\witilde{\theta}})
\Bigl( \pi_i(\bdsb{a}^*) - \pi_i\big( \witilde{b}(\bdsb{\witilde{\theta}}) \big) \Bigr)> 0
\]
for all $\bdsb{\varepsilon}_{\upsilon}$ with 
$\supnorm{\bdsb{\varepsilon}_{\upsilon}}\in (0, \bar{\varepsilon}_{\bdsb{\witilde{\theta}}'_{\bar{H}}})$,
and so we can further choose a suitable barrier $\bar{\varepsilon}\in (0,1)$ such that the inequality 
\[
\sum_{i\in N} \sum_{\witilde{\theta}_{i}\in \witilde{\varTheta}_{i}} \varepsilon_{i} \upsilon_{i}(\witilde{\theta}_{i})
\Bigl(
\varPi_{\theta_i^*}(\witilde{\mu};\witilde{b}) - \varPi_{\witilde{\theta}_i}(\witilde{\mu};\witilde{b})
\Bigr)
=
\sum_{\varnothing\neq H\varsubsetneq N} \sum_{\bdsb{\witilde{\theta}}_H\in \witilde{\varTheta}_H} S_H(\bdsb{\witilde{\theta}}_H)
+ \sum_{i\in N} \sum_{\bdsb{\witilde{\theta}}\in \witilde{\varTheta}} \witilde{\mu}(\bdsb{\witilde{\theta}})
\Bigl( \pi_i(\bdsb{a}^*) - \pi_i\big( \witilde{b}(\bdsb{\witilde{\theta}}) \big) \Bigr)> 0
\] 
holds for all $\bdsb{\varepsilon}_{\upsilon}$ with 
$\supnorm{\bdsb{\varepsilon}_{\upsilon}}\in (0,\bar{\varepsilon})$,
as desired.

Next, suppose that for every proper subset $H\subset N$, 
we always have $\witilde{b}_j(\bdsb{\witilde{\theta}}_H, \bdsb{\theta}^*_{-H}) = a_j^*$
for all $j\in H$ and for all $\bdsb{\witilde{\theta}}_H\in \witilde{\varTheta}_H$.
Then for each $i\in N$,
the post-entry average fitnesses of the incumbent $\theta^*_i$ and a mutant type $\witilde{\theta}_i\in \witilde{\varTheta}_i$ are
$\varPi_{\theta^*_i}(\witilde{\mu};\witilde{b}) = \pi_i(\bdsb{a}^*)$ and
\[
\varPi_{\witilde{\theta}_i}(\witilde{\mu};\witilde{b}) =
\Bigl( 1 - \prod_{j\neq i} \varepsilon_j \Bigr)
\pi_i(\bdsb{a}^*) +
\sum_{\bdsb{\witilde{\theta}}_{-i}\in \witilde{\varTheta}_{-i}}
\witilde{\mu}_{-i}(\bdsb{\witilde{\theta}}_{-i})
\pi_i\big( \witilde{b}(\witilde{\theta}_i, \bdsb{\witilde{\theta}}_{-i}) \big),
\]
respectively.
We define a payoff point $\bdsb{u}\in \mathcal{F}_{\co}$ by
$\bdsb{u}
= \sum_{\bdsb{\witilde{\theta}}\in \witilde{\varTheta}}
\frac{\witilde{\mu}(\bdsb{\witilde{\theta}})}{\sum_{\bdsb{\witilde{\theta}}'\in \witilde{\varTheta}}\witilde{\mu}(\bdsb{\witilde{\theta}}')}
\pi \big( \witilde{b}(\bdsb{\witilde{\theta}}) \big)$,
and for each $i\in N$, we can write
\[
\bdsb{u} = \sum_{\witilde{\theta}_{i}\in \witilde{\varTheta}_{i}} 
\upsilon_{i}(\witilde{\theta}_{i}) 
\sum_{\bdsb{\witilde{\theta}}_{-i}\in \witilde{\varTheta}_{-i}}
\frac{\witilde{\mu}_{-i}(\bdsb{\witilde{\theta}}_{-i})}{\sum_{\bdsb{\witilde{\theta}}'_{-i}\in \witilde{\varTheta}_{-i}}\witilde{\mu}_{-i}(\bdsb{\witilde{\theta}}'_{-i})}
\pi \big( \witilde{b}(\witilde{\theta}_i, \bdsb{\witilde{\theta}}_{-i}) \big).
\]
Since $\pi(\bdsb{a}^*)\in P(\mathcal{F}_{\co})$,
it follows that $\pi(\bdsb{a}^*) = \bdsb{u}$ or $\pi_i(\bdsb{a}^*)> u_i$ for some $i\in N$.
From this we conclude that
\[
\pi_i(\bdsb{a}^*) =
\sum_{\bdsb{\witilde{\theta}}_{-i}\in \witilde{\varTheta}_{-i}}
\frac{\witilde{\mu}_{-i}(\bdsb{\witilde{\theta}}_{-i})}{\sum_{\bdsb{\witilde{\theta}}'_{-i}\in \witilde{\varTheta}_{-i}}\witilde{\mu}_{-i}(\bdsb{\witilde{\theta}}'_{-i})}
\pi_i \big( \witilde{b}(\witilde{\theta}_i, \bdsb{\witilde{\theta}}_{-i}) \big)
\]
for every $i\in N$ and every $\witilde{\theta}_i\in \witilde{\varTheta}_i$, or
\[
\pi_i(\bdsb{a}^*) >
\sum_{\bdsb{\witilde{\theta}}_{-i}\in \witilde{\varTheta}_{-i}}
\frac{\witilde{\mu}_{-i}(\bdsb{\witilde{\theta}}_{-i})}{\sum_{\bdsb{\witilde{\theta}}'_{-i}\in \witilde{\varTheta}_{-i}}\witilde{\mu}_{-i}(\bdsb{\witilde{\theta}}'_{-i})}
\pi_i \big( \witilde{b}(\witilde{\theta}_i, \bdsb{\witilde{\theta}}_{-i}) \big)
\]
for some $i\in N$ and some $\witilde{\theta}_i\in \witilde{\varTheta}_i$.
This implies that for any $\bdsb{\varepsilon}_{\upsilon}$, 
either
$\varPi_{\theta_i^*}(\witilde{\mu};\witilde{b}) = \varPi_{\witilde{\theta}_i}(\witilde{\mu};\witilde{b})$
for every $i\in N$ and every $\witilde{\theta}_i\in \witilde{\varTheta}_i$, or
$\varPi_{\theta_i^*}(\witilde{\mu};\witilde{b}) > \varPi_{\witilde{\theta}_i}(\witilde{\mu};\witilde{b})$
for some $i\in N$ and some $\witilde{\theta}_i\in \witilde{\varTheta}_i$.
Thus $(\mu, b)$ is infinite-order stable in this case.
\end{proofEnd}

A strict $(n - 1)$-union equilibrium is just a strict Nash equilibrium when $n = 2$.
Hence, Theorem~\ref{prop:20141028} can also be obtained as a corollary of Theorem~\ref{prop:20250307}.


\section{Conclusion}\label{sec:conclusion}


We study evolutionarily stable preferences against multiple mutations in single- and multi-population matching settings, respectively.
A mutation distribution rather than a single mutant type is regarded as a unit of mutation.
A configuration is $r$th-order stable if for any given distribution of a sufficiently small $r$th-order mutation set, 
the post-entry aggregate outcome remains virtually unchanged and no incumbent type goes extinct.
Roughly speaking,
as the order of stability increases, 
the corresponding efficiency level also tends to rise.
Here we describe specific relations between them.
The examples and the proof techniques can help us understand how multiple mutations shape the efficiency of stable outcomes.

From a mathematical point of view,
the payoff structure of symmetric games, combined with a single-population matching setting, 
enriches the properties of the single-population case. 
We show that under single-population matching,
the $(n + 1)$th-order stability is equivalent to the infinite-order stability,
but the analogous result for the multi-population case is unknown.
In addition, 
given any $n$-player game, 
to prove that infinite-order stability implies higher-level efficiency under multi-population matching, 
we need the additional technical assumption that the coefficients can only be rational numbers. 
(However, 
this technical assumption is unnecessary when analyzing two-player games.)

Our analysis reveals that the number of players involved in a game severely affects the features of evolutionary stability.
The differences between the models of two- and $n$-player games can be observed in both single- and multi-population matching settings.
We construct examples in which novel coalitional groups in three- and four-player games
could destabilize various efficient, strictly strong Nash equilibrium outcomes.
To ensure stability in general $n$-player games,
we introduce the concept of a strict union equilibrium,
which requires that any subset of players will strictly decrease their collective payoff if any of them deviates. 
A strict union equilibrium outcome can be infinite-order stable in both of these matching settings.

It is especially worth noting that in a single-population matching setting,
the number of mutant types entering the population decides the maximum number of distinct strategies used in a coalition. 
According to this,
we define a strict $k$-type union equilibrium,
which imposes constraints more precisely on the behavior of mutants.
This allows us to derive the sharp result that
a strict $k$-type union equilibrium outcome is $k$th-order stable in the single-population case.

In this paper, 
our strong stability concept requires that a population is immune to invasions 
under all possible mutation distributions with all possible focal equilibria, 
which follows from the commonly accepted idea that mutations are random. 
Based on this setting, 
we can use indifferent types as potential entrants to prove certain results,\footnote{%
See also the remark after Example~\ref{exam:20151001}.} 
but we make it easy to destabilize outcomes. 
In essence, 
while mutations are random events, 
underlying environments may affect the likelihood of mutations. 
We may be able to ignore the mutant types which are unlikely to occur, 
if we know the distribution of mutations.


\appendix
\section{Existence of a Type-homogeneous Equilibrium}\label{sec:sym-inv} 


Consider the game induced by a given $\bdsb{\theta} = (\theta_1, \dots, \theta_n)$ 
where each $\theta_i$ is drawn from a single population. 
Suppose that $N$ can be partitioned into $m$ disjoint subsets
$N = J_1 \cup \dots \cup J_m$
where each $J_k$ consists of players having the same type.
Let $\tau$ be a permutation of $N$ which maps $J_k$ onto itself for $k = 1$, \dots,~$m$.
Then for any $i\in N$,
the two types $\theta_i$ and $\theta_{\tau(i)}$ in the $n$-tuple $\bdsb{\theta}$ are the same,
which implies that
$u^{\bdsb{\theta}}_{\tau(i)}(a_1, \dots, a_n) = u^{\bdsb{\theta}}_i(a_{\tau(1)}, \dots, a_{\tau(n)})$
for all $(a_1, \dots, a_n)\in \bar{A}^n$.
Thus every permutation of $N$ mapping each $J_k$ onto itself
is a symmetry of the normal-form game induced by $\bdsb{\theta}$.

Next,
we introduce a useful concept of symmetry imposed on strategy profiles;
see also \citet{aPla:sng}.
In a normal-form game,
a strategy profile $\bdsb{\sigma}$ is said to be \emph{symmetry-invariant} if
$(\sigma_1, \dots, \sigma_n) = (\sigma_{\tau(1)}, \dots, \sigma_{\tau(n)})$ for each symmetry $\tau$ of the game.
(Therefore, a symmetry-invariant strategy profile $\bdsb{\sigma}$ in a symmetric game is \emph{symmetric},
which means that $\sigma_1 = \dots = \sigma_n$.)
Due to Theorem~2 of \citet{jNash:ncg}, we have the following result.

\begin{theoremEnd}[]{thm}\label{prop:20191124}
Any finite normal-form game has a symmetry-invariant equilibrium.
\end{theoremEnd}

Thus, in the above normal-form game induced by $\bdsb{\theta}$,
one can find a symmetry-invariant equilibrium $\bdsb{\sigma}$ with the property that 
$\sigma_i = \sigma_j$ if $i$,~$j\in J_k$ for some $k$.\footnote{%
We can get it by letting the symmetry $\tau$ satisfy $\tau(i) = j$.} 
This means that we obtain the desired conclusion.

\begin{theoremEnd}[]{cor}
Let $\bdsb{\theta} = (\theta_1, \dots, \theta_n)$ where $\theta_i\in \spt\mu$ for all $i\in N$.
Then in the normal-form game induced by $\bdsb{\theta}$,
there exists a Nash equilibrium $b(\bdsb{\theta})$ with the property that
$b_i(\bdsb{\theta}) = b_j(\bdsb{\theta})$ whenever $\theta_i = \theta_j$.
\end{theoremEnd}


\section{Alternative Definition of Multi-mutation Stability under Single-population Matching}\label{sec:alt-def} 


Let $(\mu,b)$ be a configuration in the single-population matching environment $[(N, \bar{A}, (\pi_i)), \Gamma(\mu)]$. 
In particular, 
given an $r$th-order mutation distribution $\upsilon$ and a focal equilibrium $\witilde{b}$, 
we say that $(\mu,b)$ is \emph{strongly stable} against $(\upsilon, \witilde{b})$ 
if condition~\ref{Cs:wiped-out} in Definition~\ref{def:20200704} is satisfied for all sufficiently small $\varepsilon$. 
To define multi-mutation stability in another way, 
we must first set some notation for a player's fitness under a specific composition type of opponents.

Let the post-entry type distribution be $\witilde{\mu} = (1 - \varepsilon)\mu + \varepsilon\upsilon$. 
For each $0\leq m\leq n - 1$, we denote by $r(m)$ the integer $n - 1 - m$. 
Let $( \bdsb{\witilde{\theta}}_m, \bdsb{\theta}_{r(m)} )$ be an $(n - 1)$-tuple 
consisting of $m$ mutants and $n - 1 - m$ incumbents such as 
\[
( \bdsb{\witilde{\theta}}_m, \bdsb{\theta}_{r(m)} ) = 
( \underbrace{\witilde{\theta}_1, \dots, \witilde{\theta}_1}_{\text{$m_1$ terms}}, \dots,
\underbrace{\witilde{\theta}_s, \dots, \witilde{\theta}_s}_{\text{$m_s$ terms}},
\underbrace{\theta_1, \dots, \theta_1}_{\text{$r_1$ terms}}, \dots,
\underbrace{\theta_t, \dots, \theta_t}_{\text{$r_t$ terms}} ), 
\] 
where $m_1 + \dots + m_s = m$ and $r_1 + \dots + r_t = n - 1 - m$. 
Then the probability that a type $\theta\in \spt\witilde{\mu}$ is matched with these $n - 1$ opponents, 
regardless of their order, is 
\[
C_{[ \bdsb{\witilde{\theta}}_m, \bdsb{\theta}_{r(m)} ]}^{\witilde{\mu}} =
\frac{(n-1)!}{m_1 !\cdots m_s ! \, r_1 !\cdots r_t !} \times
\witilde{\mu}^{n-1} (\bdsb{\witilde{\theta}}_m, \bdsb{\theta}_{r(m)}), 
\] 
where $[ \bdsb{\witilde{\theta}}_m, \bdsb{\theta}_{r(m)} ]$ denotes the multiset for this opponent composition. 
If we let $\mathcal{G}_{(m, r(m))}$ be the family of multisets consisting of $m$ mutants and $n-1-m$ incumbents, 
then the post-entry average fitness of a type $\theta\in \spt\witilde{\mu}$ can be written as 
\begin{align*} 
\varPi_{\theta}(\witilde{\mu};\witilde{b}) 
&= \sum_{\bdsb{\wihat{\theta}}_{-1}\in (\spt\witilde{\mu})^{n-1}} \witilde{\mu}^{n-1}(\bdsb{\wihat{\theta}}_{-1})
\pi_1\big( \witilde{b}(\theta, \bdsb{\wihat{\theta}}_{-1}) \big)\\
&= \sum_{m=0}^{n-1}
\sum_{[\bdsb{\witilde{\theta}}_m, \bdsb{\theta}_{r(m)}]\in \mathcal{G}_{(m, r(m))}} 
C_{[\bdsb{\witilde{\theta}}_m, \bdsb{\theta}_{r(m)}]}^{\witilde{\mu}} 
\pi_1 \big( \witilde{b}(\theta, \bdsb{\witilde{\theta}}_m, \bdsb{\theta}_{r(m)}) \big) = 
\sum_{m=0}^{n-1} F^{\witilde{\mu}}_{\theta}(m),
\end{align*}
where for each $0\leq m\leq n - 1$, we define $F^{\witilde{\mu}}_{\theta}(m)$ to be the sum 
\[
F^{\witilde{\mu}}_{\theta}(m) = 
\sum_{[\bdsb{\witilde{\theta}}_m, \bdsb{\theta}_{r(m)}]\in \mathcal{G}_{(m, r(m))}} 
C_{[\bdsb{\witilde{\theta}}_m, \bdsb{\theta}_{r(m)}]}^{\witilde{\mu}} 
\pi_1 \big( \witilde{b}(\theta, \bdsb{\witilde{\theta}}_m, \bdsb{\theta}_{r(m)}) \big), 
\] 
derived from $\theta$'s fitness when facing precisely $m$ mutant opponents. 
Also notice that 
\begin{equation}\label{eq:20241128} 
\witilde{\mu}^{n-1} (\bdsb{\witilde{\theta}}_m, \bdsb{\theta}_{r(m)}) = 
\varepsilon^m (1 - \varepsilon)^{r(m)} \upsilon^m(\bdsb{\witilde{\theta}}_m) \mu^{r(m)}(\bdsb{\theta}_{r(m)}). 
\end{equation} 
Thus, for any $m$, we see that 
$F^{\witilde{\mu}}_{\theta}(m + 1) / F^{\witilde{\mu}}_{\theta}(m)$ 
converges to zero as $\varepsilon\rightarrow 0$. 
This provides an alternative way of defining multi-mutation strong stability.

\begin{theoremEnd}[normal]{thm}\label{prop:20241111}
Suppose that $(\mu,b)$ is a configuration in $[(N, \bar{A}, (\pi_i)), \Gamma(\mu)]$. 
Given an $r$th-order mutation distribution $\upsilon\in \mathcal{M}(\Theta_s)$, 
let $\witilde{\mu} = (1 - \varepsilon)\mu + \varepsilon\upsilon$, 
and let $\witilde{b}\in B(\spt\witilde{\mu};b)$. 
For any $\theta$,~$\wihat{\theta}\in \spt\witilde{\mu}$,
define
\[
U_{(\theta, \wihat{\theta})}^{\witilde{\mu}} =
\{\, 0\leq m\leq n - 1 \mid F^{\witilde{\mu}}_{\theta}(m)\neq F^{\witilde{\mu}}_{\wihat{\theta}}(m) \,\}. 
\]
Then $(\mu,b)$ is strongly stable against $(\upsilon, \witilde{b})$ if and only if
there exists $\witilde{\theta}_j\in \spt\upsilon$ such that
\begin{enumerate}[label=\upshape (\roman*)]
\item $U_{(\theta, \witilde{\theta}_j)}^{\witilde{\mu}}$ is a nonempty set for all $\theta\in \spt\mu$, and \label{ssII:not-coexist}
\item if $\min U_{(\theta, \witilde{\theta}_j)}^{\witilde{\mu}} = k$,
then $F^{\witilde{\mu}}_{\theta}(k)> F^{\witilde{\mu}}_{\witilde{\theta}_j}(k)$. \label{ssII:wiped-out}
\end{enumerate}
\end{theoremEnd}

\begin{proofEnd} 
Let us define, for each $0\leq m\leq n - 1$ and each $\theta\in \spt\witilde{\mu}$, 
$F^{\mu + \upsilon}_{\theta}(m)$ to be the sum 
\[
F^{\mu + \upsilon}_{\theta}(m) = 
\sum_{[\bdsb{\witilde{\theta}}_m, \bdsb{\theta}_{r(m)}]\in \mathcal{G}_{(m, r(m))}} 
\frac{(n-1)!}{m_1 !\cdots m_s ! \, r_1 !\cdots r_t !} \times 
\upsilon^m(\bdsb{\witilde{\theta}}_m) \mu^{r(m)}(\bdsb{\theta}_{r(m)}) 
\pi_1 \big( \witilde{b}(\theta, \bdsb{\witilde{\theta}}_m, \bdsb{\theta}_{r(m)}) \big). 
\] 
Then 
$F^{\witilde{\mu}}_{\theta}(m) = \varepsilon^m (1 - \varepsilon)^{r(m)} F^{\mu + \upsilon}_{\theta}(m)$, 
so that 
\[ 
\varPi_{\theta}(\witilde{\mu};\witilde{b}) = 
\sum_{m=0}^{n-1} F^{\witilde{\mu}}_{\theta}(m) = 
\sum_{m=0}^{n-1} \varepsilon^m (1 - \varepsilon)^{r(m)} F^{\mu + \upsilon}_{\theta}(m). 
\] 
Suppose first that there exists $\witilde{\theta}_j\in \spt\upsilon$ such that 
conditions~\ref{ssII:not-coexist} and~\ref{ssII:wiped-out} are satisfied.
For a given $\theta\in \spt\mu$,
let $k_{\theta}$ denote the smallest integer in $U_{(\theta, \witilde{\theta}_j)}^{\witilde{\mu}}$. 
Then $F^{\witilde{\mu}}_{\theta}(k_{\theta})> F^{\witilde{\mu}}_{\witilde{\theta}_j}(k_{\theta})$, so  
$F^{\mu + \upsilon}_{\theta}(k_{\theta})> F^{\mu + \upsilon}_{\witilde{\theta}_j}(k_{\theta})$. 
Thus there is a barrier $\bar{\varepsilon}_{\theta}$ such that 
\[ 
\varPi_{\theta}(\witilde{\mu};\witilde{b}) - \varPi_{\witilde{\theta}_j}(\witilde{\mu};\witilde{b}) 
= \varepsilon^{k_{\theta}} \sum_{m = k_{\theta}}^{n-1} \varepsilon^{m - k_{\theta}} (1 - \varepsilon)^{r(m)} 
\Bigl( F^{\mu + \upsilon}_{\theta}(m) - F^{\mu + \upsilon}_{\witilde{\theta}_j}(m) \Bigr) > 0
\] 
for all $\varepsilon\in (0, \bar{\varepsilon}_{\theta})$.
Choose $\bar{\varepsilon} = \min \{\, \bar{\varepsilon}_{\theta} \mid \theta\in \spt\mu \,\}$.
Then for every $\theta\in \spt\mu$,
we have $\varPi_{\theta}(\witilde{\mu};\witilde{b}) > \varPi_{\witilde{\theta}_j}(\witilde{\mu};\witilde{b})$
for all $\varepsilon\in (0, \bar{\varepsilon})$.
This means that $(\mu,b)$ is strongly stable against $(\upsilon, \witilde{b})$.

Now suppose conversely that for each $\witilde{\theta}_j\in \spt\upsilon$, 
there exists $\theta\in \spt\mu$ such that
either $U_{(\theta, \witilde{\theta}_j)}^{\witilde{\mu}} = \varnothing$,
or $U_{(\theta, \witilde{\theta}_j)}^{\witilde{\mu}}$ is a nonempty set having a smallest integer $k$
for which 
$F^{\witilde{\mu}}_{\witilde{\theta}_j}(k)> F^{\witilde{\mu}}_{\theta}(k)$. 
First, 
if $U_{(\theta, \witilde{\theta}_j)}^{\witilde{\mu}} = \varnothing$,
then
$\varPi_{\witilde{\theta}_j}(\witilde{\mu};\witilde{b}) = \varPi_{\theta}(\witilde{\mu};\witilde{b})$
for all $\varepsilon\in (0, 1)$. 
Next, if $\min U_{(\theta, \witilde{\theta}_j)}^{\witilde{\mu}} = k$ and
$F^{\witilde{\mu}}_{\witilde{\theta}_j}(k)> F^{\witilde{\mu}}_{\theta}(k)$ 
(so we have $F^{\mu + \upsilon}_{\witilde{\theta}_j}(k)> F^{\mu + \upsilon}_{\theta}(k)$),
then there is a barrier $\bar{\varepsilon}_{*}$ for the pair $(\theta, \witilde{\theta}_j)$ such that
\[
\varPi_{\witilde{\theta}_j}(\witilde{\mu};\witilde{b}) - \varPi_{\theta}(\witilde{\mu};\witilde{b})
= \varepsilon^k \sum_{m=k}^{n-1} \varepsilon^{m-k} (1 - \varepsilon)^{r(m)} 
\Bigl( F^{\mu + \upsilon}_{\witilde{\theta}_j}(m) - F^{\mu + \upsilon}_{\theta}(m) \Bigr) > 0
\]
for all $\varepsilon\in (0, \bar{\varepsilon}_{*})$.
By definition, 
we can conclude that the configuration $(\mu,b)$ is not strongly stable against $(\upsilon, \witilde{b})$. 
\end{proofEnd}

Condition~\ref{ssII:not-coexist} in Theorem~\ref{prop:20241111} excludes the possibility that
for any mutant, there is an incumbent that would earn the same average fitness as the mutant.
Condition~\ref{ssII:wiped-out} further ensures that
there is a mutant type that is strictly outperformed by all incumbent types.

\begin{Rem} 
Given an $r$th-order mutation distribution $\upsilon$ and a focal equilibrium $\witilde{b}$, 
we say that a configuration $(\mu,b)$ \emph{coexists} with $(\upsilon, \witilde{b})$ 
if condition~\ref{Cs:coexist} in Definition~\ref{def:20200704} is satisfied for all sufficiently small $\varepsilon$. 
One can use similar arguments to show that 
$(\mu,b)$ coexists with $(\upsilon, \witilde{b})$ if and only if 
for every $\theta$,~$\wihat{\theta}\in \spt\witilde{\mu}$,
the set $U_{(\theta, \wihat{\theta})}^{\witilde{\mu}} = \varnothing$. 
To summarize, 
a configuration $(\mu,b)$ is $r$th-order stable if and only if 
for any $r$th-order mutation distribution $\upsilon$ and any focal equilibrium $\witilde{b}$, 
the configuration $(\mu,b)$ is strongly stable against $(\upsilon, \witilde{b})$, 
or it coexists with $(\upsilon, \witilde{b})$. 
\end{Rem}


\section{Multi-mutation Stability of Strictly Strong Nash Equilibria}\label{sec:counterexamples} 


In Chapter~\ref{sec:single}, 
we argued informally that 
if a symmetric $n$-player game has a symmetrically strictly strong Nash equilibrium, 
then this equilibrium is the unique stable strategy profile in the game. 
Furthermore, Example~\ref{exam:20201215} shows that in a symmetric three-player game, 
such an equilibrium is not second-order stable despite having a payoff profile on the noncooperative Pareto frontier. 
However, it achieves second-order stability when its payoff profile lies on the cooperative Pareto frontier. 
We now formalize these properties below.

\begin{theoremEnd}[normal]{prop}\label{prop:20201130}
Let $(a^*, \dots, a^*)$ be a strictly strong Nash equilibrium of a symmetric $n$-player game $(N, \bar{A}, (\pi_i))$.
Then $(a^*, \dots, a^*)$ is the unique stable strategy profile.
If in addition $n = 3$ and $\pi(a^*, a^*, a^*)\in P(\mathcal{F}_{\co})$, then $(a^*, a^*, a^*)$ is second-order stable. 
\end{theoremEnd}

\begin{proofEnd}
Let $(\mu, b)$ be a monomorphic configuration in which $a^*$ is a strictly dominant strategy for $\theta$. 
Then the strictly strong Nash equilibrium $(a^*, \dots, a^*)$ is the aggregate outcome of $(\mu, b)$. 
Given a single mutant type, 
the discussion after Definition~\ref{def:20210120} makes it clear that 
the mutant type receives no higher average fitness than the incumbents as long as these entering mutants are sufficiently rare. 
Thus $(a^*, \dots, a^*)$ is stable. 
To see that $(a^*, \dots, a^*)$ is the unique stable strategy profile in $(N, \bar{A}, (\pi_i))$,
consider any other stable configuration $(\mu', b')$, 
and let $\bdsb{\sigma}$ be a strategy profile chosen by $n$ incumbents in $\spt\mu'$. 
Then by Theorem~\ref{prop:20150411}\ref{Snc:o1},
the fitness to each of the $n$ incumbents is equal to $\pi_1(a^*, \dots, a^*)$,
and hence $\bdsb{\sigma}$ is just the strictly strong Nash equilibrium $(a^*, \dots, a^*)$.

\smallskip

Next, suppose that the number $n$ of players is three,
and that $\pi(a^*, a^*, a^*)$ lies on the cooperative Pareto frontier. 
Let $\pi_1(a^*, a^*, a^*) = \pi^e$. 
Then by Lemma~\ref{prop:20200908} we have 
\begin{equation}\label{eq:20260722} 
3\pi^{e}\geq \sum_{i=1}^{3} \pi_i(\sigma_1, \sigma_2, \sigma_3) 
\end{equation} 
for all $(\sigma_1, \sigma_2, \sigma_3)\in (\Delta \bar{A})^3$. 
Consider a second-order mutation distribution $\upsilon$. 
Let $\spt\upsilon$ consist of $\witilde{\theta}_1$ and $\witilde{\theta}_2$, 
and let $\witilde{\mu} = (1 - \varepsilon)\mu + \varepsilon\upsilon$ be the post-entry type distribution. 
To verify whether the outcome $(a^*, a^*, a^*)$ is second-order stable,
it suffices to check the case where the chosen focal equilibrium $\witilde{b}$ satisfies
\[
\witilde{b}(\witilde{\theta}_i, \theta, \theta) =
\witilde{b}(\witilde{\theta}_i, \witilde{\theta}_i, \theta) =
\witilde{b}(\witilde{\theta}_i, \witilde{\theta}_j, \theta) = (a^*, a^*, a^*)
\]
for any $i$,~$j\in \{1, 2\}$ with $i\neq j$,
since if not all such conditions are satisfied,
then by $(a^*, a^*, a^*)$ being a strictly strong Nash equilibrium,
there is at least one mutant type earning a strictly lower average fitness than the incumbent type
as long as $\varepsilon$ is sufficiently small. 
This implies that the value 
$\sum_{i=1}^2 \varepsilon \upsilon(\witilde{\theta}_i) 
\bigl( \varPi_{\theta}(\witilde{\mu};\witilde{b}) - \varPi_{\witilde{\theta}_i}(\witilde{\mu};\witilde{b}) \bigr)$
is equal to 
\[ 
\sum_{i=1}^{2} \big(\varepsilon \upsilon(\witilde{\theta}_i)\big)^3 
\Bigl( \pi^e - \pi_1\big( \witilde{b}(\witilde{\theta}_i, \witilde{\theta}_i, \witilde{\theta}_i) \big) \Bigr) + 
\sum_{i=1}^{2} \varepsilon \upsilon(\witilde{\theta}_i) \big(\varepsilon \upsilon(\witilde{\theta}_j)\big)^2 
\Bigl( 3\pi^e - \sum_{k=1}^{3} \pi_k\big( \witilde{b}(\witilde{\theta}_i, \witilde{\theta}_j, \witilde{\theta}_j) \big) \Bigr),
\] 
where the index $j = 3 - i$. 
Since $(a^*, \dots, a^*)$ is a strictly strong Nash equilibrium, 
it follows that 
$\pi^e\geq \pi_1(\sigma, \sigma, \sigma)$ for all $\sigma\in \Delta \bar{A}$. 
Thus by~\eqref{eq:20260722} 
we obtain 
\[ 
\sum_{i=1}^2 \varepsilon \upsilon(\witilde{\theta}_i) 
\bigl( \varPi_{\theta}(\witilde{\mu};\witilde{b}) - \varPi_{\witilde{\theta}_i}(\witilde{\mu};\witilde{b}) \bigr)\geq 0.
\] 
Therefore, for any $\varepsilon$, 
either 
$\varPi_{\theta}(\witilde{\mu};\witilde{b}) = \varPi_{\witilde{\theta}_i}(\witilde{\mu};\witilde{b})$
for all $i = 1$,~$2$,
or
$\varPi_{\theta}(\witilde{\mu};\witilde{b}) > \varPi_{\witilde{\theta}_i}(\witilde{\mu};\witilde{b})$
for some $i = 1$,~$2$.
This shows that $(a^*, a^*, a^*)$ is a second-order stable outcome.
\end{proofEnd}

Although a symmetrically strictly strong Nash equilibrium of a symmetric three-player game is second-order stable 
when its payoff profile lies on the cooperative Pareto frontier, 
it cannot ensure higher-order stability. 
In fact, 
in games with more than three players, 
such an equilibrium may even fail to be second-order stable. 
The following simple examples illustrate these instability phenomena.

\begin{example}\label{exam:20210123}
Let $(\{1, 2, 3\}, \bar{A}, (\pi_i))$ be the symmetric underlying game with $\bar{A} = \{a_1, a_2, a_3\}$,
and let $(a_1, a_1, a_1)$ be a strictly strong Nash equilibrium of this game.
Moreover, suppose that the material payoff function $\pi$ satisfies the conditions: $\pi(a_1, a_1, a_1) = (10, 10, 10)$,
\[
\pi(a_1, a_2, a_3) = \pi(a_2, a_3, a_2) = \pi(a_3, a_2, a_3) = (5, 17, 5),
\]
and also $\sum_{i=1}^3 \pi_i(\bdsb{a})< 30$ for any other $\bdsb{a}\in \bar{A}^3$. 
Then by Lemma~\ref{prop:20240802}, we have $\pi(a_1, a_1, a_1)\in P(\mathcal{F}_{\co})$, 
and hence by Proposition~\ref{prop:20201130}, the strategy profile $(a_1, a_1, a_1)$ is second-order stable.
However, it is not third-order stable.

To prove this,
let $(\mu, b)$ be a configuration with the aggregate outcome $(a_1, a_1, a_1)$.
Then it is clear that 
$b(\theta, \theta', \theta'') = (a_1, a_1, a_1)$ for all $\theta$, $\theta'$,~$\theta''\in \spt\mu$. 
Consider a third-order mutation distribution $\upsilon$ with equal probabilities 
assigned to indifferent types $\witilde{\theta}_1$, $\witilde{\theta}_2$, and $\witilde{\theta}_3$ in $\spt\upsilon$. 
Let $\witilde{\mu} = (1 - \varepsilon)\mu + \varepsilon\upsilon$, 
and suppose that the chosen focal equilibrium $\witilde{b}$ satisfies: 
\[
\witilde{b}_2(\theta, \witilde{\theta}_1, \witilde{\theta}_2) = \witilde{b}_2(\theta, \witilde{\theta}_2, \witilde{\theta}_3)
= \witilde{b}_2(\theta, \witilde{\theta}_3, \witilde{\theta}_1) = a_2
\quad \text{and} \quad
\witilde{b}_3(\theta, \witilde{\theta}_1, \witilde{\theta}_2) = \witilde{b}_3(\theta, \witilde{\theta}_2, \witilde{\theta}_3)
= \witilde{b}_3(\theta, \witilde{\theta}_3, \witilde{\theta}_1) = a_3
\]
for any $\theta\in \spt\mu$,
and the mutants play $a_1$ in all other interactions; 
in addition,
the incumbents of any type choose a fixed best reply against the mutant opponents' actions $a_2$ and $a_3$, 
and they continue to play $a_1$ when facing the mutant opponents who also play $a_1$. 
Then for $i = 1$,~$2$,~$3$, 
the difference between the post-entry average fitnesses of $\witilde{\theta}_i$ and $\theta\in \spt\mu$ is always equal to
\[
\varPi_{\witilde{\theta}_i}(\witilde{\mu};\witilde{b}) - \varPi_{\theta}(\witilde{\mu};\witilde{b}) = 
\frac{4\varepsilon(1 - \varepsilon)}{3} + \frac{10 \varepsilon^2}{3} > 0, 
\] 
regardless of the strategy an incumbent $\theta$ chooses against $a_2$ and $a_3$. 
Thus, although $\pi(a_1, a_1, a_1)\in P(\mathcal{F}_{\co})$,
this strictly strong Nash equilibrium $(a_1, a_1, a_1)$ cannot be a third-order stable outcome.
\end{example}

\begin{example}\label{exam:20210124}
Consider a symmetric four-player underlying game $(\{1, 2, 3, 4\}, \bar{A}, (\pi_i))$ with $\bar{A} = \{a_1, a_2, a_3\}$.
Suppose that the material payoff function $\pi$ is defined such that
$(a_1, a_1, a_1, a_1)$ is a strictly strong Nash equilibrium.
Then Proposition~\ref{prop:20201130} assures us that this strategy profile is stable. 
Moreover, suppose that 
$\pi(a_1, a_1, a_1, a_1) = (10, 10, 10, 10)$,
\[
\pi(a_1, a_2, a_2, a_3) = \pi(a_3, a_2, a_2, a_3) = (3, 15, 15, 3),
\]
$\pi(a_2, a_2, a_2, a_3) = (3, 3, 3, 27)$,
and also $\sum_{i=1}^4 \pi_i(\bdsb{a})< 40$ for any other $\bdsb{a}\in \bar{A}^4$.
Then by Lemma~\ref{prop:20240802},
we have $\pi(a_1, a_1, a_1, a_1)\in P(\mathcal{F}_{\co})$.
Even so,
we will show that $(a_1, a_1, a_1, a_1)$ is not second-order stable.

Let $(\mu, b)$ be a configuration with the aggregate outcome $(a_1, a_1, a_1, a_1)$.
Then for all incumbents $\theta$, $\theta'$, $\theta''$, and $\theta'''$,
we have $b(\theta, \theta', \theta'', \theta''') = (a_1, a_1, a_1, a_1)$. 
Consider a second-order mutation distribution $\upsilon$ with equal probabilities 
assigned to indifferent types $\witilde{\theta}_1$ and $\witilde{\theta}_2$ in $\spt\upsilon$. 
Let $\witilde{\mu} = (1 - \varepsilon)\mu + \varepsilon\upsilon$, 
and let $\witilde{b}\in B(\spt\witilde{\mu};b)$ be chosen. 
Suppose that for any $\theta\in \spt\mu$, 
the mutants' strategies satisfy 
\[
\witilde{b}_i(\theta, \witilde{\theta}_1, \witilde{\theta}_1, \witilde{\theta}_2) =
\witilde{b}_i(\theta, \witilde{\theta}_2, \witilde{\theta}_2, \witilde{\theta}_1) = a_2
\quad \text{for $i = 2$,~$3$,}
\] 
and $\witilde{b}_4(\theta, \witilde{\theta}_1, \witilde{\theta}_1, \witilde{\theta}_2) =
\witilde{b}_4(\theta, \witilde{\theta}_2, \witilde{\theta}_2, \witilde{\theta}_1) = a_3$, 
and the mutants play $a_1$ in all other interactions.
In addition, 
suppose that the incumbents of any type choose a fixed best reply against the mutant opponents' actions $a_2$, $a_2$, and $a_3$, 
and that they continue to play $a_1$ when facing the mutant opponents who also play $a_1$. 
Then for $i = 1$,~$2$, 
the difference between the post-entry average fitnesses of $\witilde{\theta}_i$ and $\theta\in \spt\mu$ is 
\[
\varPi_{\witilde{\theta}_i}(\witilde{\mu};\witilde{b}) - \varPi_{\theta}(\witilde{\mu};\witilde{b}) = 
\frac{9(1 - \varepsilon)\varepsilon^2}{4} + \frac{21\varepsilon^3}{4} > 0, 
\] 
regardless of the strategy an incumbent $\theta$ chooses against $a_2$, $a_2$, and $a_3$. 
Thus this strictly strong Nash equilibrium $(a_1, a_1, a_1, a_1)$ is not second-order stable, 
although its payoff profile $\pi(a_1, a_1, a_1, a_1)\in P(\mathcal{F}_{\co})$. 
\end{example}


\section{Proofs of Theorems}\label{sec:proofs}


\printProofs


\bibliographystyle{plainnat}
\bibliography{EvlPrfMut}

@article{rAum:ccgwsp, 
 author = {Robert J. Aumann},
 journal = {Transactions of the American Mathematical Society},
 number = {3},
 pages = {539-552},
 publisher = {American Mathematical Society},
 title = {The Core of a Cooperative Game Without Side Payments}, 
 volume = {98},
 year = {1961}
}

@ARTICLE{hBes-wGut:iaes,
  author = {Helmut Bester and Werner G{\"{u}}th},
  title = {Is Altruism Evolutionarily Stable?},
  journal = {Journal of Economic Behavior \& Organization},
  year = {1998},
  volume = {34},
  pages = {193-209}
}

@ARTICLE{pDas-eMas:eedeg,
  author = {Partha Dasgupta and Eric Maskin},
  title = {The Existence of Equilibrium in Discontinuous Economic Games, {I}: Theory},
  journal = {Review of Economic Studies},
  year = {1986},
  volume = {53},
  pages = {1-26}
}

@ARTICLE{eDek-jEly-oYil:ep,
  author = {Eddie Dekel and Jeffrey C. Ely and Okan Yilankaya},
  title = {Evolution of Preferences},
  journal = {Review of Economic Studies},
  year = {2007},
  volume = {74},
  pages = {685-704}
}

@ARTICLE{wGut:eaecbri,
  author = {Werner G{\"{u}}th},
  title = {An Evolutionary Approach to Explaining Cooperative Behavior by Reciprocal Incentives},
  journal = {International Journal of Game Theory},
  year = {1995},
  volume = {24},
  pages = {323-344}
}

@ARTICLE{yHel-eMoh:cdp,
  author = {Yuval Heller and Erik Mohlin},
  title = {Coevolution of Deception and Preferences: {Darwin} and {Nash} Meet {Machiavelli}},
  journal = {Games and Economic Behavior},
  year = {2019},
  volume = {113},
  pages = {223-247}
}

@ARTICLE{fHer-cKuz:esdo,
  author = {Florian Herold and Christoph Kuzmics},
  title = {Evolutionary Stability of Discrimination under Observability},
  journal = {Games and Economic Behavior},
  year = {2009},
  volume = {67},
  pages = {542-551}
}

@ARTICLE{sHuc-jOec:ieaefa,
  author = {Steffen Huck and J{\"{o}}rg Oechssler},
  title = {The Indirect Evolutionary Approach to Explaining Fair Allocations},
  journal = {Games and Economic Behavior},
  year = {1999},
  volume = {28},
  pages = {13-24}
}

@ARTICLE{Monroe-etal:mbrnsat,
  author = {J. Grey Monroe and Thanvi Srikant and Pablo Carbonell-Bejerano and Claude Becker and Mariele Lensink and Moises Exposito-Alonso and Marie Klein and Julia Hildebrandt and Manuela Neumann and 
  Daniel Kliebenstein and Mao-Lun Weng and Eric Imbert and Jon Ågren and Matthew T. Rutter and Charles B. Fenster and Detlef Weigel},
  year = {2022}, 
  pages = {101-105},
  title = {Mutation Bias Reflects Natural Selection in Arabidopsis thaliana},
  volume = {602},
  journal = {Nature}
}

@ARTICLE{jNash:ncg,
  author = {John Nash},
  title = {Non-Cooperative Games},
  journal = {Annals of Mathematics},
  year = {1951},
  volume = {54},
  pages = {286-295}
}

@ARTICLE{eOst:caesn,
  author = {Elinor Ostrom},
  title = {Collective Action and the Evolution of Social Norms},
  journal = {Journal of Economic Perspectives},
  year = {2000},
  volume = {14},
  pages = {137-158}
}

@article{aPla:sng,
author = {Asaf Plan},
title = {Symmetry in {$n$}-Player Games},
journal = {Journal of Economic Theory},
year = {2023},
volume = {207},
pages = {105549}
}

@ARTICLE{aPos:tsespsg,
  author = {Alex Possajennikov},
  title = {Two-Speed Evolution of Strategies and Preferences in Symmetric Games},
  journal = {Theory and Decision},
  year = {2005},
  volume = {57},
  pages = {227-263}
}

@ARTICLE{dRay-rVoh:eba,
  author = {Debraj Ray and Rajiv Vohra},
  title = {Equilibrium Binding Agreements},
  journal = {Journal of Economic Theory},
  year = {1997},
  volume = {73},
  pages = {30-78}
}

@ARTICLE{aRob:eegdnsh,
  author = {Arthur J. Robson},
  title = {Efficiency in Evolutionary Games: {Darwin}, {Nash} and the Secret Handshake},
  journal = {Journal of Theoretical Biology},
  year = {1990},
  volume = {144},
  pages = {379-396}
}

@ARTICLE{rSel:nessaac,
  author = {Reinhard Selten},
  title = {A Note on Evolutionarily Stable Strategies in Asymmetric Animal Conflicts},
  journal = {Journal of Theoretical Biology},
  year = {1980},
  volume = {84},
  pages = {93-101}
}

@ARTICLE{rSel:eleb,
  author = {Reinhard Selten},
  title = {Evolution, Learning, and Economic Behavior},
  journal = {Games and Economic Behavior},
  year = {1991},
  volume = {3},
  pages = {3-24}
}

@ARTICLE{pTay:essttp,
  author = {Peter D. Taylor},
  title = {Evolutionarily Stable Strategies with Two Types of Player},
  journal = {Journal of Applied Probability},
  year = {1979},
  volume = {16},
  pages = {76-83}
}

@ARTICLE{ysTu-wtJuang:epmp,
       author = {Yu-Sung Tu and Wei-Torng Juang},
        title = {Evolution of Preferences in Multiple Populations},
      journal = {International Journal of Game Theory},
         year = {2024},
       volume = {53},
        pages = {211-259}
}

@ARTICLE{mvV:rii,
  author = {van Veelen, Matthijs},
  title = {Robustness against Indirect Invasions},
  journal = {Games and Economic Behavior},
  year = {2012},
  volume = {74},
  pages = {382-393}
}

@ARTICLE{2017arXiv170501454T,
       author = {Yu-Sung Tu and Wei-Torng Juang},
        title = {The Payoff Region of a Strategic Game and Its Extreme Points},
      journal = {arXiv e-prints},
         year = {2018},
archivePrefix = {arXiv},
       eprint = {1705.01454},
 primaryClass = {cs.GT},
          url = {https://arxiv.org/abs/1705.01454}
}

@BOOK{rCre:scegt,
  AUTHOR =    {Ross Cressman},
  TITLE =     {The Stability Concept of Evolutionary Game Theory: A Dynamic Approach},
  PUBLISHER = {Springer-Verlag Berlin Heidelberg},
  YEAR =      {1992},
  series =    {Lecture Notes in Biomathematics},
  VOLUME =    {94}
}

@BOOK{jMay:etg,
   AUTHOR = {Maynard Smith, John},
    TITLE = {Evolution and the Theory of Games},
publisher = {Cambridge University Press},
     YEAR = {1982}
}

@BOOK{wSan:pged,
  author = {William H. Sandholm},
  title = {Population Games and Evolutionary Dynamics},
  publisher = {The MIT Press},
  year = {2010}
}

@BOOK{jWei:egt,
  author={J{\"{o}}rgen W. Weibull},
  title={Evolutionary Game Theory},
  publisher={The MIT Press},
  year={1995}
}

@BOOK{svW:eneup,
  author={von Widekind, Sven},
  title={Evolution of Non-Expected Utility Preferences},
  publisher={Springer-Verlag Berlin Heidelberg},
  year={2008}
}

@INCOLLECTION{rAum:apgcng,
   AUTHOR = {Robert J. Aumann},
   EDITOR = {A. W. Tucker and R. D. Luce},
    TITLE = {Acceptable Points in General Cooperative $n$-Person Games},
BOOKTITLE = {Contributions to the Theory of Games, Volume IV},
    PAGES = {287--324},
PUBLISHER = {Princeton University Press},
     YEAR = {1959}
}

@INCOLLECTION{wGut-mYaa:erbssg,
   AUTHOR = {Werner G\"{u}th and Menahem E. Yaari},
   EDITOR = {Ulrich Witt},
    TITLE = {Explaining Reciprocal Behavior in Simple Strategic Games: An Evolutionary Approach},
BOOKTITLE = {Explaining Process and Change: Approaches to Evolutionary Economics},
    PAGES = {23--34},
PUBLISHER = {The University of Michigan Press},
     YEAR = {1992}
}



\end{document}